\documentclass[nofootinbib,aps,twocolumn,superscriptaddress]{revtex4}
\usepackage{amsmath}
\usepackage{citesort}
\usepackage{epsfig}

\newcommand{\beqa}{\begin{eqnarray}}
\newcommand{\eeqa}{\end{eqnarray}}
\newcommand{\beq}{\begin{equation}}
\newcommand{\eeq}{\end{equation}}
\newcommand{\bal}{\begin{align}}
\newcommand{\eal}{\end{align}}
\newcommand{\Br}{{\rm Br}}
\renewcommand{\Re}{\operatorname{Re}}
\renewcommand{\Im}{\operatorname{Im}}
\newcommand{\Tr}{\operatorname{Tr}}

\def\sla{\negthinspace\not\negmedspace}
\def\gsim{\ \rlap{\raise 3pt \hbox{$>$}}{\lower 3pt \hbox{$\sim$}}\ }
\def\lsim{\ \rlap{\raise 3pt \hbox{$<$}}{\lower 3pt \hbox{$\sim$}}\ }

\begin{document}

\preprint{\vbox{
\hbox{}
\hbox{hep-ph/0601214}
\hbox{October 2005}
}}

\vspace*{18pt}

\title{Two body $B$ decays with isosinglet final states in SCET}

\def\addcmu{
Department of Physics, Carnegie Mellon University,
Pittsburgh, PA 15213}

\def\addIJS{J.~Stefan Institute, Jamova 39, P.O. Box 3000, 1001
Ljubljana, Slovenia}

\author{Alexander R. Williamson}\affiliation{\addcmu} 
\author{Jure Zupan} \affiliation{\addcmu}\affiliation{\addIJS}

\begin{abstract} \vspace*{18pt}
Expressions for decay amplitudes of $\bar B^0, B^-$ and $\bar B_s^0$ mesons
to two light pseudoscalar or vector mesons, including isosinglet mesons $\eta, \eta', \omega, \phi$, are obtained using
Soft Collinear Effective Theory at LO in $1/m_b$. These are then used to predict unmeasured branching ratios, direct and indirect CP asymmetries in $\bar B^0, B^-$ and $\bar B_s^0$ decays into two light pseudoscalars, following a determination of nonperturbative SCET parameters from existing data
using a $\chi^2$-fit. A separate discussion of indirect CP asymmetries in penguin dominated $\bar B^0\to\eta^{(')}K_{S,L},
\pi^0 K_{S,L}$ decays is given. 
\end{abstract}

\maketitle

\section{Introduction}
Hadronic two-body $B$ decays constitute both a probe of the electroweak structure of the Standard Model as well as a 
testing ground for our understanding of QCD dynamics. In $B\to M M'$ decays with two energetic
light final mesons $M,M'$ going back-to-back in opposite directions, the amplitude factorizes 
in the limit $m_b\gg \Lambda_{\rm QCD}, m_{M, M'}$.  This had made possible the great advancements in our theoretical understanding of $B\to M M'$ decays over the past several years \cite{Beneke:1999br,Szczepaniak:1990dt,Bauer:2000ew,Bauer:2004tj,Bauer:2005kd,Bauer:2005wb,Bauer:2001cu,Bauer:2002nz,Bauer:2001yt,Beneke:2001ev,Beneke:2002jn,Beneke:2003zv,Beneke:2004bn,Chay:2003zp,Chay:2003ju,Burrell:2001pf,Becher:2001hu,Neubert:2002ix,Keum:2000ph,Feldmann:2004mg,Buchalla:2004tw,Kagan:2004uw}. The observation of factorization initially relied 
on a 2-loop proof \cite{Beneke:1999br} (for earlier attempts see \cite{Szczepaniak:1990dt}). 
Following the advent of Soft Collinear Effective Theory (SCET)
\cite{Bauer:2000ew} the factorization was then shown to hold to all orders in $\alpha_S$ at leading order in $1/m_b$ 
\cite{Bauer:2001yt,Bauer:2002nz,Bauer:2001cu,Chay:2003zp,Chay:2003ju,Bauer:2004tj},  
with a possible exception of the contributions
from intermediate charm quark states \cite{Bauer:2004tj,Bauer:2005kd,Bauer:2005wb,Ciuchini:1997hb}. 

In this paper we present the first SCET analysis 
of two body $B$ decays  where one of the final
mesons contains an isospin singlet admixture to the wave function. This includes all the decays with 
$\eta,\eta'$ or $\phi, \omega$ mesons in the final state. As we will show the factorization  in the limit $m_b\to \infty$
which was obtained for nonisosinglet final states in Refs. \cite{Chay:2003zp,Chay:2003ju,Bauer:2004tj,Bauer:2005kd}, generalizes to the case of isospin singlet final states.
The expression for the $B\to M M'$ amplitude is then schematically
\beq\label{schem}
A \propto 
T_{\zeta}\otimes \phi_{M} \otimes \zeta^{BM'}+ T_{J} \otimes \phi_{M} \otimes \zeta_J^{BM'} 
  +\dots,
\eeq
with $\otimes$ denoting the convolutions over momenta fractions, $\phi_{M}$ the light-cone distribution amplitude (LCDA) of a meson that does not absorb the spectator quark, $\zeta^{BM'}$ the soft overlap function,   
$\zeta_J^{BM'}$ the function describing the completely factorizable contribution, and $T_{\zeta,J}$ the corresponding hard scattering
kernels,
while the ellipsis denotes the charming penguin contributions.

Eq.~\eqref{schem} already contains additional contributions arising from purely gluonic configurations that are specific to decays into pseudoscalar isosinglet states. 
These for instance 
lead to new jet functions in the SCET$_{\rm I}\to$SCET$_{\rm II}$ matching 
in addition to the ones found in \cite{Bauer:2004tj}, so that both $\zeta^{B\eta^{(')}}$ and  
$\zeta_J^{B\eta^{(')}}$ receive leading order gluonic contributions.
The phenomenological importance of $\eta'$ gluonic  content was 
emphasized  by a number of authors \cite{Atwood:1997bn,Hou:1997wy,Fritzsch:1997ps} after a discovery of surprisingly 
large branching ratio $\sim 10^{-4}$ for semi-inclusive $B\to \eta'X$ decays some years ago
\cite{Browder:1998yb}. These 
expectations are at least partially confirmed by our SCET analysis.  
The gluonic contributions to the amplitude where the spectator light quark in the $B$ meson is annihilated in the 
weak vertex are of leading order in the $1/m_b$ and $\alpha_S(m_b)$ expansions. This result also confirms the discussion of
gluonic contributions presented in the 
QCD factorization calculation of $B\to \eta' K$ modes \cite{Beneke:2002jn,Beneke:2003zv}, 
where a similar contribution was taken into account as part of $B\to \eta'$ form factor.
Unlike the authors of Ref. \cite{Beneke:2002jn,Beneke:2003zv}, which were forced to 
assign a rather arbitrary size of 
either $0\%$ or $\sim 40\%$
for the gluonic contribution to the $B\to \eta'$ form factor, we are able to use the wealth of new data from the $B$-factories
and fit for the corresponding SCET nonperturbative parameters. 

Following \cite{Bauer:2004tj} we do not expand the jet functions in terms of $\alpha_S(\sqrt{\Lambda m_b})$
and treat the corresponding functions as nonperturbative parameters that are determined from data
along with the charming penguins. 
Although BBNS \cite{Beneke:1999br,Beneke:2001ev,Beneke:2004bn} have argued that the 
charming penguins are perturbatively calculable, the more conservative approach of determining the charming penguins
from data avoids any inherent uncertainties associated with possibly leading order long distance effects \cite{Bauer:2004tj,Bauer:2005kd,Bauer:2005wb,Ciuchini:1997hb}.
Taking a conservative approach is especially important, if one aims at interpreting hints of beyond the Standard Model
contributions to the observables in $\Delta S=1$ processes: the $\pi K$ puzzle \cite{Gronau:2003kj,Buras:2003yc,Baek:2004rp} 
and the deviations of
$S$ parameters in penguin dominated modes $K_S\eta'$, $K_S\pi^0$, $K_S\phi$ from the naive expectation of 
$S\sim \sin 2\beta$ \cite{Grossman:1996ke,London:1997zk,Silvestrini:2005zb}. We devote the second part of our paper to these phenomenological considerations. 
In the numerical studies we assume that $1/m_b$ corrections are of typical size $\sim 20\%$ \cite{Bauer:2005wb}. 
 
In this paper we provide expressions for $B\to PP, PV$ and $VV$ decays at LO in the $1/m_b$ and $\alpha_S(m_b)$ expansions, while in the 
numerical estimates we restrict the analysis to the decays into two pseudoscalars. Similar numerical analyses
of $B\to \pi\pi, \pi K $ and $B\to K K$ decays using SCET were performed previously in Refs. \cite{Bauer:2004tj,Bauer:2005kd}. In addition to these modes we also discuss decays into final states with $\eta$ and $\eta'$ mesons, using a $\chi^2$-fit to extract the SCET parameters. Since in \cite{Bauer:2005kd} a subset of measured observables rather than a $\chi^2$-fit was used to fix the values of SCET parameters, the numerical results of the two analyses do not match exactly, but they do agree within the errors quoted. Furthermore SU(3) flavor symmetry for the SCET parameters is needed at present
due to limited data for decays into $\eta$ and $\eta'$ mesons.

Assuming only isospin symmetry, the amplitudes for $\Delta S=0$
decays with isosinglets in the final states, i.e. the decays $B\to \pi\eta^{(')}$ and $B\to \eta^{(')}\eta^{(')}$, 
are written in terms of eight new real nonperturbative SCET parameters beyond those describing the $B\to \pi\pi$ system (see Eqs.
\eqref{zetaetarel}-\eqref{AccRel} below)
at leading order in $1/m_b$.
 In total there are nineteen observables in $B\to \pi\eta^{(')}$ and $B\to \eta^{(')}\eta^{(')}$ decays, only four of which have been measured so far. Therefore, we are forced to use 
SU(3) flavor symmetry to reduce the number of unknowns. The situation is very similar for $\Delta S=1$ $B\to K\eta^{(')}$
decays with a total of ten observables, seven of which have been measured so far, while the amplitudes are
expressed in terms of eight new real parameters beyond the ones already present in $B\to \pi\pi$ and $B\to \pi K$ decays.
Using SU(3) flavor symmetry greatly reduces the number of parameters. In this limit all the $B\to PP$ decays
are described at LO in $1/m_b$ in terms of only eight real unknowns (four without isosinglet final states). This should be contrasted
with the conventional use of SU(3) decomposition that leads to eighteen reduced matrix elements (nine for each of the two independent CKM elements combination, see Appendix \ref{SU3decomposition}) and therefore to 
35 real unknowns and one unobservable overall phase. 
We thus choose to 
work in the SU(3) flavor limit and perform an analysis of all presently available data on $B\to PP$ decays
from which we predict the values of yet unmeasured observables in both $\Delta S=0$ and $\Delta S=1$
decays of $\bar B^0$, $B^-$ and $\bar B_s^0$.

The paper is organized as follows: in Section \ref{twobody} the matching of QCD$\to$ SCET$_{\rm I}\to$SCET$_{\rm II}$ and the resulting
amplitudes for two-body decays into light pseudoscalar and vector mesons (including decays into flavor singlet mesons) at leading order in $1/m_b$ 
are derived. The phenomenological implications of these results for $B^-, \bar B^0$ and $\bar B_s^0$ decays into two light pseudoscalar mesons are then developed in Section \ref{pheno}. Notation used throughout the paper is collected in Appendix \ref{Notation}, 
while
Appendix \ref{app:jet} deals with the Dirac structure of the operators multiplying jet functions. Finally, in Appendix
\ref{SU3decomposition} our results are rewritten in terms of SU(3) reduced matrix elements and a translation to 
the diagrammatic notation is made.
 
\section{Two-body $B$ decays in SCET}\label{twobody}
The starting point is the effective weak Hamiltonian at the scale $\mu \sim m_b$ for the $\Delta S=1$ two body $B$
decays \cite{Buchalla:1995vs}
\beq\label{HW}
H_W=\frac{G_F}{\sqrt{2}}\sum_{p=u,c}\lambda_p^{(s)} \Big(C_1 O_1^p+C_2 O_2^p+\sum_{i=3}^{10, 7\gamma, 8g}C_iO_i\Big),
\eeq
where the CKM factors are $\lambda_p^{(s)}=V_{pb} V_{ps}^*$ and the standard basis of four-quark operators is
\beq\label{Oi}
\begin{split}
O_1^p=(\bar pb)_{} (\bar s p)_{-}, \quad &O_2^p=(\bar p_\beta b_\alpha)_{} (\bar s_\alpha p_\beta)_{-},
\\
O_{3,5}=(\bar s b)_{} (\bar q q)_{\mp },\quad &O_{4,6}=(\bar s_\alpha b_\beta)_{} (\bar q_\beta q_\alpha)_{\mp},
\\
O_{7,9}=\frac{3e_q}{2}(\bar s b)_{} (\bar q q)_{\pm },\quad
&O_{8,10}=\frac{3e_q}{2}(\bar s_\alpha b_\beta)_{} (\bar q_\beta q_\alpha)_{\pm},
\end{split}
\eeq
with the abbreviation $(\bar q_1\gamma^\mu(1-\gamma_5) q_2)(\bar q_3\gamma^\mu(1\mp\gamma_5) q_4)\equiv (\bar q_1 q_2)
(\bar q_3 q_4)_{\mp}$. The color indices $\alpha, \beta$  are displayed only when the sum is over fields in 
different brackets. In the definition of the penguin operators $O_{3-10}$ in \eqref{Oi}
there is also an implicit sum over $q=\{u,d,s,c,b\}$. The electromagnetic and chromomagnetic operators are
\beq\label{magnetic}
O_{\{7\gamma,8g\}}=-\frac{m_b}{4\pi^2}\bar s \sigma^{\mu\nu}\{eF_{\mu\nu},g G_{\mu\nu}\}P_R b.
\eeq
The weak Hamiltonian for $\Delta S=0$ decays is obtained from \eqref{HW}-\eqref{magnetic}
 through the replacement $s\to d$.
 We will be working to leading order in $\alpha_S(m_b)$ so the values for the Wilson coefficients in \eqref{HW} are given
at LL order in the NDR scheme for $\alpha_S(m_Z)=0.119$, $\alpha^{\rm em}=1/128$, $m_t=174.3$, even though NLL values are available \cite{Buchalla:1995vs}.
At the scale $\mu=m_b=4.8$ GeV the Wilson 
coefficients $C_i$ for tree and QCD penguin operators \eqref{HW} are  
\cite{Buchalla:1995vs}
\beq
\begin{split}\label{Cvalues}
C_{1-6}(m_b)=&\{1.110,-0.253,0.011,-0.026,0.008,-0.032\},
\end{split}
\eeq
while for electroweak penguin (EWP) operators \cite{Buchalla:1995vs,Beneke:2001ev}
\beq
C_{7-10}(m_b)=\{0.09, 0.24, -10.3,2.2\}\times 10^{-3},
\eeq
and for the magnetic operators $C_{7\gamma}(m_b)=-0.315$, $C_{8g}(m_b)=-0.149$.

The effective weak Hamiltonian \eqref{HW} of full QCD is matched to the corresponding weak Hamiltonian in SCET.
In the two-body $B\to M_1 M_2$ decays there are three distinct scales, 
the hard scale $\sim m_b$ due to the energy available
to the decay products in the  $B$ rest frame, the typical hadronic soft scale $\Lambda$, and the 
hard collinear scale $\sqrt{m_b \Lambda}$. This last scale corresponds to the typical momentum transfer needed to boost the spectator
quark in the $B$ meson with soft momentum $k\sim \Lambda$ so that it ends up in the final meson $M_1$ with a momentum
$p\sim m_b(\lambda^2, 1, \lambda)$, where $\lambda=\Lambda/m_b$. The notation here is 
$p^\mu=(n\cdot p, \bar n \cdot p, p_\perp)$,
with the meson $M_1$ containing the spectator quark going
 in the $n^\mu=(1,0,0,-1)$ direction, while $M_2$ goes in the opposite $\bar n^\mu=(1,0,0,1)$ direction. 
The presence of three distinct scales leads to a two-step matching procedure \cite{Bauer:2002aj}. 
First hard scale $m_b$ is integrated
out so that  the effective weak Hamiltonian \eqref{HW} in full QCD
is matched to the Hamiltonian in SCET$_{\rm I}$ where the scaling of jet momenta in $n$ and $\bar n$ directions are 
$m_b(\lambda, 1, \sqrt \lambda)$  and $m_b(1, \lambda, \sqrt \lambda)$ respectively. This is then matched to 
a set of nonlocal operators in SCET$_{\rm II}$ with jet functions as Wilson coefficients \cite{Bauer:2002aj}. In the 
following two subsections we closely follow the matching at LO in $1/m_b$ performed in Ref. \cite{Bauer:2004tj}, extending it at the same time to a larger set of operators including the operators that contribute only for the isosinglet final states.

\subsection{Matching to SCET$_{\rm I}$}
\begin{figure}
\begin{center}
\epsfig{file=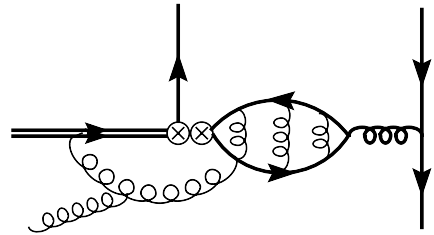, width=5.5cm}
\caption{The configuration with intermediate on-shell charm quarks annihilating into two collinear quarks
going in the opposite directions can lead to nonperturbative long distance contributions due to
soft gluon exchanges \cite{Bauer:2004tj}. The double line
denotes heavy quark, spectator quark is not shown.} \label{ggs}
\end{center}
\end{figure}

To have a complete set of leading order contributions in the second step, we keep in the ${\rm SCET}_I$
Hamiltonian \cite{Bauer:2004tj}
\beq\label{HWSCETI}
\begin{split}
H_W=\frac{2G_F}{\sqrt{2}}\sum_{n,\bar n}\Big\{&
\sum_i\int\big[d\omega_j\big]_{j=1}^3 c_i^{(f)}(\omega_j)Q_{if}^{(0)}(\omega_j)\\
+\sum_i\int[d&\omega_j]_{j=1}^4 b_i^{(f)}(\omega_j)Q_{if}^{(1)}(\omega_j)+Q_{\bar c c}\Big\},
\end{split}
\eeq
both the leading order operators $Q_{if}^{(0)}$ as well as a subset of relevant subleading operators $Q_{if}^{(1)}$ 
in the $\sqrt\lambda$ expansion, where $f=d$ $(f=s)$ for $\Delta S=0$ $(\Delta S=1)$ processes. In \eqref{HWSCETI} the  charming penguin contributions  \cite{Ciuchini:1997hb}
are isolated in the operator $Q_{\bar c c}$. Since $2 m_c\sim m_b$ the intermediate charm quarks annihilating 
into two collinear quarks (see Fig. \ref{ggs})
is a configuration where the intermediate on-shell charm quarks have 
a small relative velocity $v$ and can lead to long distance nonperturbative effects due to the exchange of soft gluons.
These contributions are $\alpha_S(2 m_c) f(2m_c/m_b) v$ 
parametrically suppressed \cite{Bauer:2004tj,Bauer:2005kd,Bauer:2005wb}, with $f(2m_c/m_b)$ a factor encoding that only in 
part of the phase 
space the charm quarks have small relative velocities. The view of BBNS \cite{Beneke:1999br,Beneke:2004bn} 
is that this phase space suppression of the threshold region is strong enough so that nonperturbative contributions are 
subleading. Bauer et al. \cite{Bauer:2004tj,Bauer:2005kd,Bauer:2005wb} on the other hand 
argue that since $2 m_c/m_b\sim O(1)$ then also $f(2m_c/m_b)\sim 1$.  
Sizable charming penguin contributions have also been found in recent light-cone calculation \cite{Khodjamirian:2003eq}.  
The most conservative approach is to introduce new unknown parameters describing charming penguin contributions that are 
then fit from data, so this is the approach we will follow. 

In \eqref{HWSCETI} the same notation as in Refs. \cite{Bauer:2004tj,Bauer:2005kd} is used. The set of leading order operators in this
notation is 
\begin{align}
Q_{1s}^{(0)}&=[\bar u_{n, \omega_1}\sla \bar n P_L b_v] [\bar s_{\bar n,\omega_2} \sla n P_L u_{\bar n, \omega_3}],
\nonumber
\\
Q_{2s,3s}^{(0)}&=[\bar s_{n, \omega_1}\sla \bar n P_L b_v] [\bar u_{\bar n,\omega_2} \sla n P_{L, R} u_{\bar n, \omega_3}],
\nonumber
\\
Q_{4s}^{(0)}&=[\bar q_{n, \omega_1}\sla \bar n P_L b_v] [\bar s_{\bar n,\omega_2} \sla n P_L q_{\bar n, \omega_3}],
\nonumber
\\
Q_{5s,6s}^{(0)}&=[\bar s_{n, \omega_1}\sla \bar n P_L b_v] [\bar q_{\bar n,\omega_2} \sla n P_{L, R} q_{\bar n, \omega_3}],
\label{QZero}
\end{align}
and the gluonic operator
\beq\label{QZeroGlue}
Q_{gs}^{(0)}= m_b[\bar s_{n, \omega_1}\sla \bar n P_L b_v]  \Tr[{\cal B}^{\perp \mu}_{\bar n, \omega_2}{\cal B}^{\perp \nu}_{\bar n, \omega_3}]i\epsilon_{\perp \mu\nu},
\eeq
where the trace is over color indices.  Above, an implicit summation over light quark flavors $q=\{u,d,s\}$ is understood. 
The operators $Q_{5s, 6s, gs}^{(0)}$ contribute only to isospin singlet mesons and were not needed in \cite{Bauer:2004tj,Bauer:2005kd}.
The quark fields in Equations \eqref{QZero} and \eqref{QZeroGlue} 
already contain the Wilson line together 
with the collinear quark field
\beq
q_{n,\omega}=[\delta(\omega -\bar n \cdot {\cal P}) W_n^\dagger \xi_n^{(q)}],
\eeq
with the usual bracket prescription that ${\cal P}$ operates only inside the square brackets \cite{Bauer:2000ew}. We also define a
 purely gluonic operator related to the $(\bar n, \perp)$ component of the gluon field strength
\beq\label{Bdef}
i g {\cal B}_{n, \omega}^{\perp \mu} =\frac{1}{(-\omega)} \big[W_n^\dagger [i \bar n \cdot D_{c,n},i D_{n, \perp}^\mu]W_n 
\delta(\omega -\bar n \cdot {\cal P}^\dagger)\big].
\eeq
 In \eqref{QZero} the operators
with $T^A\otimes T^A$ color structure were not listed, since they do not contribute to color singlet final states. Similarly, 
 a gluonic operator with $\epsilon_{\perp\mu\nu}\to g_{\perp \mu\nu}$ in \eqref{QZeroGlue}  is not considered, as it does not contribute to $P,V$ final states due to parity.  The $Q_{id}^{(0)}$ operators for the $\Delta S=0$ weak Hamiltonian are obtained from 
\eqref{QZero}, \eqref{QZeroGlue} with the replacement $s\to d$. 
\begin{figure}
\begin{center}
\epsfig{file=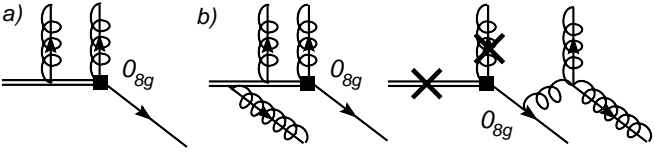, width=8.7cm}
\caption{
Matching to $Q_{gs}^{(0)}$ vanishes at LO in $1/m_b$ with a representative diagram given in a). In b) 
nonzero
diagrams that contribute at LO in $1/m_b$ to $Q_{gs}^{(1)}$ matching are given.  The square represents insertion of $O_{8g}$. The up- (down-) going wiggly-solid lines denote $\bar n$ ($n$) collinear gluons. The offshell gluon (wiggly line) in the third diagram can also attach to any of the cross vertices. } \label{08g}
\end{center}
\end{figure}

The relevant $O(\sqrt\lambda)$ operators are \cite{Bauer:2004tj}
\beq\label{Q1}
\begin{split}
Q_{1s}^{(1)}&=\frac{-2}{m_b}[\bar u_{n,\,\omega_1} i g 
\sla {\cal B}_{n,\,\omega_4}^\perp P_L b_v][\bar s_{\bar n, \,\omega_2}
\sla n P_L u_{\bar n, \,\omega_3}],
\\
Q_{2s,3s}^{(1)}&=\frac{-2}{m_b}[\bar s_{n,\,\omega_1} i g 
\sla {\cal B}_{n,\,\omega_4}^\perp P_L b_v][\bar u_{\bar n,\, \omega_2}
\sla n P_{L,R} u_{\bar n, \,\omega_3}],
\\
Q_{4s}^{(1)}&=\frac{-2}{m_b}[\bar q_{n,\,\omega_1} i g 
\sla {\cal B}_{n,\,\omega_4}^\perp P_L b_v][\bar s_{\bar n, \,\omega_2}
\sla n P_L q_{\bar n, \,\omega_3}],
\\
Q_{5s, 6s}^{(1)}&=\frac{-2}{m_b}[\bar s_{n,\,\omega_1} i g \sla {\cal B}_{n,\,\omega_4}^\perp P_L b_v]
[\bar q_{\bar n, \,\omega_2}
\sla n P_{L, R} q_{\bar n, \,\omega_3}],
\\
Q_{7s}^{(1)}&=\frac{-2}{m_b}[\bar u_{n,\,\omega_1} i g 
{\cal B}_{n,\,\omega_4}^{\perp\mu} P_L b_v][\bar s_{\bar n, \,\omega_2}
\sla n \gamma_\mu^\perp P_{R} u_{\bar n, \,\omega_3}],
\\
Q_{8s}^{(1)}&=\frac{-2}{m_b}[\bar q_{n,\,\omega_1} i g {\cal B}_{n,\,\omega_4}^{\perp\mu} P_L b_v]
[\bar s_{\bar n, \,\omega_2}
\sla n \gamma_\mu^\perp P_{R} q_{\bar n,\, \omega_3}],
\end{split}
\eeq
and the operator with additional $\bar n$ collinear gluon fields
\beq
\begin{split}\label{Q9s10s}
Q_{gs}^{(1)}&=-2[\bar s_{n,\,\omega_1} i g \sla {\cal B}_{n,\,\omega_4}^{\perp} P_L b_v] 
\Tr[{\cal B}^{\perp \mu}_{\bar n, \omega_2}{\cal B}^{\perp \nu}_{\bar n, \omega_3}]i\epsilon_{\perp \mu\nu},
\end{split}
\eeq
while, quite similarly to \eqref{QZeroGlue},  we do not consider the additional gluonic operator with $g_{\perp \mu\nu}$ instead of $\epsilon_{\perp \mu\nu}$ in \eqref{Q9s10s}, since it does not contribute to $P,V$ final states due to parity conservation. 
As before $Q_{id}^{(1)}$ are obtained by making the replacement $s\to d$ in the above definitions. 

\begin{figure}
\begin{center}
\epsfig{file=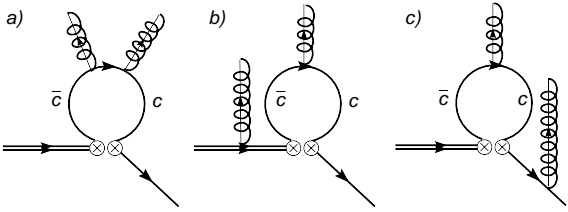, width=7.5cm}
\caption{The power suppressed contributions with off-shell charm quarks emitting collinear gluons in 
$\bar n$ direction.} \label{charm_gluon}
\end{center}
\end{figure}

The Wilson coefficients $c_{i}^{(f)}$ for the operators in \eqref{QZero} and $b_{1s,2s}^{(1)}$ for the operators in \eqref{Q1} are known to NLO in $\alpha_S(m_b)$ \cite{Beneke:1999br,Chay:2003ju,Beneke:2005vv}, while the rest of the Wilson coefficients $b_{i}^{(f)}$ for the operators in \eqref{Q1} are at present known only to leading order. 
In the phenomenological part of our paper, Section \ref{pheno}, we work at LO in the $\alpha_S(m_b)$ expansion. At this order there are no contributions from $Q_{gs}^{(0)}$ and $Q_{gs}^{(1)}$ operators \eqref{QZeroGlue}, \eqref{Q9s10s} since the tree level matching shown in Fig. \ref{08g} and given explicitly below, is already $\alpha_S(m_b)$ suppressed.

We also work at leading order in $\Lambda/m_b$. In particular this means that we systematically neglect the contributions to $B\to P\eta^{(')}$ amplitudes with two collinear gluons being emitted from the charm loop
(see Fig. \ref{charm_gluon} a)). 
Here the intermediate charm quarks are off-shell, unlike the charming penguin configuration, Fig.
\ref{ggs}. The diagram in Fig. \ref{charm_gluon} a) matches onto the $(\sqrt \lambda)^2$ suppressed operator
\beq\label{suppr}
[\bar s_{n, \omega_1}\sla \bar n P_L b_v]  \Tr[{\cal B}^{\perp \mu}_{\bar n, \omega_2} i \bar n\cdot D_{\bar n}{\cal B}^{\perp \nu}_{\bar n, \omega_3}]\epsilon_{\perp \mu\nu},
\eeq
with a Wilson coefficient that is a function of $2 m_c/m_b$ and can be found from results in \cite{Beneke:2002jn}. 
Similarly, the remaining diagrams of Fig. \ref{charm_gluon} with one
$\bar n$ gluon emitted from $b$ or $n$ collinear quark line are part of the matching onto at least $\sqrt{\lambda}$ suppressed operators.

This leaves us with the set of operators \eqref{QZero} and \eqref{Q1} that contribute at LO in $1/m_b$ and $\alpha_S(m_b)$. Note that the Wilson coefficients in \eqref{HWSCETI} already contain the CKM elements. Furthermore, the basis of operators
\eqref{QZero} and \eqref{Q1} is a minimal choice, where the operator relation 
$O_{9,10}=\frac{3}{2}[O_{2,1}^u+O_{2,1}^c] -\frac{1}{2} O_{3,4}$, valid after off-shell $b$ quarks are integrated out, was utilized. This leads to the following 
  tree level matching for the four-quark operators
\beq
\begin{split}\label{cmatch}
c_{1,2}^{(f)}&=\lambda_u^{(f)}\Big[C_{1,2}+\frac{1}{N}C_{2,1}\Big]-\lambda_t^{(f)}\frac{3}{2}\Big[\frac{1}{N}C_{9,10}
+C_{10,9}\Big],\\
c_3^{(f)}&=-\frac{3}{2}\lambda_t^{(f)}\Big[C_7+\frac{1}{N}C_8\Big],\\
c_{4,5}^{(f)}&=-\lambda_t^{(f)}\Big[\frac{1}{N}C_{3,4}+C_{4,3}-\frac{1}{2N}C_{9,10}-\frac{1}{2}C_{10,9}\Big],\\
c_6^{(f)}&=-\lambda_t^{(f)}\Big[C_5+\frac{1}{N}C_6-\frac{1}{2}C_7-\frac{1}{2N}C_8\Big].
\end{split}
\eeq
while NLO matching can be obtained from \cite{Beneke:1999br,Chay:2003ju}.
The tree level matching of the dimension seven operators leads to
\begin{align}
\begin{split}
b_{1,2}^{(f)}&=\lambda_u^{(f)}\Big[C_{1,2}+\frac{1}{N}\Big(1-\frac{m_b}{\omega_3}\Big)C_{2,1}\Big]-
\\
&\qquad \qquad\lambda_t^{(f)}\frac{3}{2}\Big[C_{10,9}+\frac{1}{N}\Big(1-\frac{m_b}{\omega_3}\Big)C_{9,10}\Big],
\end{split}
\nonumber
\\
b_3^{(f)}&=-\lambda_t^{(f)}\frac{3}{2}\Big[C_7+\Big(1-\frac{m_b}{\omega_2}\Big)\frac{1}{N}C_8\Big],
\nonumber
\\
\begin{split}
b_{4,5}^{(f)}&=-\lambda_t^{(f)}\Big[C_{4,3}+\frac{1}{N}\Big(1-\frac{m_b}{\omega_3}\Big)C_{3,4}\Big]+\\
&\qquad \qquad\lambda_t^{(f)}\frac{1}{2}\Big[C_{10,9}+\frac{1}{N}\Big(1-\frac{m_b}{\omega_3}\Big)C_{9,10}\Big],
\end{split}\nonumber
\\
\begin{split}
b_6^{(f)}&=-\lambda_t^{(f)}\Big[C_5+\frac{1}{N}\Big(1-\frac{m_b}{\omega_2}\Big)C_6\Big]+\\
&\qquad\qquad\lambda_t^{(f)}\frac{1}{2}\Big[C_{7}+\frac{1}{N}\Big(1-\frac{m_b}{\omega_2}\Big)C_{8}\Big],
\end{split}\nonumber
\\
b_7^{(f)}&=-\lambda_t^{(f)}\Big(C_5-\frac{1}{2}C_7\Big)\frac{1}{N}\Big(\frac{m_b}{\omega_1}-
\frac{1}{2}\frac{m_b}{\omega_3}\Big),
\nonumber
\\
b_8^{(f)}&=-\lambda_t^{(f)}\frac{3}{2}C_7\frac{1}{N}\Big(\frac{m_b}{\omega_1}-\frac{1}{2}\frac{m_b}{\omega_3}\Big),\label{b8}
\end{align}
while the NLO contributions from tree operators only was recently obtained in  \cite{Beneke:2005vv}.

For completeness we also list the result of tree level matching for $Q_{gs}^{(0)}$ and $Q_{gs}^{(1)}$ operators \eqref{QZeroGlue}, \eqref{Q9s10s} following from the diagrams in Fig. \ref{08g}
\beq\label{cg}
c_{g}^{(f)}=0,
\eeq
and 
\begin{align}
\begin{split}\label{bg}
b_{g}^{(f)}=&\lambda_t^{(f)} C_{8g} \frac{\alpha_S(m_b)}{16C_F} \Big(\frac{1}{\bar u}-\frac{1}{u}\Big)\Big[ \frac{2+z}{1-z}\\
&+2\left(1-\frac{1}{N^2}\right)\frac{ u\bar u}{(1-z u)(1-z \bar u)} \Big],
\end{split}
\end{align}
where  $z=\omega_1/m_b$ and $u=\omega_3/m_b$, with $\omega_{1,3}$ the fraction momenta in \eqref{Q9s10s}, while 
$C_F=(N^2-1)/2N$ and $N=3$. Even though 
the matching result \eqref{cg}, \eqref{bg}, which is already NLO in $\alpha_S(m_b)$, will not be needed for our phenomenological discussions in Section \ref{pheno}, it does 
represent first nonzero contribution from $O_{8g}$ weak operator to the matching. Since this chromomagnetic operator may be enhanced in new physics models  (in MSSM this is due to the fact that the chirality flip can be performed on gluino line instead on the $b$ line \cite{Silvestrini:2005zb,Khalil:2002fm}),
we provide the matching as an important input for future studies of such new physics effects in two body $B$ decays using SCET.

\begin{figure*}
\begin{center}
\epsfig{file=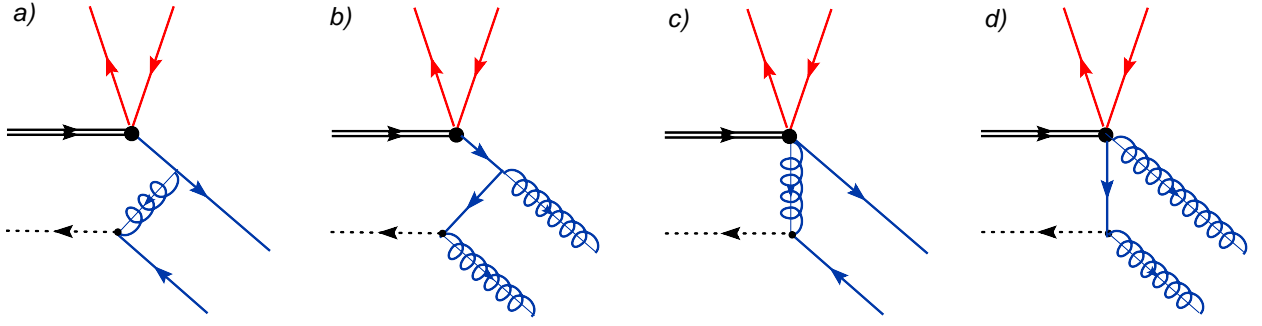}
\caption{The T-products of weak Hamiltonian operators (large black dot) with ${\cal L}_{\xi_n q}^{(1)}$ \eqref{Lsoftcoll} that boosts soft quark (dashed line) to $n$ collinear quark (blue/dark gray
 solid line). Two 
representatives of soft overlap contribution diagrams \eqref{T1i} 
with a) outgoing $n$ collinear quarks and
b) outgoing $n$ collinear gluons are shown. Diagrams c) and d) give factorizable contributions corresponding to \eqref{T2i}. The red/light gray
 solid lines denote $\bar n$ collinear quarks, the blue/dark gray 
wiggly and solid lines the $n$ collinear gluons, while double lines denote
heavy quarks. The diagrams b) and d) contribute only to $\eta,\eta'$ going in $n$ direction. 
The $\alpha_S(m_b)$ suppressed diagrams with $\bar n$ collinear gluons are not shown.
} \label{matching}
\end{center}
\end{figure*}

\subsection{Matching to SCET$_{\rm II}$}
The matching 
of SCET$_{\rm I}$ onto SCET$_{\rm II}$ is performed by integrating out the degrees of freedom with $p^2\sim \Lambda m_b$. To do so  
it is useful to perform a 
redefinition of fields $\xi_{\tilde n}\to Y_{\tilde n }\xi_{\tilde n}$, $A_{\tilde n}\to
Y_{\tilde n} A_{\tilde n} Y_{\tilde n}^\dagger$, where $\tilde n$ denotes either $n$ or $\bar n$ direction, while
$Y_{\tilde n}$ are Wilson lines of $\tilde n \cdot A_{us}$ SCET$_{\rm I}$ usoft gluon fields. After the redefinition
the usoft fields decouple from  collinear fields in the leading order SCET$_{\rm I}$ Lagrangian \cite{Bauer:2002nz,Bauer:2004tj,Bauer:2001yt}. They still appear in
the operators $Q_{is}\in \{Q_{1s}^{(0)},\dots , 
Q_{6s}^{(0)},Q_{gs}^{(0)}, Q_{1s}^{(1)}, \dots , Q_{8s}^{(1)}, Q_{gs}^{(1)}\}$, but only in the combination $Y_{n}^\dagger b_v$. Thus $n$ collinear and $\bar n$ collinear sectors decouple \cite{Bauer:2002nz} and the operators $Q_{is}$ factor into\footnote{Operators $Q_{4s}^{(0)}, Q_{4s, 8s}^{(1)}$ factorize into a sum
$\sum_q Q_{q,is}^n Q_{q,is}^{\bar n}$, but we suppress this sum in the notation.}
 \cite{Bauer:2004tj}
\beq\label{nnbar}
Q_{is}=Q_{is}^n Q_{is}^{\bar n}.
\eeq
Since the $n$ and $\bar n$ operators do not communicate through usoft gluons, we can focus only on the matching of $Q_{is}^n$ operators onto 
SCET$_{\rm II}$ operators. The soft spectator quark in the $B$ meson is boosted to a collinear quark in the $n-$direction through an application of the $O(\sqrt{\lambda})$ suppressed Lagrangian 
\cite{Beneke:2002ph}
\beq\label{Lsoftcoll}
{\cal L}_{\xi_n q}^{(1)}=\bar q_{us} Y_n i g \sla {\cal B}_n^\perp W_n^\dagger \xi_n + h.c.,
\eeq
in the $T$ products  \cite{Bauer:2004tj} 
(see also Fig. \ref{matching})
\beq\label{T1i}
\begin{split}
T_{1,if}&=\int d^4y \;d^4 y' \;T[Q_{if}^{n (0)}(0), i {\cal L}_{\xi_n\xi_n}^{(1)}(y')+i {\cal L}_{cg}^{(1)}(y'),\\
&\qquad\quad i {\cal L}_{\xi_n q}^{(1)} (y)]+\int d^4 y T[Q_{if}^{n (0)}(0), i {\cal L}_{\xi_n q}^{(1,2)} (y)],
\end{split}
\eeq
and 
\beq\label{T2i}
T_{2,if}=\int d^4y T[Q_{if}^{n (1)}(0), i {\cal L}_{\xi_n q}^{(1)} (y)].
\eeq
The explicit forms of subleading SCET Lagrangians  ${\cal L}_{\xi_n\xi_n}^{(1)}$, 
${\cal L}_{cg}^{(1)}$, ${\cal L}_{\xi_n q}^{(2)}$  can 
be found in \cite{Bauer:2002aj}. Note that all the terms in $T_{1,if}$ and $T_{2,if}$ have the same $\sqrt\lambda$ scaling.
The superficially enhanced $T$ product $T[Q_{if}^{n (0)}(0), i {\cal L}_{\xi_n q}^{(1)} (y)]$ gets suppressed
once the external momenta are restricted to SCET$_{\rm II}$ scaling, since it involves  an odd number of $D_c^\perp$\cite{Bauer:2002aj}. 

The matrix elements $\langle M|T_{1,if}|B\rangle$ lead to
endpoint singularities, signaling a presence of soft-overlap contributions both for nonisosinglet final states \cite{Bauer:2002aj,Beneke:2003pa,Beneke:2000wa,Beneke:2005gs,Hill:2004if,Lange:2003pk,Chay:2002vy,Beneke:2004rc,Pirjol:2002km} as well as for isosinglet $M$. Following \cite{Bauer:2004tj}
we therefore define the matrix elements of $T_{1,if}$ as new nonperturbative 
functions to be determined from data
\beq\label{zeta}
\langle M |T_{1,if}|B\rangle=C_{if}^{BM} \zeta^{BM}.
\eeq
The final meson 
$M$ can be either $P$, or $V_\parallel$, while $\langle V_\perp |T_{1,if}|B\rangle$ vanish at leading order \cite{Bauer:2004tj}. 
The coefficients $C_{if}^{BM}=1, \pm 1/\sqrt{2}$ describe the flavor content of the final meson.

To facilitate the calculation of $T_{2,if}$ the $Q_{if}^{n(1)}$ operators are stripped of the soft quark fields 
$(\gamma^\alpha Y_n^\dagger b_v)^{ia}$ ($(Y_n^\dagger b_v)^{ia}$ in the case of $Q_{7s,8s}^{n(1)}$) and similarly ${\cal L}_{\xi_n q}^{(1)}$ 
is stripped of $(\bar q_{us}' Y_n)^{jb}$. The common matching result is then (see appendix \ref{app:jet}) 
\beq\label{jet-funcs}
\begin{split}
T&\big[(\bar \xi_n^{(q)}W_n)_{z \omega}i g {\cal B}_{n,\,\bar z \omega}^{\perp\,\alpha}
P_{R,L}\big]^{ia}(0)\big[i g\sla{\cal B}_n^\perp W_n^\dagger \xi_n^{(q')}\big]_0^{jb}(y)=
\\
&i\delta^{ab}\delta(y_+)\delta^{(2)}(y_\perp)\int_0^1 dx \int \frac{dk_+}{2\pi}e^{ik_+y_-/2}
\\
\times \Big\{
&\frac{1}{\omega}
J(z, x, k_+)\big({\sla n}P_{L,R}\gamma_\perp^\alpha\big)^{ji}\big[\bar 
q_{n,\, x\omega}\sla \bar nP_{L,R}\;q'_{n,\, -\bar x\omega}\big]\\
-&\frac{1}{\omega}
J_\perp(z, x, k_+)\Big(\frac{\;\sla n}{2}P_{R,L}\gamma_\perp^\alpha\gamma_\perp^\beta\Big)^{ji}
\big[\bar q_{n,\, x\omega}\sla \bar n\gamma_\beta^\perp q'_{n,\, -\bar x\omega}\big]
\\
+&\delta_{q,q'}J_g(z, x, k_+)\big(\sla n P_{L,R}\gamma^\perp_\mu \big)^{ji}\Tr[ {\cal B}_{n, x\omega}^{\perp\mu}  
{\cal B}_{n,\bar x\omega}^{\perp\alpha}]\\
+&\delta_{q,q'}J_g'(z, x, k_+)\big(  \sla n P_{L,R}\gamma_\perp^\alpha\big)^{ji}
\Tr[{\cal B}_{n, x\omega}^\perp \cdot 
{\cal B}_{n,\bar x\omega}^\perp]\Big\}+\dots,
\end{split}
\eeq
where the trace is over color indices, while ellipses denote operators with nontrivial color structure that do not contribute to color singlet initial state. This generalizes the result of Ref. \cite{Bauer:2004tj} to the case of 
isosinglet final state, where also the operators with two interpolating gluon fields contribute, leading to two additional jet functions $J_g(z,x,k_+)$ and $J_g'(z,x,k_+)$.
At tree level 
$J(z,x,k_+)=J_\perp(z,x,k_+)=\delta(z-x)\alpha_S \pi C_F/(N_c \bar x k_+)$, with 1-loop corrections calculated in \cite{Hill:2004if,Beneke:2005gs}, and 
$J_g(z,x,k_+)=\delta(z-x)\alpha_S 2 \pi/(N_c k_+)$, $J_g'(z,x,k_+)=0$. 

The matching of different $Q_{is}^{n(1)}$ operators is obtained from \eqref{jet-funcs} by multiplying with 
$(P_R\gamma^\alpha Y_n^\dagger b_v)^{ia}$ (by $(P_L Y_n^\dagger b_v)^{ia}$ in the case of $Q_{7s,8s}^{n(1)}$) and  with $(\bar q_{us}' Y_n)^{jb}$. The final result for $B\to M_1M_2$ amplitude 
depends on only two jet functions, $J$ and $J_g$. 
The operators multiplying  $J_g'$ do not contribute since the matrix elements of $\Tr[{\cal B}_{n, x\omega}^\perp \cdot 
{\cal B}_{n,\bar x\omega}^\perp]$ between vacuum and pseudoscalar meson or between vacuum and vector meson vanish due to parity conservation. That $J_\perp$ jet function does not appear in $B\to M_1M_2$ amplitudes when $M_{1,2}$ are nonisosinglet mesons was already shown in \cite{Bauer:2004tj}, and is true also when either of the two outgoing light mesons $M_{1,2}$ is an isosinglet meson. Namely, the operator multiplying 
$J_\perp$ 
cancels in the matching of 
$Q_{1s}^{n(1)}\dots Q_{6s}^{n(1)}, Q_{gs}^{n(1)}$ operators
because $\gamma_\perp^\alpha \gamma_\perp^\beta \gamma_{\perp\alpha}=0$. It also cancels
in the matching of the $Q_{7s, 8s}^{n(1)}$ operators, since  
$\sla n P_L \gamma_\perp^\alpha\gamma_\perp^\beta=(g_\perp^{\alpha\beta}
+i\epsilon_\perp^{\alpha\beta})\sla n P_L$ 
leads to a vanishing term
\beq
(g_\perp^{\alpha\beta}
+i\epsilon_\perp^{\alpha\beta})[\bar q_{\bar n} \sla n P_L \gamma_\alpha^\perp q'_{\bar n}]=0,
\eeq
where for the last equality the relation $\gamma^\alpha_\perp \gamma_\perp^\beta \gamma_{\perp\alpha}=0$ was used once again. 
Furthermore, matching of the $Q_{7s, 8s}^{n(1)}$ operators leads to a soft operator 
$[\bar q_{us} {\sla n}P_{R}\gamma_\perp^\alpha b_v]$ multiplying $J,J_g$ or $J_g'$ jet functions. 
The matrix element of this operator between vacuum and pseudoscalar $|B\rangle$ 
states vanishes. Thus $Q_{7s, 8s}^{n(1)}$ operators contribute only to $B^*$ and not at all to $B\to M_1 M_2$ decays. 
 
\begin{figure}
\begin{center}
\epsfig{file=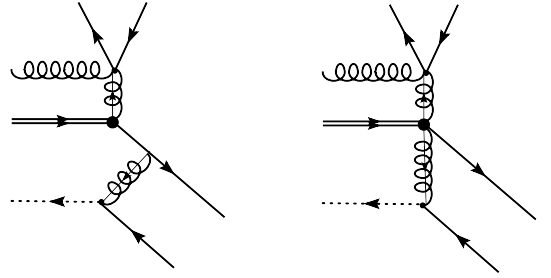, width=7.cm}
\caption{The SCET$_{\rm I}\to$SCET$_{\rm II}$ matching onto operators contributing only to $B^*$ decays with isosinglet mesons in $\bar n$ direction. The weak SCET$_{\rm I}$ operator arising at tree level from $O_{8g}$ insertion (large black dot) is enhanced due to one less $\bar n$ gluon field compared to operators in Fig. \ref{matching}, which is compensated by
additional soft gluon in the initial state. Additional diagrams similar to Fig. \ref{matching} b), d) are
possible for isosinglet final state in $n$ direction.} \label{Bstar_contribs}
\end{center}
\end{figure}

We note on passing that $B^*$ decays into two light mesons would also receive, at leading order in $1/m_b$, contributions from non-valence Fock state of the B meson with an additional soft gluon, i.e. from the configurations with incoming heavy quark, soft spectator quark and soft gluon. The SCET$_{\rm I}\to$SCET$_{\rm II}$ matching is shown on 
Fig. \ref{Bstar_contribs}. Since the 
weak SCET$_{\rm I}$ operators are $1/\sqrt \lambda$ enhanced compared to $T_{1,if}, T_{2,if}$ (the $T$ products in $n$ direction are the same as in \eqref{T1i}, \eqref{T2i}, while in $\bar n$ direction the weak operator has only one ${\cal B}_{\bar n}^\perp$ field), this allows for another external $A_{us}$ field. Due to Dirac structure these operators 
do not contribute to $B$ decays, quite similar to equivalent contributions found in $B^*\to P\gamma$ \cite{Becher:2005fg}, with $P$ an isosinglet.
Note that these contributions would not be present in $B^*\to M$ form factors. 

We are now ready to write down the final result for the $B\to M_1 M_2$ amplitude at LO in $1/m_b$ 
in SCET that will extend the result of Ref \cite{Bauer:2004tj} to the decays including isosinglet mesons. This result exhibits two levels of factorization: 
the $\bar n$ direction collinear modes decouple from the rest, and for the $T_{2,if}$ term the 
$n$ collinear fields decouple from soft fields.  Because of the factorization, the amplitude has a very simple form 
\eqref{schem} expressible in terms of several universal nonperturbative functions such as the $\zeta^{BM}$ \eqref{zeta} function
and the  twist-2 light-cone 
distribution amplitudes (LCDA) \cite{Lepage:1980fj} that are for pseudoscalar final state defined through (see e.g. \cite{Hardmeier:2003ig})
\beq
\begin{split}\label{pseudoscalar}
\langle P^a(p)|\bar q_{n,\omega_1} \lambda^a \sla \bar n \gamma_5 q'_{n, \omega_2}|0\rangle&=
- i f_P 2 E \phi^{P^a}(u),
\end{split}
\eeq
with $M=\lambda^a P^{a\dagger}$ a $3\times 3$ matrix of light pseudoscalar fields $P^a$ given in \eqref{M},  
and for the longitudinally polarized vector meson final state through
\beq
\begin{split}
\langle V_\parallel^a(p)|\bar q_{n,\omega_1} \lambda^a \sla\bar n  q'_{n, \omega_2}|0\rangle&=
i f_V 2 E \phi^{V_\parallel^a}(u).
\end{split}
\eeq
Amplitudes for decays to transversely polarized vector mesons do not receive  leading order contributions from 
operators \eqref{QZero}, \eqref{Q1} (see also discussion around Eq. \eqref{gammaOp}).

\begin{table*}
\caption{Table of hard kernels for $\Delta S=0$ decays of $B^-, \bar B^0$ and $\bar B^0_s$ into isosinglet
mesons, separated respectively by horizontal lines. The superscript $(d)$ in the coefficients $c_{i}^{(d)}$ is not displayed. $T_{1J}$ and $T_{2J}$ are obtained with the replacement
$c_i\to b_i$.}\label{table:zeta0}
\begin{ruledtabular}
\begin{tabular}{ccc} 
Mode  & $T_{1\zeta}$  & $T_{2\zeta}$\\ \hline
$\pi^-\eta_q, \rho^-\eta_q$ & $\frac{1}{\sqrt{2}}(c_{1}+c_{4})$ & ${\frac{1}{\sqrt{2}}(c_{2}-c_{3}+c_{4})}{+\sqrt{2}(c_{5}-c_{6})}$ 
\\
$\pi^-\eta_s, \rho^-\eta_s$ & $c_1+c_4$ & $c_{5}-c_{6}$  \\ 
$\pi^-\phi, \rho^-\phi$ & $0$ & $c_{5}+c_{6}$  \\ 
$\pi^-\omega, \rho^-\omega$ & $\frac{1}{\sqrt{2}}(c_{1}+c_{4})$ & $\frac{1}{\sqrt{2}}(c_{2}+c_{3}+c_{4})+\sqrt{2}(c_{5}+c_{6})$ 
\\
\hline
$\pi^0\eta_s$,$\rho^0\eta_s$ & $\frac{1}{\sqrt 2}(c_2\mp c_3-c_4)$ & $-\frac{1}{\sqrt{2}}(c_{5}-c_{6})$  \\ 
$\pi^0\eta_q$, $\rho^0\eta_q$ & $\frac{1}{2}(c_{2}\mp c_{3}-c_{4})$ & $-\frac{1}{2}(c_{2}-c_{3}+c_{4})-c_{5}+c_{6}$   \\ 
$\pi^0\phi, \rho^0\phi$ & $0$ & $-\frac{1}{\sqrt{2}}(c_{5}+c_{6})$  \\ 
$\pi^0\omega$,$\rho^0\omega$ & $\frac{1}{2}(c_{2}\mp c_{3}-c_{4})$ & $-\frac{1}{2}(c_{2}+c_{3}+c_{4})-c_{5}-c_{6}$  
\\
$\eta_q\phi, \omega\phi$ & $0$ & $\frac{1}{\sqrt{2}}(c_{5}+c_{6})$  \\ 
$\eta_q\omega$, $\omega\omega$ & $\frac{1}{2}(c_{2}\mp c_{3}+c_{4})+c_{5}\mp c_{6}$ & $\frac{1}{2}(c_{2}+c_{3}+c_{4})+c_{5}+c_{6}$ 
\\
$\eta_q\eta_q$ & $\frac{1}{2}(c_{2}-c_{3}+c_{4})+c_{5}-c_{6}$ & $\frac{1}{2}(c_{2}-c_{3}+c_{4})+c_{5}-c_{6}$ 
\\
$\eta_q\eta_s$ & $\frac{1}{\sqrt 2}(c_2-c_3+c_4)+\sqrt 2(c_5-c_6)$ & $\frac{1}{\sqrt{2}}(c_{5}-c_{6})$ 
\\
$\eta_s\eta_s$ & $c_5-c_6$ & $c_{5}-c_{6}$ 
\\
\hline
${K^0}^{(*)} \phi$, ${K^0}^{(*)}\eta_s$& $\{0, c_4\}$& $c_5\pm c_6$  \\ 
${K^0}^{(*)}\omega$, ${K^0}^{(*)}\eta_q $ & 
$\{0, \frac{1}{\sqrt{2}}\;c_4\}$ & $\frac{1}{\sqrt{2}}(c_{2}\pm c_{3}+c_{4})+\sqrt{2}(c_{5}\pm c_{6})$ 
\\ 
\end{tabular}
\end{ruledtabular}
\end{table*}

For the gluonic operator on the other hand \cite{Kroll:2002nt}
\beq\label{gluonicwave}
i\epsilon_{\perp \mu \nu} \langle P(p) | \Tr[{\cal B}_{n,\omega_1}^{\perp \mu} 
{\cal B}_{n,- \omega_2}^{\perp \nu}]|0\rangle=\frac{i}{4}\sqrt{C_F} f_P^1  \bar \Phi_P^g(u),
\eeq 
where $f_{\eta_q}^1=\sqrt{2/3}f_{\eta_q}$, $f_{\eta_s}^1=f_{\eta_s}/\sqrt{3}$ and $f_P^1=0$ for other pseudoscalars. Here $f_{\eta_q}$ and $f_{\eta_s}$ are the decay constants corresponding
to $\bar q q=(\bar u u+\bar d d)/\sqrt{2}$ and $\bar s s$ axial currents in \eqref{pseudoscalar} respectively.  The matrix elements of gluonic operators between vacuum and vector meson states vanish.
In all of the above definitions the integration $\int_0^1 du \delta(\omega_1 -u \bar n \cdot p)
\delta (\omega_2+\bar u \bar n \cdot p)$ on the r.h.s. is implicitly assumed. 
The LCDA $\bar \Phi_P^g(u)$ is related to the LCDA $\phi_{Pg}(u)$ used in \cite{Kroll:2002nt} through
\beq
\bar \Phi_P^g(u)=\frac{\phi_{Pg}(u)}{u (1-u)},
\eeq
and coincides with the definition in \cite{Blechman:2004vc}.
It has the symmetry property  $\bar \phi_P^g(u)=-\bar \phi_P^g(1-u)$, as can be easily seen from \eqref{gluonicwave}.  Finally, 
the relevant $B$ meson LCDA  is \cite{Grozin:1996pq,Beneke:2000wa}
\beq
\begin{split}
\langle 0 |(\bar q_s Y_n)|_{x_-} &\sla n (1-\gamma_5) (Y_n^\dagger b_v)|_0 |B\rangle =\\
&-i f_B m_B \int d r_+ e^{-i r_+ x_-/2}
\phi_B^+(r_+).
\end{split}
\eeq

Combining the SCET$_{\rm I}\to$ SCET$_{\rm II}$ matching \eqref{jet-funcs} with the definitions of the 
nonperturbative matrix elements and using the relation
\beq
\begin{split}
\sla n P_{L} \sla {\cal B}_{n,x\omega}^\perp \sla {\cal B}_{n,\bar x\omega}^{\perp}=
\big(g_{\perp\mu\nu}+ i \epsilon_{\perp \mu\nu}\big) {\cal B}_{n,x\omega}^{\perp\mu}  {\cal B}_{n,\bar x\omega}^{\perp\nu}\sla n P_{L},
\end{split}
\eeq
for the gluonic operator in \eqref{jet-funcs},
we arrive at the expression for the decay amplitude into pseudoscalar or longitudinally polarized vector mesons $M_{1,2}$
valid at leading order in $1/m_b$ and to all orders in $\alpha_S(m_b)$ 
\beq\label{Az}
\begin{split}
&A(B\to M_1 M_2)=\\
&\frac{G_F}{\sqrt{2}} m_B^2 \Big\{ f_{M_1}\int\negthickspace du\phi_{M_1}(u)
\int\negthickspace dz T_{1J}(u,z) \zeta_J^{BM_2}(z) \\
&\qquad\quad + f_{M_1} \int\negthickspace du  \phi_{M_1}(u)  
T_{1\zeta}(u)\zeta^{BM_2} 
\\
&\qquad\quad + f_{M_1}^1 \int\negthickspace du  \bar \Phi_{M_1}^g(u) \int\negthickspace dz 
T_{1J}^g(u,z)\zeta_J^{BM_2}(z) 
\\
&\qquad\quad + f_{M_1}^1 \int\negthickspace du  \bar \Phi_{M_1}^g(u)  
T_{1\zeta}^g(u)\zeta^{BM_2} 
\\
&\qquad\quad +1\leftrightarrow 2+ \lambda_c^{(f)} A^{M_1 M_2}_{cc}\Big\},
\end{split}
\eeq
where $A^{M_1 M_2}_{cc}$ is a term denoting the nonperturbative charming penguin contributions to be discussed in more detail below, see Eqs. \eqref{Accpipieta}-\eqref{Accpieta}, $\zeta^{BM}$
was defined in \eqref{zeta}, while $\zeta_J^{BM}(z)$ is given in terms of the jet functions \eqref{jet-funcs} as
\beq
\begin{split}\label{zetaJ}
\zeta_J^{BM}(z)=
&\frac{f_Bf_M}{m_B}\negthickspace
\int\negthickspace dk_+ \phi_B^+(k_+)\Big[ C_J^{BM}\negthickspace\negthickspace
\int\negthickspace dx  \phi_{M}(x)  J(z, x, k_+)\\
&+ C_g^{BM}\frac{1}{4} \sqrt{\frac{C_F}{3}}\int\negthickspace dx \bar\Phi_{M}^g(x)  J_g(z, x, k_+)\Big],
\end{split}
\eeq
for both isosinglet and nonisosinglet mesons $M$.
The term with gluonic jet function $J_g(z, x, k_+)$ contributes only when $M=\eta_{q},\eta_{s}$. The coefficients 
$C_{J,g}^{BM}$ parameterizing the relative weights of the two contributions in $\zeta_J^{BM}(z)$ are 
\begin{center}
\begin{tabular}{c|c|c} 
\label{table:deltas1}
$BM$ & \quad $C_J^{BM}$\quad  & \quad $C_g^{BM}$ \\ \hline
$\bar B^0\eta_q, B^-\eta_q$ & $1$ & $ 2  $
\\ 
$\bar B^0\eta_s, B^-\eta_s$ & $0$ & $1$
\\
$\bar B^0_s\eta_s$ & $1$ & $1  $
\\
$\bar B^0_s \eta_q$ & $0$ & $2$
\end{tabular}
\end{center}
In all other cases, when $M$ is either an isospin nonsinglet pseudoscalar meson or a vector meson, we have simply
$C_J^{BM}=1$ and $C_g^{BM}=0$. Because $\eta_q\sim(u\bar u+d\bar d)/\sqrt{2}$  contains two quark flavors the gluonic coefficient $C_g^{BM}$ is twice as large as for $\eta_s$ (the normalization factor $1/\sqrt 2$ is absorbed in hard kernels). 

The hard kernels $T_{1J}(u,z)$ in \eqref{Az} are
\beq
\begin{split}
T_{1J}(u,z)=\sum_i& b_i(u,z) C_{i,\bar n}^{M_1}C_{i,n}^{M_2} \times\\
\times & \Big(\delta_{V M_1}+\delta_{PM_1}\big(1-2 \delta_{i,3}-2  \delta_{i,6}\big)\Big)
\end{split}
\eeq
where $C_{i,\bar n}^{M_1}, C_{i,n}^{M_2}$ are coefficients describing the content of final state mesons going in the 
$n$, $\bar n$
directions respectively, while $\delta_{i,j}$ are Kronecker deltas. 
The hard kernels $T_{2J}(u,z)$ are obtained from the above expression with the
replacement $1 \leftrightarrow 2$, while the hard kernels $T_{1\zeta}(u), T_{2\zeta}(u)$ 
are obtained from $T_{1J}(u,z), 
T_{2J}(u,z)$ by replacing $b_i\to c_i$. For the reader's convenience, explicit formulas for the hard kernels are provided in 
Tables \ref{table:zeta0}-\ref{table:piK} for the $\Delta S=0$ and $\Delta S=1$ decays of $\bar B^0, B^-$ and 
$\bar B^0_s$ mesons 
(FKS mixing scheme is used for treatment of $\eta,\eta'$ final states, see discussion above and below Eq. \eqref{eta'mix}).
Note that our phase convention for final states, $\pi^+ \sim u \bar d$, 
$\pi^0 \sim (u \bar u- d\bar d)/\sqrt{2} $, $\pi^- \sim d \bar u$, $\bar K^0 \sim s \bar d $, $K^0\sim 
d \bar s $, $K^+ \sim u \bar s $, $K^-\sim s \bar u $, $\eta_q\sim  (u \bar u+ d\bar d)/\sqrt{2}$, $\eta_s\sim s\bar s$, differs from the one used in \cite{Bauer:2004tj,Bauer:2005kd}.

The terms in the third and fourth line of Eq. \eqref{Az} are coming from $Q_{gs}^{(1)}$ and $Q_{gs}^{(0)}$ operators
respectively. They contribute only if $M_1$ is an isosinglet pseudoscalar meson, with hard kernels $T_{1J}^g(u,z)$, $T_{1\zeta}^g(u)$
that start at NLO in $\alpha_S(m_b)$ \eqref{cg}, \eqref{bg} 
\beq
T_{1J}^g=\frac{\sqrt{C_F}}{2}C_{BM_2}^{(f)} b_{g}^{(f)},\qquad T_{1\zeta}^g=\frac{\sqrt{C_F}}{2}C_{BM_2}^{(f)} c_{g}^{(f)},
\eeq
where $C_{B M_2}^{(f)}$ is a coefficient describing the $\bar q f$ flavor content of $M_2$, with $q$ the flavor of
spectator quark. For instance $C_{B^- \pi^-}^{(d)}=1$, $C_{\bar B^0 \pi^0}^{(d)}=-1/\sqrt{2}$, while $C_{B^- \pi^-}^{(s)}=0$. 

The discussion of $B$ decays into transversely polarized vector mesons is complicated by the presence of $m_b$ enhanced
electromagnetic operator, as pointed out recently in Ref. \cite{Beneke:2005we}. Namely, for the decays into neutral
$V_\perp$ mesons the $1/\lambda$ enhanced operator
\beq\label{gammaOp}
Q_\gamma^{(-1)}=m_b^2[\bar s_{n,\omega_1}\sla \bar n \gamma_\perp^\mu P_R b_v][{\cal B}^{\perp\mu}_{\gamma, \bar n, \omega_2}],
\eeq
also contributes. Here $i e {\cal B}^{\perp\mu}_{\gamma, \bar n, \omega}$ is a purely electromagnetic operator related to 
the $(\bar n,\perp)$ component of the electromagnetic field strength, defined in the same way as the gluonic counterpart in \eqref{Bdef}, but with the replacement of QCD Wilson lines and covariant derivatives with the QED ones, $W_n\to W_{\gamma n}, i D_{c\mu}\to i\partial_\mu+e A_{\mu}$. At leading order in the electromagnetic coupling the operator \eqref{gammaOp} is a result of a tree level matching of $O_{7\gamma}$ \eqref{magnetic} and of four quark operators $O_{1,\dots, 10}$ with a photon emitted from  a closed quark loop \cite{Beneke:2005we}. 

The operator \eqref{gammaOp} leads to contributions that are $m_b/\Lambda$ enhanced  compared to the amplitudes for $B\to V_\parallel V_\parallel$ in \eqref{Az}, but which are, on the other hand, also $\alpha^{\rm em}$ suppressed due to the exchanged photon. Numerically thus the contributions from 
\eqref{gammaOp} can be expected to be smaller than the $O(m_b^0)$ terms in \eqref{Az}. Thus at 
leading order the only contributions to $B\to V_\perp V_\perp$ can arise from nonperturbative charming penguins $A_{cc}$ \cite{Bauer:2004tj}, possibly explaining large transversely polarized amplitudes in $B\to \phi K^*$ decays \cite{Chen:2005zv}, while the other terms are either $1/m_b$ or $\alpha_0^{\rm em}m_b/\Lambda$ suppressed (for 
alternative explanations see \cite{Kagan:2004uw,Chen:2006jz}). 
Therefore, a complete first order treatment of observables, most notably all the CP asymmetries, in $B\to V_\perp V_\perp$ decays requires an inclusion 
of $1/m_b$ operators, which is beyond the scope of the present paper and will not be pursued further. 

\begin{table}
\caption{Table of hard kernels for $\Delta S=1$ decays of  
$B^-, \bar B^0$ and $\bar B^0_s$ into isosinglet mesons (separated by horizontal lines in the table).
 The superscript $(s)$ on coefficients 
$c_{i}^{(s)}$ is not displayed. $T_{1J}$ and $T_{2J}$ follow from the replacement
$c_i\to b_i$.}
\begin{ruledtabular}
\begin{tabular}{ccc} 
Mode  & $T_{1\zeta}$  & $T_{2\zeta}$ \\ \hline
$K^{-(*)} \eta_q, K^{-(*)} \omega$ & $\frac{1}{\sqrt{2}}(c_1+c_{4})$ & 
$\frac{1}{\sqrt{2}}(c_{2}\mp c_{3}+2c_{5}\mp 2 c_{6})$ 
\\
$K^{-(*)} \eta_s,  K^{-(*)} \phi$ & $\{ c_1+c_4, 0\}$ & $c_4+c_{5}\mp c_{6}$  \\ \hline
$\bar K^{0(*)} \eta_q$, $\bar K^{0(*)} \omega$ & $\frac{1}{\sqrt{2}}c_{4}$ & 
$\frac{1}{\sqrt{2}}(c_{2}\mp c_{3}+2c_{5}\mp 2c_{6})$ 
\\
$\bar K^{0(*)} \eta_s$, $\bar K^{0(*)} \phi$ & $\{c_4, 0\}$ & $c_4+c_{5}\mp c_{6}$  \\ \hline
$ \eta_{s}\pi^0, \phi \pi^0$ & $0$& $\frac{1}{\sqrt 2} (c_2-c_3)$ 
\\
$ \eta_{s}\rho^0, \phi \rho^0$& $0$ & $\frac{1}{\sqrt 2} (c_2+c_3)$  
\\
$ \eta_{q}\pi^0, \eta_q \rho^0$& $0$ & $\frac{1}{2} (c_2\mp c_3)$  
\\
$ \eta_q\eta_q$ & $\frac{c_2}{2}-\frac{c_3}{2}+c_5- c_6$ & $\frac{c_2}{2}-\frac{c_3}{2}+c_5- c_6$ 
\\
$ \eta_s\eta_q, \phi \eta_q$ & $\frac{1}{\sqrt 2}(c_4+c_5\mp c_6)$ & $\frac{1}{\sqrt2}(c_{2}-c_{3}+2c_{5}-2c_{6})$ 
\\
$ \eta_s\eta_s,  \phi\phi$ & $c_4+c_{5}\mp c_{6}$& $c_4+c_{5}\mp c_{6}$  \\ 
$ \phi\eta_s$ & $c_4+c_{5}+c_{6}$ & $c_4+c_{5}-c_{6}$ \\ 
$\eta_s\omega, \phi\omega$ & $0$ & $\frac{1}{\sqrt{2}}(c_2+c_3+2c_5+2c_6)$ 
\\ 
\end{tabular}
\end{ruledtabular}
\end{table}

The amplitude \eqref{Az} has the form of  a convolution of nonperturbative light-cone wave functions $\phi_M(x)$, 
$\bar \Phi_M^g(x)$, $\phi_B^+(k^+)$ and the perturbative hard kernel and jet functions. With prior knowledge of the light-cone
wave function from a fit to an unrelated experiment, a prediction for $B\to M_1M_2$ decays can be made using
a perturbative expansion 
in $\alpha_S(\sqrt{m_b\Lambda})$ for the jet functions and in $\alpha_S(m_b)$ for the hard kernels. Alternatively,
the nonperturbative parameters can be fit from observables in $B\to M_1M_2$ decays. This approach is especially useful
at leading order in $\alpha_S(m_b)$, since then the hard kernels $T_{1\zeta,2\zeta}(u)$
are constants, while $T_{1J, 2J}(u,z)$ are 
functions of $u$ only. 
Furthermore, at this order the hard kernels $T_{1J,2J}^g(u,z)$, $T_{1\zeta,2\zeta}^g(u)$
do not contribute at all.
At LO in $\alpha_S(m_b)$ thus
\beq\label{LO}
\begin{split}
&A_{B\to M_1 M_2}=\\
&\frac{G_F}{\sqrt{2}} m_B^2 \Big\{f_{M_1} \zeta_J^{BM_2} \int\negthickspace du \phi_{M_1}(u)  T_{1J}(u)
 \\
&\qquad\quad + f_{M_1}\zeta^{BM_2}T_{1\zeta}  
+1\leftrightarrow 2+ \lambda_c^{(f)} A^{M_1 M_2}_{cc}\Big\},
\end{split}
\eeq
where 
$\zeta^{BM_1}$ and
\beq
\zeta_J^{BM_1}=\int dz  \zeta_J^{BM_1}(z),
\eeq
are treated as
nonperturbative parameters to be fit from experiment. Note that in this way no perturbative expansion in 
$\alpha_S(\sqrt{m_b\Lambda})$ is needed. 
Eq. \eqref{LO} has exactly the same form as a factorization formula obtained in Refs. \cite{Bauer:2004tj,Bauer:2005kd} for decays into nonisosinglet mesons, but is now valid also for decays into isosinglets. 
We will use this form for the factorized amplitudes in the phenomenological
analyses in Section \ref{pheno}.

\begin{table}
\caption{Table of hard kernels for $\Delta S=0$ decays  of
$B^-, \bar B^0$ and $\bar B^0_s$ mesons respectively (separated by horizontal lines in the table)
without isosinglets in the final 
state.  The superscripts on $c_i^{(d)}$ are not displayed for brevity. 
$T_{2J}$ and $T_{1J}$ are obtained with the replacement
$c_i\to b_i$.}\label{table:pipi}
\begin{ruledtabular}
\begin{tabular}{ccc} 
Mode  & $T_{1\zeta}$  & $T_{2\zeta}$ \\ \hline
$\pi^-\pi^0$, $\rho^-\pi^0$ & $\frac{1}{\sqrt 2} \big(c_1+c_4\big)$ & $\frac{1}{\sqrt 2}
\big(c_2 -c_3-c_4\big)$
\\
$\pi^-\rho^0$, $\rho^-\rho^0$ & $\frac{1}{\sqrt 2} \big(c_1+c_4\big)$ & $\frac{1}{\sqrt 2}
\big(c_2 +c_3-c_4\big)$
\\
$K^{0(*)}  K^{-(*)}$ & $c_4$ & $0$
\\
\hline
$\pi^+\pi^- $, $\rho^+\pi^-$ & $0$ & $c_{1}+c_{4}$  
\\ 
$\pi^+\rho^- $, $\rho^+\rho^-$ & $0$ & $c_{1}+c_{4}$  
\\ 
$\pi^0\pi^0$, $\pi^0\rho^0$ & $\frac{1}{2} \big(-c_2 +c_3+c_4\big)$ & $\frac{1}{2} 
\big(-c_2 \pm c_3+c_4\big)$
\\
$\rho^0\rho^0$& $\frac{1}{2} \big(-c_2 -c_3+c_4\big)$ & $\frac{1}{2} 
\big(-c_2 - c_3+c_4\big)$
\\
$K^{0(*)} \bar K^{0(*)}$ & $c_4$ & $0$
\\
\hline
$\pi^-K^{+(*)}$, $\rho^- K^{+(*)}$ & $c_1+c_4$ & $0$
\\
$\pi^0 K^{0(*)}$, $\rho^0 K^{0(*)}$ & $\frac{1}{\sqrt 2}\big(c_2\mp c_3- c_4\big)$ & $0$
\end{tabular}
\end{ruledtabular}
\end{table}

We turn now to the treatment of $\eta$ and $\eta'$ states, where we use the FKS mixing scheme 
\cite{Feldmann:1998vh}. 
An arbitrary isospin zero biquark operator $O$ 
can be written as a linear combination of $O_{q}\sim (u\bar u+d\bar d)/\sqrt{2}$ and $O_s\sim s\bar s$ operators with well defined flavor
structure (here Dirac structure is ignored for simplicity). The matrix elements of $O=c_q O_q+c_s O_s$ between $\eta^{(')}$ states
and the vacuum can then 
be parameterized in a completely general way  by
\begin{align}
\langle0|O|\eta\rangle&=c_q \cos \phi_q  \langle O_q\rangle  - c_s \sin \phi_s \langle O_s \rangle,\\
\langle0|O|\eta'\rangle&=c_q \sin \phi_q  \langle O_q \rangle + c_s \cos \phi_s \langle O_s \rangle,
\end{align}
where the four matrix elements $\langle 0|O_{q,s}|\eta^{(')}\rangle$ have been traded for two angles $\phi_{q,s}$ and
two reduced matrix elements $\langle O_{q,s}\rangle$, all of which in principle depend on the structure of operator $O$. Phenomenologically,  $\phi_q=\phi_s=\phi$ to a very good degree, with 
$\phi=(39.3 \pm 1.0)^\circ$ irrespective of the operator $O$ \cite{Feldmann:1998vh}.  In other words, if the mass eigenstates
$\eta$, $\eta'$ are related to the flavor basis through
\beq
\begin{split}
\eta&=\eta_q \cos \phi -\eta_s \sin \phi,\\
\eta'&=\eta_q \sin \phi +\eta_s \cos \phi, \label{eta'mix}
\end{split}
\eeq
then to a very good approximation (i) the matrix elements corresponding to OZI suppressed processes vanish, $\langle 0|O_q|\eta_s\rangle=
\langle 0|O_s|\eta_q\rangle=0$, and (ii) the shapes of $\eta_{q,s}$ components of the wave functions do not depend on 
the mass eigenstate  so that the reduced matrix elements $\langle O_q\rangle_\eta=\langle0|O_q|\eta\rangle/\cos \phi  $, $ \langle O_q\rangle_{\eta'}=\langle0|O_q|\eta'\rangle/\sin \phi $ (and similarly for $O_s$) are independent of the final state particle, $\langle O_q\rangle_\eta=\langle O_q\rangle_{\eta'}$.

\begin{table}
\caption{Table of hard kernels for $\Delta S=1$ decays  of
$B^-, \bar B^0$ and $\bar B^0_s$ mesons respectively (separated by horizontal lines in the table)
without isosinglets in the final 
state.  
$T_{2J}$ and $T_{1J}$ are obtained with the replacement
$c_i\to b_i$.}\label{table:piK}
\begin{ruledtabular}
\begin{tabular}{ccc} 
Mode  & $T_{1\zeta}$  & $T_{2\zeta}$ \\ \hline
$K^{-(*)}\pi^0$, $K^{-(*)}\rho^0$ & $\frac{1}{\sqrt 2}\big(c_1^{(s)}+c_4^{(s)}\big) $& 
$\frac{1}{\sqrt 2} \big(c_2^{(s)}\mp c_3^{(s)}\big)$
\\ 
$\bar K^{0(*)} \pi^-$, $\bar K^{0(*)} \rho^-$ & $c_4^{(s)}$ & $0$
\\
\hline
$\bar K^{0(*)} \pi^0 $, $\bar K^{0(*)} \rho^0 $ & 
$-\frac{1}{\sqrt 2}c_4^{(s)}$ & $\frac{1}{\sqrt 2}\big(c_2^{(s)}\mp c_3^{(s)}\big)$
\\
$K^{-(*)}\pi^+$, $K^{-(*)}\rho^+$ &$c_1^{(s)}+c_4^{(s)}$&$ 0$
\\ 
\hline
$K^{-(*)}K^{+(*)}$ &$c_1^{(s)}+c_4^{(s)}$&$ 0$
\\ 
$\bar K^{0(*)}K^{0(*)}$ & $c_4^{(s)}$ & $0$ 
\end{tabular}
\end{ruledtabular}
\end{table}

We will make one further assumption
\beq\label{phirel}
\phi_\pi(u,\mu)=\phi_{\eta_q}(u,\mu),
\eeq
that is well respected by data. The relation \eqref{phirel} is true for asymptotic forms of LCDA, where for $\mu\to \infty$ 
one has $\phi_\pi(u)=\phi_{\eta_q}(u)=6u\bar u$. It can be, however, only approximately true for all other scales, since $\phi_{\eta_q}(u,\mu)$ mixes with gluonic LCDA $\bar \Phi_{\eta_q}^g(u,\mu)$, while $\phi_{\pi}(u,\mu)$ does not. Even so, 
for $\mu$ as low as $\mu=1$ GeV the relation \eqref{phirel} is very well respected \cite{Kroll:2002nt}. The inverse moments
or $\pi$ and $\eta_q$ LCDA for instance agree within experimental errors, which are at the level of $3-5\%$.

Eq. \eqref{phirel} leads to relations between the nonperturbative function $\zeta_{(J)}$ and $A_{cc}$ for decays into $\pi,\eta_q$, so that one can write 
\beq\label{zetaetarel}
\zeta_{(J)}^{B\eta_q} = \zeta_{(J)}^{B\pi} + 2\zeta_{(J)g},
\eeq
with $\zeta_{(J)g}$ new nonperturbative functions that are entirely due to contributions from interpolating gluons (the
function $\zeta_{Jg}$ for instance is equal to the second term in \eqref{zetaJ}). The decomposition \eqref{zetaetarel} is useful
only in the limit of exact flavor SU(3) symmetry, when
\beq\label{zetaetas}
\zeta_{(J)}^{B\eta_s}=\zeta_{(J)g}.
\eeq
Similar relations for $A_{cc}$ will be given below. 

\begin{figure}
\begin{center}
\epsfig{file=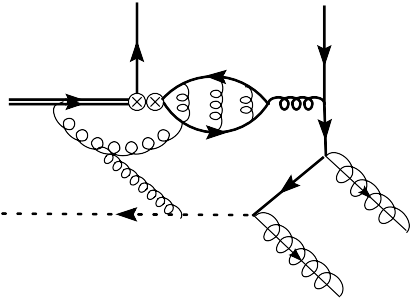, width=5.5cm}
\caption{The gluonic charming penguin contributions with intermediate on-shell charm quarks annihilating into two collinear quarks
going in the opposite directions, with $n$ collinear quark annihilating with spectator quark and producing two $n$ collinear gluons (compare also with diagrams b) and d) of Fig. \ref{matching}).} \label{ggsgluon}
\end{center}
\end{figure}

Using SU(3) symmetry further relations are possible. In the exact SU(3) limit only two $\zeta$ functions are needed
for the decays without isosinglet mesons
\beq\label{zetaSU3}
\zeta_{(J)} \equiv \zeta_{(J)}^{B\pi}= \zeta_{(J)}^{BK}=\zeta_{(J)}^{B_sK}.
\eeq
Furthermore, to describe all the decays into isosinglet mesons only the two new functions $\zeta_{(J)g}$ defined in \eqref{zetaetarel}, are needed. Namely, in exact SU(3) one has (cf. Eq. \eqref{zetaJ})
\beq\label{zetaSU3eta}
\zeta_{(J)}^{B_s\eta_q}=2 \zeta_{(J)g},\qquad \zeta_{(J)}^{B_s\eta_s}=\zeta_{(J)}+\zeta_{(J)g},
\eeq
in addition to the relations \eqref{zetaetarel}, \eqref{zetaetas}.

Let us now discuss the nonperturbative charming penguin contributions $A_{cc}^{M_1M_2}$ \eqref{LO} in the isospin limit assuming FKS mixing along with relation \eqref{phirel}. The charming 
penguins in $\bar B^0, B^-$ decays into $\pi\eta_q$, $\pi\eta_s$ and $\pi\pi$ final states are parameterized in terms of four complex parameters 
\begin{align}
\label{Accpipieta}
A_{cc}^{\pi\pi}&\equiv A_{cc}^{\pi^+\pi^-}=A_{cc}^{\pi^0\pi^0}, \nonumber \\
A_{cc,g}^{\pi\eta_s}&\equiv  A_{cc}^{\pi^-\eta_s}=-\sqrt{2}A_{cc}^{\pi^0\eta_s}=\sqrt{2}A_{cc}^{\eta_q\eta_s},
\end{align}
and $A_{cc}^{\pi\eta}$, $A_{cc,g}^{\pi\eta_q}$ in terms of  which 
\begin{align}
A_{cc}^{\pi^- \eta_q}&={\sqrt{2}} \big(A_{cc}^{\pi\eta} +  A_{cc,g}^{\pi\eta_q}\big), \nonumber \\
A_{cc}^{\pi^0 \eta_q}&=- A_{cc}^{\pi\eta} -  A_{cc,g}^{\pi\eta_q}, \nonumber \\
A_{cc}^{\eta_q \eta_q}&=A_{cc}^{\pi\pi} + 2 A_{cc,g}^{\pi\eta_q},
\end{align}
and $A_{cc}^{\pi^- \pi^0}=0$. Here $A_{cc,g}^{\pi\eta_{q,s}}$ describes the charming penguin contributions, where the $n$ collinear quark coming from the annihilation of charm quarks annihilates the 
spectator quark and produces two $n$ collinear gluons, Fig. \ref{ggsgluon}. At LO in $1/m_b$ there is one additional 
relation
\beq\label{AccRel}
A_{cc}^{\pi\pi}=A_{cc}^{\pi\eta}.
\eeq
The amplitude $A_{cc}^{\pi\pi}$ receives contributions from SCET operators of higher order in $1/m_b$, where the spectator quark directly attaches to the weak vertex. These higher order corrections correspond to penguin
annihilation in the diagrammatic language and do not contribute to $A_{cc}^{\pi\eta}$.

At LO in $1/m_b$ one further parameter is introduced for $\Delta S=0$ decays into two kaons
\beq
A_{cc}^{KK}\equiv A_{cc}^{K^0K^-}=A_{cc}^{K^0\bar K^0},
\eeq 
while higher order penguin annihilation contributions to $A_{cc}^{K^0\bar K^0}$ distinguish between the two amplitudes. 

Three additional complex parameters describe charming penguins in $\Delta S=1$ decays of $\bar B^0$, $B^-$ 
\begin{align}
A_{cc}^{K\pi}&\equiv A_{cc}^{K^-\pi^+}=A_{cc}^{\bar K^0 \pi^-}=-\sqrt{2}A_{cc}^{\bar K^0\pi^0} = \sqrt{2}A_{cc}^{K^-\pi^0}, \nonumber \\
\sqrt{2} A_{cc,g}^{K\eta_q}&+\frac{1}{\sqrt{2}}A_{cc}^{K\pi}\equiv A_{cc}^{\bar K^0\eta_q}=A_{cc}^{K^-\eta_q}, \nonumber \\
A_{cc,g}^{K\eta_s}&+A_{cc}^{K\eta_s}\equiv A_{cc}^{K^-\eta_s}=A_{cc}^{\bar K^0\eta_s}, \label{AccKetas}
\end{align}
where the gluonic component $A_{cc,g}^{K\eta_s}$ has been pulled out for later convenience.
An additional six complex parameters describe charming penguin contributions in $\bar B_s^0$ decays
\begin{align}
A_{cc}^{\pi K}(s)&\equiv A_{cc}^{\bar B_s^0\to \pi^-K^+}=-\sqrt{2}A_{cc}^{\bar B_s^0\to \pi^0K^0} \nonumber\\
A_{cc}^{K K}(s)&\equiv A_{cc}^{\bar B_s^0\to K^-K^+}=A_{cc}^{\bar B_s^0\to K^0 \bar K^0},\nonumber\\
\sqrt{2} A_{ccg}^{\eta_s\eta_q}(s)&\equiv A_{cc}^{\bar B_s^0\to \eta_s\eta_q},\nonumber\\
2A_{cc}^{\eta_s\eta_s}(s)+2A_{ccg}^{\eta_s\eta_s}(s)&\equiv A_{cc}^{\bar B_s^0\to \eta_s\eta_s},\nonumber\\
\frac{1}{\sqrt 2}A_{cc}^{\pi K}(s)+\sqrt{2} A_{ccg}^{K\eta_q}(s)&\equiv A_{cc}^{\bar B_s^0\to K^0\eta_q},\nonumber\\
A_{cc}^{K\eta_s}(s)+A_{ccg}^{K\eta_s}(s)&\equiv A_{cc}^{\bar B_s^0\to K^0\eta_s},\label{Acc(s)}
\end{align}
where the subscript $g$ again denotes gluonic contributions as before. Note that the above relations are valid to all orders in the $\alpha_S(m_b)$ and $1/m_b$ expansions, under the assumptions leading
to FKS mixing along with relation \eqref{phirel}. 

In the limit of exact SU(3) and at LO in $1/m_b$ the above seventeen complex parameters in \eqref{Accpipieta}-\eqref{Acc(s)} are related to only two complex parameters
\beq\label{AccSU3}
\begin{split}
A_{cc}&=A_{cc}^{\pi\pi}=A_{cc}^{\pi\eta}=A_{cc}^{K\pi}=A_{cc}^{K\eta_s}=A_{cc}^{KK}=\\
&=A_{cc}^{\pi K}(s)=A_{cc}^{K K}(s)=A_{cc}^{\eta_s\eta_s}(s)=A_{cc}^{K\eta_s}(s),
\end{split}
\eeq
and 
\beq\label{Accpieta}
\begin{split}
A_{ccg}&=A_{cc,g}^{\pi\eta_q}=A_{cc,g}^{\pi\eta_s}=A_{cc,g}^{K\eta_q}=A_{cc,g}^{K\eta_s}\\
&=A_{ccg}^{\eta_s\eta_q}(s)=A_{ccg}^{\eta_s\eta_s}(s)=A_{ccg}^{K\eta_q}(s)=A_{ccg}^{K\eta_s}(s),
\end{split}
\eeq
The same relations also apply to $B$ decays into two vector mesons, with the replacements $\eta_q\to \omega$, 
$\eta_s\to \phi$, $\pi\to \rho$, $K\to K^*$, but with additional simplification since the gluonic contributions \eqref{Accpieta} vanish. The relations apply for each polarization of the vector mesons separately, with transversely polarized vector mesons receiving only contributions from nonperturbative charming penguins \cite{Bauer:2004tj}. 
The relations between $A_{cc}$ for $B\to PV$ decays are slightly more complicated, because separate nonperturbative
parameters are needed if pseudoscalar or vector meson absorb the spectator quark. Since we will not perform the 
phenomenological analysis of $B\to PV$ decays we do not display these relations.

\subsection{Semileptonic $B$ decays into isosinglet mesons}
Using the above derivations, it is fairly straightforward to obtain the results for vector and axial vector form factors in semileptonic $B$ 
decays to light pseudoscalar or vector mesons at $q^2=0$. The $V-A$ current $\bar q\gamma^\mu (1-\gamma_5) b$ is 
matched to SCET$_{\rm I}$ LO and NLO currents \cite{Bauer:2000ew,Pirjol:2002km,Beneke:2003pa,Chay:2002vy,Beneke:2002ph}
\beq
\int d\omega_1 c_0(\omega_1) j_0^\mu(\omega_1)+\sum_{\tilde n=n, \bar n}\int [d\omega_{1,2}]c_{1\tilde n}(\omega_{1,2}) j_{1\tilde n}^\mu(\omega_{1,2})+\dots,
\eeq
where
\begin{align}
j_0^\mu=&n^\mu [\bar q_{n, \omega_1}\sla  \bar n P_L b_v],\\
j_{1n}^\mu=&\frac{-2}{m_B}n^\mu[\bar q_{n,\omega_1} i g \sla {\cal B}_{n, \omega_2}^\perp P_L b_v],\\
j_{1\bar n}^\mu=&\frac{-2}{m_B}{\bar n}^\mu[\bar q_{n,\omega_1} i g \sla {\cal B}_{n, \omega_2}^\perp P_L b_v],
\end{align}
with the ellipses denoting the remaining SCET$_{\rm I}$ currents that either receive contributions only at NLO in $\alpha_S(m_b)$ or contribute to the form factors at subleading order in $1/m_b$.  
At leading order in $\alpha_S(m_B)$ 
\beq
c_0=c_{1n}=c_{1\bar n}=1.
\eeq
Since $j_0^\mu$, $j_{1 n}^\mu$ are equal to $n^\mu Q_{1d, 2d, 2s}^{n(0,1)}$ in \eqref{nnbar} for $b\to u,d,s$ 
transitions respectively, while $j_{1 \bar n}^\mu$ is equal to 
$\bar n^\mu Q_{1d, 2d, 2s}^{n(1)}$,
we can readily obtain the result for the SCET$_{\rm I}$ to SCET$_{\rm II}$ matching of the V-A current from the results obtained in the previous subsection for the four quark
operators. The soft overlap contribution is proportional to $n^\mu$, while the hard scattering contributions contribute equaly to terms with $n^\mu$ and $\bar n^\mu$ Lorentz structure.  
More precisely, the soft overlap for $b\to d$ decay equals $n^\mu T_{1,2d}$, leading to a contribution of
$m_B n^\mu \zeta^{BM}$ to the matrix element  $\langle M| \bar d\gamma^\mu (1-\gamma_5) b|B\rangle$ at $q^2=0$, with similar results for the hard scattering contributions. 
Using a definition of form factors
\beq
\begin{split}\label{BPform}
\langle P| \bar q\gamma^\mu (1-\gamma_5) b|B\rangle=&C^{BP}\big[m_B n^\mu f_+^{BP}(0) \\
&+\frac{m_B}{2} \bar n^\mu \big(f_+^{BP}(0)+f_-^{BP}(0)\big)\big],
\end{split}
\eeq
where the relations $p_B^\mu=m_B v^\mu$, and $p_P^\mu=m_B n^\mu/2$ have been used, this gives for the 
form factors of $B\to P$ transition at maximal recoil at LO in $1/m_b$ and $\alpha_S(m_b)$
\begin{align}
\label{f_+}
f_+^{BP}(0)= \zeta^{BP}+\zeta_J^{BP},\\
f_+^{BP}(0)+f_-^{BP}(0)=2 \zeta_J^{BP}.
\end{align}
Quite similarly one obtains for the form factors at $q^2=0$ in $B\to V$ transition
\beq
\begin{split}
A_0(0)&=\frac{m_B}{2 m_V} (A_1(0)-A_2(0))=A_3(0)=\zeta^{BV}+\zeta_J^{BV},
\end{split}
\eeq
and
\beq
\begin{split}
A_2(0)+2m_Vm_B\left.\frac{d(A_3-A_0)}{d q^2}\right|_{q^2=0}=-\frac{4m_V}{m_B}\zeta_J^{BV},
\end{split}
\eeq
where the standard definition of form factors has been used (and can be found e.g. in \cite{Ball:2004rg}).  The coefficients $C^{BP}$ in the definition of $B\to P$ form factors \eqref{BPform} and equivalent coefficients $C^{BV}$ in the definition of $B\to V$ form factors, take care of the flavor content of the final state $M$. For instance for $\eta_q,\omega$ these are
$C^{B^-M}=C^{\bar B^0 M}=1/\sqrt{2}$, while for $\eta_s, \phi$, they are $C^{B^-M}=C^{\bar B^0 M}=1$. 
The expressions for $\zeta_{J}$ parameters in terms of jet functions are given in Eq. 
\eqref{zetaJ}. Note that the derived relations between form factors and the SCET nonperturbative functions are valid also 
for isosinglet mesons and in this sense extend the previous discussions of $B\to P, V$ form factors \cite{Bauer:2002aj,Beneke:2003pa,Beneke:2000wa,Beneke:2005gs,Hill:2004if,Lange:2003pk,Chay:2002vy,Pirjol:2002km,Beneke:2004rc}. 
In particular, the 
$B\to \eta_q$ form factors $f_\pm(0)$ receive the gluonic contribution as can be seen from \eqref{zetaJ}.


\section{Phenomenology}\label{pheno} 
We now apply the LO (in $1/m_b$ and $\alpha_S(m_b)$ expansions) factorization formula Eq. \eqref{LO}  to $B$ decays into pseudoscalar mesons. 

For the nonperturbative 
parameters $A_{cc}^{M_1M_2}$ and $\zeta_{(J)}^{BM_i}$ exact SU(3) relations \eqref{zetaetarel}-\eqref{zetaSU3eta}, \eqref{AccSU3}, \eqref{Accpieta} will be used, leading to four independent real  parameters $\zeta_{(J)}$, $\zeta_{(J)g}$ and 
two complex parameters $A_{cc}$, $A_{ccg}$ describing nonperturbative charming penguins. The parameters $\zeta_{(J)g}$ and 
$A_{ccg}$ correspond to gluonic contributions and are specific to the decays into isosinglet final states. We will thus 
first determine $\zeta_{(J)}$ and $A_{cc}$ using a $\chi^2$-fit to observables in $B\to \pi\pi$, $B\to \pi K$ decays and then fix the remaining parameters $\zeta_{(J)g}$ and 
$A_{ccg}$ from a separate $\chi^2$-fit to observables in $B\to\eta^{(')}\pi$, $B\to\eta^{(')}K$ decays. 

The data used in the analysis are from HFAG, Summer  2005 compilation \cite{hfag}, apart from the observables where different experiments do not agree, in which case the errors are inflated according to the PDG prescription \cite{Eidelman:2004wy}.  The observables used throughout the analysis are the CP averaged decay widths 
\beq\label{Gammadef}
\overline \Gamma(B\to f)=\frac{|\vec{p}\;|}{8 \pi m_B} \frac{1}{2}\big(|\bar{A_f}|^2+|A_f|^2\big),
\eeq
with $|\vec p\;|=m_B/2+O(m_M^2/m_B^2)$ the three momentum of light mesons in the $B$ rest frame, and an additional factor of $1/2$ on the right hand side, if  two identical mesons are in the final state, while the abbreviations are $\bar A_f=A_{\overline{B} \to f}$, $A_f=A_{B\to \overline{f}}$.  Then the direct CP asymmetries
\beq
{\cal A}_f^{\rm CP}=\frac{|\bar{ A_f}|^2-|A_f|^2}{|\bar{ A_f}|^2+|A_f|^2},
\eeq
and the additional observables, $S_f$ and $H_f$, that can
be extracted from time dependent decays of neutral $B$ mesons are
(restricting $f$ to have definite CP)
\beq
\begin{split}
\Gamma&(B^0_q(t)\to f)=
e^{-\Gamma t}\; \overline \Gamma(B_q\to f) \Big[\cosh \Big(\frac{\Delta \Gamma t}{2}\Big)+\\
&+H_f \sinh \Big(\frac{\Delta \Gamma t}{2}\Big)-{\cal A}^{\mathrm{CP}}_f\cos( \Delta m t)- S_f \sin (\Delta m t) \Big],
\end{split}
\eeq
where $\Delta m=m_H-m_L>0$, $\Gamma$ is the average decay width and $\Delta \Gamma=\Gamma_H-\Gamma_L$ the difference
of decay widths for heavier and lighter $B_q^0$ mass eigenstates. The time dependent decay width $\Gamma(\bar B^0_q(t)\to f)$ is obtained from the above expression by flipping the signs of the $\cos(\Delta m t)$ and $\sin(\Delta m t)$ terms.

Since $(\Delta \Gamma/\Gamma)_{B_d}\ll 1$ in the $B_d^0$ system, only measurements of the parameter
\beq\label{Sf}
S_f= 2 \frac{\Im\Big[ e^{-i 2\beta} \bar A_{f}({A_{f}})^*\Big]}{|\bar A_{f}|^2+|{A_{f}}|^2},
\eeq
are experimentally feasible in the foreseeable future. In the above relation \eqref{Sf} the  phase convention with
$\arg V_{cb}=\arg V_{cs}=\arg(- V_{cd})=0$, so that  $\beta=\arg(-V_{cb}^* V_{td}^* V_{cd} V_{tb})=\arg(V_{td}^* V_{tb})$
  was employed \cite{Branco:1999fs}.

In the $B_s$ system, on the other hand, we expect a much larger decay width difference $(\Delta \Gamma/\Gamma)_{B_s}=-0.12\pm0.05$ within the Standard Model \cite{Lenz:2004nx}, while experimentally $(\Delta \Gamma/\Gamma)_{B_s}=-0.33^{+0.09}_{-0.11}$ \cite{hfag},
so that both 
\beq\label{SfBs}
(S_f)_{B_s}= 2 \frac{\Im\Big[ e^{+i 2 \epsilon} \bar A_{f}({A_{f}})^*\Big]}{|\bar A_{f}|^2+|{A_{f}}|^2},
\eeq
and
\beq\label{Hf}
(H_f)_{B_s}= 2 \frac{\Re\Big[ e^{+i 2 \epsilon} \bar A_{f}({A_{f}})^*\Big]}{|\bar A_{f}|^2+|{A_{f}}|^2},
\eeq
might be experimentally accessible \cite{Dunietz:1995cp}. Thus predictions for both \eqref{SfBs} and \eqref{Hf}
will be given in Section \ref{Bs}.
In writing \eqref{SfBs} and \eqref{Hf} the same phase convention as in
\eqref{Sf} was used, with $\epsilon=\arg(-V_{cb}V_{ts} V_{cs}^* V_{tb}^*)$ \cite{Branco:1999fs,Aleksan:1994if}.

In our numerical estimates we use preferred values of Ref. 
\cite{Ball:2004ye} for inverse moments of LCDA at 1 GeV $\langle x^{-1} \rangle_\pi
=3.3$ and $\langle x_s^{-1} \rangle_K=2.79$, $\langle x_q^{-1} \rangle_K=\langle(1- x_s)^{-1} \rangle_K=3.81$, and take $\langle x^{-1} \rangle_{\eta_q}=\langle x^{-1} \rangle_{\eta_s}=\langle x^{-1} \rangle_{\pi}$. Note
that this choice respects the sum rule for inverse moments \cite{Chen:2003fp}
\beq
\langle x^{-1} \rangle_{\pi} +3 \langle x^{-1} \rangle_{\eta}=2 
(\langle x_s^{-1} \rangle_{K}+\langle x_q^{-1} \rangle_{K}).
\eeq
The inverse moment of the pion LCDA also agrees with the experimental determination $\langle x^{-1}\rangle_\pi=2.91\pm0.54$ obtained from CLEO data on the pion's electromagnetic form factor in Ref. \cite{Bakulev:2003cs}. The values of the CKM elements and the CKM unitarity triangle angles $\gamma$ and $\beta$ are taken 
from CKM fitter, Summer 2005 update \cite{CKMfitter}. In particular, $|V_{ub}|= (3.899\pm0.10)\times 10^{-3}$, 
$\gamma=58.6^\circ\pm6.4^\circ$ and $\beta=23.22^\circ\pm 0.75^\circ$, while for the weak phase in the $B_s^0-\bar B_s^0$ mixing $\epsilon=1.04^\circ\pm0.07^\circ$ \cite{CKMfitter}. These values agree with the ones obtained by the UTFit 
collaboration \cite{Bona:2005vz}. Note that the constraints on CKM angle $\gamma$ ($\alpha$) are obtained using an isospin 
decomposition of time dependent $B\to \pi\pi, \rho\pi, \rho\rho$ decays \cite{Gronau:1990ka,Aubert:2005av} (for a discussion of isospin violating effects see \cite{Gronau:2005pq}) and from $B\to D K$ decays \cite{Gronau:1991dp,Abe:2004gu}. No use of theoretical inputs from a $1/m_b$ expansion has been made at this point (such a possibility for using SCET to facilitate extraction of $\alpha$ from $B\to \pi\pi$ has been discussed in \cite{Bauer:2004dg}). Also, $B\to \pi\pi$ data are not restrictive at present in the determination of $\alpha$, so
essentially no use of $B\to PP$ data has been made to fix the above CKM parameters to be used in our analysis.
 For the decay constants we use $f_\pi=131$~MeV, $f_B=218\pm23$~MeV \cite{Stewart-LP2005} and $f_{\eta_q}=140\pm 3$~MeV, $f_{\eta_s}=176\pm8$~MeV \cite{Feldmann:1998vh}.

Since our phase convention for the $\pi$ states, 
$\pi^+ \sim u \bar d$, 
$\pi^0 \sim 1/\sqrt{2} (u \bar u- d\bar d)$, $\pi^- \sim d \bar u$ and the kaon states $\bar K^0 \sim \bar d s $, $K^0\sim 
\bar s d$, $K^+ \sim \bar s u$, $K^-\sim \bar u s$ 
differs from the one used in Refs. \cite{Bauer:2004tj,Bauer:2005kd}  the corresponding hard kernels are gathered in Tables \ref{table:pipi} and \ref{table:piK}. 

\subsection{Analysis of $B\to \pi \pi$ and $B\to \pi K$ decays}
Let us start the analysis with a discussion of the $\Delta S=0$ $B\to\pi\pi$ and $\Delta S=1$ $B\to \pi K$ decays. Our
analysis  differs in the treatment of errors and in the way the SCET nonperturbative parameters $\zeta_{(J)}$, $A_{cc}$ are determined from LO SCET analysis of the same decay modes in Refs. \cite{Bauer:2004tj,Bauer:2005kd}. 
Here a combined $\chi^2$-fit to available experimental information will be made to determine the SCET parameters, while in \cite{Bauer:2005kd} only a 
subset of modes was used for this purpose. The values of the SCET parameters that will be determined in this 
section will then be used in subsequent sections;  in the discussion of decays into final states with $\eta^{(')}$ mesons in Section \ref{isossect}, for predictions on $S$ parameters in penguin dominated modes in Section \ref{Ssubsection}, and for prediction of observables in $\bar B_s^0$ decays in Section \ref{Bs}. 

The majority of the differences between  $B\to \pi\pi$ and $B\to \pi K$ decays can be explained by a CKM hierarchy of different contributions.
Here and in the rest of the analysis we will split the amplitudes into ``tree" and ``penguin" contributions according to 
the CKM elements. Using the unitarity relation $\lambda_t^{(d,s)}=-\lambda_c^{(d,s)}-\lambda_u^{(d,s)}$ we define 
\beq\label{deftree}
A_{\bar B\to f}= \lambda_u^{(d,s)}T_{\bar B\to f}+\lambda_c^{(d,s)}P_{\bar B\to f},
\eeq
for any process $\bar B\to f$. In addition we will also use the nomenclature where tree contributions (without quotation marks) will denote insertions
of $O_{1,2}^{u}$ \eqref{HW} operators and will be present only in ``tree" amplitudes $T_{B\to f}$, charming penguin contributions will denote insertions of $O_{1,2}^c$ and will be present only in ``penguin" amplitudes $P_{B\to f}$, while
QCD penguin and EWP contributions will denote insertions of the operators $O_{3,\dots,6}$ and $O_{7,\dots,10}$ respectively
and will contribute to both $T_{B\to f}$ and $P_{B\to f}$.

In $\Delta S=1$ decays, such as $B\to \pi K$, there is a strong CKM
hierarchy between the two terms in \eqref{deftree} since $|\lambda_u^{(s)}|\sim 0.02 |\lambda_c^{(s)}|$. No such CKM hierarchy is present in $\Delta S=0$ decays, where both terms are of the same order in the Wolfenstein expansion $|\lambda_u^{(d)}|\sim |\lambda_c^{(d)}|\sim \lambda^3$ \cite{Wolfenstein:1983yz}. One therefore expects sizeable 
direct CP asymmetries in $\Delta S=0$ processes and much smaller ones in  $\Delta S=1$ decays. 

There are a number of other observations that can be made using the LO SCET expression \eqref{LO} even before the nonperturbative parameters of SCET are determined from the data. Because of the intriguing discrepancies between theoretical expectations
and experiment in the $\Delta S=1$ transition, we will focus in this section mainly on $B\to \pi K$ decays.
We start with the isospin decomposition of $B\to \pi K$ decay amplitudes \cite{Nir:1991cu,Lipkin:1991st,Gronau:1991dq}
\begin{align}\label{isospinKpi-start}
A_{\bar B^0\to \pi^+ K^-}=&A_{1/2}^0- A_{3/2}^1-A_{1/2}^1,\\
A_{B^-\to \pi^0 K^-}=&\frac{1}{\sqrt 2} A_{1/2}^0- \sqrt{2} A_{3/2}^1+\frac{1}{\sqrt{2}}A_{1/2}^1,\\
A_{\bar B^0\to \pi^0 \bar K^0}=&-\frac{1}{\sqrt 2} A_{1/2}^0-\sqrt{2} A_{3/2}^1+\frac{1}{\sqrt{2}}A_{1/2}^1,\\
A_{B^-\to \pi^- \bar K^0}=&A_{1/2}^0+A_{3/2}^1+A_{1/2}^1,\label{isospinKpi-end}
\end{align}
with  $I$ and $\Delta I$ in the reduced matrix elements $A_{I}^{\Delta I}$ denoting the isospin $I=3/2,1/2$ of the final states and
$\Delta I=1,0$  denoting the isospin content of the weak Hamiltonian. 
If electroweak penguin contributions are neglected, the reduced matrix elements $A_{3/2}^{1}$ and $A_{1/2}^{1}$ receive contributions only from 
tree operators $O_{1,2}^u$, while $A_{1/2}^0$ receives both penguin and tree contributions. 
 Explicitly, at LO in SCET 
\begin{align}
\begin{split}
A_{1/2}^0 &=m_B^2\frac{G_F}{\sqrt 2} \Big\{ \lambda_c^{(s)} A_{cc}^{K\pi}+\frac{f_K}{2}
\Big[\zeta^{B\pi}\big(c_1^{(s)}+2c_4^{(s)}\big)\\
&\qquad\quad+\zeta_{J}^{B\pi}\big(b_1^{(s)}+2b_4^{(s)}\big)\Big]\Big\},\label{A1/20}
\end{split}
\\
\begin{split}
A_{1/2}^1&=- \frac{m_B^2}{6}\frac{G_F}{\sqrt 2}  \Big\{ f_K\Big[\zeta^{B\pi}c_1^{(s)}+\zeta_{J}^{B\pi}b_1^{(s)}\Big]
\\
&+2 f_\pi
\Big[\zeta^{BK}\big(c_3^{(s)}-c_2^{(s)}\big)+\zeta_{J}^{BK}\big(b_3^{(s)}-b_2^{(s)}\big)\Big]\Big\},\label{A1/2}
\end{split}
\end{align}
\begin{align}
\begin{split}
A_{3/2}^1&=-  \frac{m_B^2}{3}\frac{G_F}{\sqrt 2} \Big\{ f_K\Big[\zeta^{B\pi}c_1^{(s)}+\zeta_{J}^{B\pi}b_1^{(s)}\Big]\\
&- f_\pi
\Big[\zeta^{BK}\big(c_3^{(s)}-c_2^{(s)}\big)+\zeta_{J}^{BK}\big(b_3^{(s)}-b_2^{(s)}\big)\big]\Big\},\label{A3/2}
\end{split}
\end{align}
where SCET$_{\rm}$ Wilson coefficients $c_i^{(s)}$, $b_i^{(s)}$ are to be understood as already convoluted with LCDA. This
amounts to a replacement $\omega_{2,3}\to \pm m_B/\langle x_{s,q}^{-1}\rangle_K$,  in $b_i^{(s)}$ that are multiplied by $f_K$ and a replacement $\omega_{2,3}\to \pm m_B/\langle x^{-1}\rangle_\pi$ in $b_i^{(s)}$ that are multiplied by $f_\pi$.

The dominant term in $B\to K\pi$ decays is the charming penguin term $A_{cc}^{K\pi}$, which is $\lambda_c^{(s)}/\lambda_u^{(s)}$ enhanced over tree contributions. Since it arises from insertions of $O_{1,2}^c$ operators it is also larger than QCD penguins by a factor $C_1\alpha_S(2m_c)/\max(C_3,\dots,C_6)\sim 10$. The charming penguin contribution has 
$\Delta I=0$ and
is thus present only in $A_{1/2}^0$ \eqref{A1/20}. Similarly, the reduced matrix elements $A_{3/2,1/2}^1$ do not receive contributions from QCD penguins since these are $\Delta I=0$ operators.
Note also, that the presence of the inverse moment $\langle x^{-1}\rangle_{\pi,K}\sim 3$ in $b_i$ lifts the 
color suppression of tree operators in $\bar B^0\to \bar K^0\pi^0$. Similarly the color suppression of EWP contributions to $\bar B^0\to\pi^+K^-$ and $B^-\to \pi^-\bar K^0$ is lifted as well. 
Isospin decomposition also leads to the relation
\beq
\begin{split}
A_{B^-\to K^-\pi^0}=& \frac{1}{\sqrt 2}\big(A_{\bar B^0\to K^-\pi^+}+
A_{B^-\to \bar K^0\pi^-}\big)\\
&+A_{\bar B^0\to \bar K^0\pi^0},\label{simplpiK}
\end{split}
\eeq
valid to all orders in $1/m_b$ and $\alpha_S$ \cite{Nir:1991cu,Lipkin:1991st,Gronau:1991dq}. 
As pointed out recently in Ref. \cite{Gronau:2006eb} this relation furthermore receives corrections from 
isospin breaking which are of only second order numerically, with corrections to penguins that are linear in $m_{u,d}/\Lambda$  canceling exactly.

\begin{table}
\caption{Predicted CP averaged branching ratios ($\times 10^{-6}$, first row) and direct CP asymmetries (second row in each mode) for $\Delta S=0$ and $\Delta S=1$ $B$ decays (separated by 
horizontal line) to two nonisosinglet pseudoscalar mesons. The errors on the predictions are estimates of SU(3) breaking, $1/m_b$ corrections and due to errors on SCET parameters, respectively.}\label{table:predPP}
\begin{ruledtabular}
\begin{tabular}{lll} 
Mode  & Exp  & Theory \\ \hline
$\bar B^0\to \pi^-\pi^+$& $5.0\pm0.4$ 	& $5.4\pm1.3\pm1.4\pm0.4$\\
						& $0.37\pm0.23^{\;a}$	& $0.20\pm0.17\pm0.19\pm0.05$\\
$\bar B^0\to \pi^0\pi^0$& $1.45\pm0.52^{\;b}$ 	& $0.84\pm0.29\pm0.30\pm0.19$\\
						& $0.28\pm0.40$	& $-0.58\pm0.39\pm0.39\pm0.13$\\
$B^-\to \pi^0\pi^-$& $5.5\pm0.6$ 	& $5.2\pm1.6\pm2.1\pm0.6$\\
					& $0.01\pm0.06$ 	& $<0.04$\\
$B^-\to K^0K^-$				& $ 1.2\pm0.3$ 		& $1.1\pm0.4\pm1.4\pm0.03$\\
							& $-$				& $-$\\	
$\bar B^0\to K^0 \bar K^0$	& $0.96\pm0.25$ 	& $1.0\pm0.4\pm1.4\pm0.03$\\
							& $-$				& $-$\\	\hline	
$\bar B^0\to \pi^0\bar K^0$	& $11.5\pm1.0$ 	& $9.4\pm3.6\pm0.2\pm0.3$\\
							& $0.02\pm0.13$	& $0.05\pm0.04\pm0.04\pm0.01$\\	
$\bar B^0\to K^-\pi^+$	 	& $18.9\pm0.7$ 	& $20.1\pm7.4\pm1.3\pm0.6$\\
							& $-0.115\pm0.018$	& $-0.06\pm0.05\pm0.06\pm0.02$\\
$B^-\to K^-\pi^0$			& $12.1\pm0.8$ 		& $11.3\pm4.1\pm1.0\pm0.3$\\
							& $0.04\pm0.04$	& $-0.11\pm0.09\pm0.11\pm0.02$\\
$B^-\to \bar K^0\pi^-$		& $24.1\pm1.8^{\;c}$ 		& $20.8\pm7.9\pm0.6\pm0.7$\\
							& $-0.02\pm0.05^{\;d}$	& $<0.05$\\		
\end{tabular}
\end{ruledtabular}
\begin{flushleft}
PDG scaled errors $^a(S=2.3)$, $^b(S=1.8)$, $^c(S=1.4)$, $^d(S=1.5)$.
\end{flushleft}
\end{table}

The dominance of the $A_{cc}^{K\pi}$ term in $B\to K\pi$ amplitudes leads to the approximate relation between branching ratios
\beq
Br_{\pi^+ K^-}\simeq Br_{ \pi^- \bar K^0}\simeq 2 Br_{\pi^0 K^-}\simeq 2Br_{ \pi^0 \bar K^0},
\eeq
that is well obeyed by the data. The corrections to these relations come from $A_{1/2,3/2}^1$ reduced matrix elements \eqref{A1/2}, \eqref{A3/2} that receive only contributions from $|\lambda_{u}^{(s)}|\sim 0.02 |\lambda_c^{(s)}|$ suppressed tree operators or from EWP. To study them it is useful to construct ratios of CP averaged 
decay widths in which the dependence on $A_{cc}^{\pi K}$ cancels to first approximation.  
Following the notation in the literature we define
\cite{Fleischer:1995cg,Buras:1998rb}
\begin{align}
R&=\frac{\bar \Gamma(\bar B^0\to K^-\pi^+)}{\bar \Gamma( B^-\to \bar K^0\pi^-)}
\stackrel{\textrm{\scriptsize Exp.}}{=}0.84\pm0.07 ,\label{R}
\\
R_c&=2 \frac{\bar \Gamma(B^-\to K^-\pi^0)}{\bar \Gamma(B^-\to \bar K^0\pi^-)}
\stackrel{\textrm{\scriptsize Exp.}}{=}1.00\pm0.10 ,\label{Rc}
\\
R_n&=\frac{1}{2}\frac{\bar \Gamma(\bar B^0\to K^-\pi^+)}{\bar \Gamma( \bar B^0\to \bar K^0\pi^0)}
\stackrel{\textrm{\scriptsize Exp.}}{=}0.82\pm0.08 ,\label{Rn}
\end{align}
where the experimental information on the ratios has also been displayed. For the reader's convenience we will also discuss the ratio 
$R_{00}={R}/{R_n}$ proposed in \cite{Beneke:2003zv}
\beq\label{Roo}
R_{00}= 2 \frac{\bar \Gamma(\bar B^0\to \bar K^0\pi^0)}{\bar \Gamma(B^-\to \bar K^0\pi^-)}
\stackrel{\textrm{\scriptsize Exp.}}{=} 1.03\pm0.12 .
\eeq
The deviations from $1$ are experimentally only at the level of at most $2\sigma$. Also, of the CP asymmetries only ${\cal A}_{K^-\pi^+}^\mathrm{CP}$ is well measured, so that the values of the SCET parameters $\zeta_{(J)}^{BK},\zeta_{(J)}^{B\pi}$ and the phase of $A_{cc}^{\pi K}$ cannot at
present be reliably extracted from $B\to K \pi$ experimental data alone. We thus impose SU(3) symmetry $A_{cc}=A_{cc}^{\pi\pi}=A_{cc}^{\pi K}$,  $\zeta_{(J)}=\zeta_{(J)}^{BK}=\zeta_{(J)}^{B\pi}$ and construct
$\chi^2$ from observables in $B\to \pi\pi$ and $B\to \pi K$ decays. 

In $B\to \pi\pi$ decays there is experimental information on seven observables: the time dependent CP asymmetry $S_{\pi^+\pi^-}$,
the three CP averaged decay widths $\bar\Gamma(B\to \pi^+\pi^- ), \bar\Gamma(B\to \pi^0\pi^0 ),\bar \Gamma(B\to \pi^-\pi^0 )$ and 
three direct CP asymmetries ${\cal A}_{\pi^+\pi^-}^\mathrm{CP}, {\cal A}_{\pi^0\pi^0 }^\mathrm{CP}$ ${\cal A}_{\pi^-\pi^0}^\mathrm{CP}$. This latter is not used in the fit since $B^-\to \pi^-\pi^0$ is a $\Delta I=3/2$ process and thus does not receive QCD or charming penguin contributions, so that strong phases are generated only at NLO in $\alpha_S(m_b)$, while at LO the asymmetry is zero irrespective of the SCET parameters. 

In addition, the following  observables in $B\to K\pi$ decays are used in the $\chi^2$-fit: the four CP averaged decay widths $\bar\Gamma(B\to \bar K^0 \pi^0), \bar\Gamma(B\to K^-\pi^+),\bar\Gamma(B\to K^-\pi^0), \bar\Gamma(B\to \bar K^0\pi^-)$, but only three direct CP asymmetries ${\cal A}_{\bar K^0\pi^0}^\mathrm{CP}, {\cal A}_{K^-\pi^+}^\mathrm{CP}, {\cal A}_{K^-\pi^0}^\mathrm{CP}$. The prediction for the remaining direct CP asymmetry ${\cal A}_{\bar K^0\pi^-}^{\rm CP}$ can receive large corrections at NLO in $\alpha_S(m_b)$ from terms of the form $\lambda_u^{(s)} C_{1,2} \alpha_S(m_b)$. These can be comparable in size to LO terms proportional to $\lambda_u^{(s)}$ which come entirely
from QCD penguin operators. Also, the experimental information on $S_{K_S\pi^0}$ is not used in the $\chi^2$-fit, and  will
be discussed separately in Section \ref{Ssubsection}.
 
From the $\chi^2$-fit to the $B\to \pi\pi, K\pi$ data we then obtain
\beq\label{zetafit}
\begin{split}
\zeta&=(7.3\pm1.8)\times 10^{-2},\\
\zeta_J&=(10.3\pm1.6)\times 10^{-2},
\end{split}
\eeq
and 
\beq\label{Accfit}
\begin{split}
|A_{cc}|&=(46.8\pm0.8)\times 10^{-4}~{\rm GeV},\\
 \arg(A_{cc})&=156^\circ\pm 6^\circ,
\end{split}
\eeq
with $\chi^2/{\rm d.o.f.}=44.6/(13-4)$, where the largest discrepancies are in ${\cal A}_{\pi^0K^-}^{\rm CP}$, ${\cal A}_{\pi^-K^+}^{\rm CP}$ as can be seen from Table \ref{table:predPP}. This very high value of $\chi^2$ predominantly reflects the fact that the expected theory errors coming from NLO $1/m_b$ and $\alpha_S(m_b)$ terms and from SU(3) breaking are larger than experimental errors. If the estimates for these errors, to be discussed below, that are given as
first and second errors on the theoretical values in Table \ref{table:predPP}, are added quadratically to experimental errors in the definition of $\chi^2$, the resulting value is $\chi^2/{\rm d.o.f.}=8.9/(13-4)$ ($\chi^2/{\rm d.o.f.}=15.3/(13-4)$ if SU(3) breaking errors are taken to be correlated).
The extracted values of SCET parameters \eqref{zetafit}, \eqref{Accfit} agree within errors with similar extractions from only $\pi\pi$ data or a combination of $\pi\pi$ and 
$\pi K$ data without modes that depend on $\zeta_{(J)}^{BK}$ that were performed in Ref. \cite{Bauer:2005kd}. 

\begin{table}
\caption{Predictions for the CP violating $S$ parameters. The errors on the predictions are estimates of SU(3) breaking, $1/m_b$ corrections and errors
due to SCET parameters, respectively.}\label{table:predSPP}
\begin{ruledtabular}
\begin{tabular}{ccc} 
Mode  & Exp  & Theory \\ \hline
$\bar B^0\to \pi^-\pi^+$ & $-0.50\pm0.19$\footnote[1]{Error scaled according to PDG (S=1.5).} & $-0.86\pm0.07\pm0.07\pm0.02$\\
$\bar B^0\to \pi^0\pi^0$& $-$ 	& $0.71\pm0.34\pm0.33\pm0.10$\\\hline	
$\bar B^0\to \pi^0 K_S$	& $0.31\pm0.26$ 	& $0.80\pm0.02\pm0.02\pm0.01$\\
\end{tabular}
\end{ruledtabular}
\end{table}

Using the above values for the SCET parameters one can predict CP averaged decay widths and direct CP asymmetries
with the results listed in Tables \ref{table:predPP} and \ref{table:predSPP}.
These results were obtained in the limit of exact SU(3) and as such an error of $20\%$ is introduced as an estimate of SU(3) breaking effects in the relation $\zeta_{(J)}=\zeta_{(J)}^{BK}=\zeta_{(J)}^{B\pi}$.
Similarly an additional $20\%$ error on the magnitude and $20^\circ$ error on the strong phase is introduced due to SU(3) breaking 
and $1/m_b$ corrections
in the relation $A_{cc}=A_{cc}^{\pi K}=A_{cc}^{\pi\pi}$.  
These variations result in the first theoretical error estimate in Tables \ref{table:predPP}, \ref{table:predSPP}.
The second error is 
 an estimate of the remaining $1/m_b$ and $\alpha_S(m_b)$ corrections which we take to have a magnitude of $20\%$ of
the leading order contributions 
proportional to $\lambda_{u}^{(f)}$ and $\lambda_{t}^{(f)}$ CKM elements 
and assume that they introduce an error of $20^\circ$ on the strong phase (with the exception of $A_{cc}$ in $\Delta S=0$ decays, where the error of $5\%$ on the magnitude 
is assigned due to insertions of $O_{3,\dots, 10}$ operators leading to nonperturbative charm contributions proportional to $\lambda_u^{(d)}$ instead of $\lambda_c^{(d)}$, while in $\Delta S=1$ decays this effect is negligible). 

A different approach is used for the observables in which the subleading corrections are expected to be anomalously large.
This can happen in the decay modes in which the
leading contributions are not proportional to
$c_1^{(f)}, b_1^{(f)}$ or $c_2^{(f)}, b_2^{(f)}$, so that $1/m_b$ or $\alpha_S(m_b)$ suppressed contributions may lead to $O(1)$ corrections.  For instance, 
\beq
A_{B^-\to \bar K^0\pi^-}=\frac{G_F}{\sqrt 2} m_B^2 \Big\{ f_K \Big[\zeta^{B\pi}c_4^{(s)}+\zeta_J^{B\pi}b_4^{(s)}\Big]+\lambda_c^{(s)}A_{cc}^{\pi K}\Big\},
\eeq
where the meaning of SCET$_{\rm I}$ Wilson coefficients $c_{i}^{(s)}, b_i^{(s)}$ is the same as in Eqs. \eqref{A1/20}-\eqref{A3/2}.
The $\lambda_u^{(s)}$ part of the $B^-\to \bar K^0\pi^-$ can receive
corrections  of the form $C_{1,2}\alpha_S(m_b)$ which can compete in size with $\lambda_u^{(s)}$ part of $c_4^{(s)}$, $b_4^{(s)}$ in LO hard kernel. Thus only a conservative upper bound on the CP asymmetry is given in Table \ref{table:predPP}. Similar reasoning holds for $B^-\to K^0K^-$ decay 
\beq
\begin{split}
A_{B^-\to \bar K^0 K^-}=&\frac{G_F}{\sqrt 2} m_B^2 \Big\{ f_K \Big[\zeta^{BK}c_4^{(d)}+\zeta_J^{BK}b_4^{(d)}\Big]\\
&+\lambda_c^{(d)}A_{cc}^{KK}\Big\},\label{AKK}
\end{split}
\eeq
and for $\bar B^0\to K^0\bar K^0$ decays, whose amplitude is equal to \eqref{AKK} at LO in $1/m_b$ and $\alpha_S(m_b)$. Again, for the $\lambda_u^{(d)}$ ``tree" part of
the amplitude we can expect large NLO corrections from $C_{1,2}\alpha_S(m_b)$ terms. Because there is no CKM hierarchy between ``tree" and ``penguin" amplitudes the predictions for these two modes are even more uncertain, so that
we do not give any bound on CP asymmetry, 
while the estimate of $\alpha_S(m_b)$ corrections to branching ratios is 
taken to be the same as for $\bar B^0\to \pi^-\pi^+$.

Many of the errors cancel to a large extent in the ratios of the decay widths, which are then predicted more precisely
than the individual rates
\begin{align}
R-1&\stackrel{\rm{\scriptsize Th.}}{=}(3.7\pm1.5\pm3.9\pm2.1)\times 10^{-2},\label{RTh}\\
R_c-1&\stackrel{\rm{\scriptsize Th.}}{=}(8.8\pm2.3\pm6.9\pm1.2)\times 10^{-2},\\
R_n-1&\stackrel{\rm{\scriptsize Th.}}{=}(6.9\pm2.0\pm7.5\pm0.8)\times 10^{-2},\label{RnTh}\\
R_{00}-1&\stackrel{\rm{\scriptsize Th.}}{=}(-3.0\pm0.9\pm3.2\pm1.4) \times 10^{-2},
\end{align}
with the errors estimating the SU(3) breaking, $1/m_b$ and $\alpha_S(m_b)$ corrections, and the errors due to uncertainties on SCET
parameters respectively. Note that even though the SCET parameters were determined using these data as well, the agreement between predicted
ratios and the experimental values \eqref{R}-\eqref{Roo} is far from impressive. Experimentally $R,R_n<1$ with about a $2\sigma$ difference from the above expectations. 

An important input to the above theoretical predictions was provided by $\pi\pi$ data using SU(3) symmetry.  If instead no $\pi\pi$ data is used and the SCET parameters are determined solely from 
the $B\to \pi K$ decays, the theoretical and experimental values of the $R$ ratios \eqref{R}-\eqref{Rn} would agree within experimental errors, but with values of SCET parameters that differ significantly from \eqref{zetafit}, with $\zeta$ a factor of 4 larger, while $\zeta_J$ even flips sign!\footnote{The result of this fit is $\zeta=0.26\pm0.05$, $\zeta_J=-0.19\pm0.05$, $A_{cc}=(48.8\pm1.1)\times 10^{-2}$, $\arg A_{cc}=131\pm13^\circ$, which gives unacceptably small $Br(B^-\to \pi^-\pi^0)=(0.4^{+0.9}_{-0.4})\times 10^{-6}$.} This leads us to two conclusions, that (i)
the SCET expansion on itself is not in contradiction with $\pi K$ data and (ii) without permitting extremely large SU(3)
breakings that allow even for a change of sign for the nonperturbative parameters, there is a discrepancy between data and theoretical expectations. 
An independent check on the validity of the SCET $1/m_b$ expansion can be provided with a phenomenological analysis
of $B\to \pi K$ data using diagrammatic decomposition, where no SU(3) is assumed but annihilation topologies are
neglected \cite{CKMfitter,Imbeault:2003it}.

Especially interesting is 
the difference between  $R_n$ and $R_c$ \cite{Buras:1998rb,Fleischer:2002zv}. Defining the ``tree" and ``penguin" contents of $A_I^{\Delta I}$ analogously to the general decomposition Eq. \eqref{deftree}
\beq\label{deftree2}
A_I^{\Delta I} = \lambda_u^{(s)}T^{\Delta I}_I+\lambda_c^{(s)}P^{\Delta I}_I,
\eeq
and using the fact that $A_{cc}^{\pi K}$ dominates the amplitudes so that $\lambda_c^{(s)} P_{1/2}^{0}\gg \lambda_u^{(s)}T_{3/2,1/2}^{1},\lambda_u^{(s)}T_{1/2}^{0}$ and   $\lambda_c^{(s)} P_{1/2}^{0}\gg \lambda_c^{(s)} P_{3/2,1/2}^{1}$ (this latter 
hierarchy follows from the fact that $P_{3/2,1/2}^{1}$ receive only EWP contributions and are thus smaller than charming penguin
contributions in $P_{1/2}^{0}$ that arise from insertions of $O_{1,2}^c$), then completely model independently to first order in small parameters
\beq\label{Rc=Rn}
\begin{split}
R_c=R_n=&1-6\left[\Re\left(\frac{\lambda_u^{(s)}}{\lambda_c^{(s)}}\right)
\Re\left(\frac{T_{3/2}^1}{P_{1/2}^0}\right)+\Re\left(\frac{P_{3/2}^1}{P_{1/2}^0}\right)\right].
\end{split}
\eeq
The higher order corrections to the above relation amount to  a difference
\beq
(R_c-R_n)\stackrel{\rm{\scriptsize Th.}}{=}(1.8\pm0.9\pm0.9\pm0.4)\times 10^{-2},
\eeq
in the theoretical 
predictions \eqref{Rc}, \eqref{Rn}. This is to be contrasted with the difference between experimental values 
$(R_c-R_n)_{\rm Exp.}=0.18\pm0.13$. 
If this difference persists 
with reduced experimental errors, it will be very difficult to explain in 
the Standard Model. A possible explanation due to
 isospin violating new physics that changes EWP contributions and thus enhances the $P_{3/2}^1$ term in \eqref{Rc=Rn}
has been extensively discussed 
in the literature 
\cite{Gronau:2003kj,Buras:2003yc,Baek:2004rp,Agashe:2005hk,Grossman:1999av,Atwood:2003tg}

That the $\pi K$ data alone are not inconsistent with the theory is well demonstrated for instance by the sum of CP averaged decay widths 
\beq\label{LipkinSumRule}
\begin{split}
&2  \bar\Gamma(\bar B^0\to {\bar K^0\pi^0})- \bar\Gamma(\bar B^0\to {K^-\pi^+})+\\
 &+2 \bar\Gamma(B^-\to {K^-\pi^0})-
\bar\Gamma(B^-\to {\bar K^0\pi^-}),
\end{split}
\eeq
first discussed  by Lipkin \cite{Lipkin:1998ie} and by Gronau and Rosner \cite{Gronau:1998ep}. The sum \eqref{LipkinSumRule} does not depend on $A_{1/2}^0$ and thus on $A_{cc}^{K\pi}$. In terms of the $R$ ratios
the sum \eqref{LipkinSumRule} is
\beq
\Delta L=R_{00}-R+R_c-1.
\eeq
Using the expansion to first order in small parameters $P_{3/2,1/2}^1/P_{1/2}^0$, $T_{3/2,1/2}^{1,0}/P_{1/2}^0$ as in \eqref{Rc=Rn} for $R_{00}$ and $R$
\begin{align}
\begin{split}
R_{00}=&1+2\left[\Re\left(\frac{\lambda_u^{(s)}}{\lambda_c^{(s)}}\right)
\Re\left(\frac{T_{3/2}^1-2T_{1/2}^1}{P_{1/2}^0}\right)\right.\\
&\left.+\Re\left(\frac{P_{3/2}^1-2P_{1/2}^1}{P_{1/2}^0}\right)\right],
\end{split}
\\
\begin{split}
R=&1-4\left[\Re\left(\frac{\lambda_u^{(s)}}{\lambda_c^{(s)}}\right)
\Re\left(\frac{T_{3/2}^1+T_{1/2}^1}{P_{1/2}^0}\right)\right.\\
&\left.+\Re\left(\frac{P_{3/2}^1+P_{1/2}^1}{P_{1/2}^0}\right)\right],
\end{split}
\end{align}
it is easy to check that $\Delta L$ is only of second order in small parameters in accordance with the fact that all interference terms with $A_{1/2}^0$ in \eqref{LipkinSumRule} cancel.
In $\Delta L$ the dependence on $A_{cc}^{K\pi}$ drops out completely leading to the value of  $\Delta L$  that is $|\lambda_u^{(s)2}/\lambda_c^{(s)2})|\sim \lambda^4=2 \cdot 10^{-3}$ CKM suppressed. 
Experiment at present is consistent with vanishing $\Delta L$ at one $\sigma$
\beq
\Delta L\stackrel{\rm{\scriptsize Exp.}}{=}0.19\pm0.14,
\eeq
while the theoretical prediction using LO SCET expressions and \eqref{zetafit}, \eqref{Accfit}, is 
\beq
\Delta L\stackrel{\rm{\scriptsize Th.}}{=}(2.0\pm0.9\pm0.7\pm0.4)\times 10^{-2}.
\eeq

Another very precisely predictable quantity is the sum of partial decay differences $\Delta \Gamma=\Gamma(\bar B\to f)-\Gamma(B\to \bar f)$
\beq
\begin{split}
\Delta_{\sum}=&2 \Delta\Gamma(B^-\to {K^-\pi^0})-\Delta\Gamma(B^-\to {\bar K^0\pi^-})\\
&+2  \Delta\Gamma(\bar B^0\to {\bar K^0\pi^0})- \Delta\Gamma(\bar B^0\to {K^-\pi^+}) .
\end{split}
\eeq
In the limit of exact isospin and no EWP $\Delta_{\sum}$ vanishes \cite{Atwood:1997iw,Gronau:2005gz,Gronau:2005kz}. However, 
even in the presence of EWP, the corrections are subleading in the $1/m_b$ expansion \cite{Gronau:2005kz}. Using 
isospin decomposition \eqref{A1/20}-\eqref{A3/2} and defining ``tree" and ``penguin" terms of the corresponding
reduced matrix elements \eqref{deftree2} one has
\beq
\Delta_{\sum}=- 24 \Im\big[\lambda_u^{(s)}\lambda_c^{(s)*}\big]
\Im\big[\big(T_{1/2}^1-T_{3/2}^1\big)P_{3/2}^{1*} +T_{3/2}^1P_{1/2}^{1*}\big].
\eeq
The ``penguin" terms $P_{3/2,1/2}^1$ receive only EWP contributions, while the ``tree" terms $T_{3/2,1/2}^1$ 
are a sum of tree and EWP contributions, with tree contributions dominating due to larger Wilson coefficients. 
At leading order in $1/m_b$ and $\alpha_S(m_b)$ expansion the strong phase is nonzero only due to $A_{cc}^{K\pi}$. 
Thus at this order  $T_{3/2,1/2}^1$ and $P_{3/2,1/2}^1$ are real (in our phase convention) so that $\Delta_{\sum}=0$ to the
order that we are working.

\subsection{Decays into isosinglet states}\label{isossect}
At present the experimental data on the decays with $\eta^{(')}$ are not
abundant. Of the 55 observables describing the complete set of $\bar B^0, B^-$ and $\bar B_s^0$ decays to two body final states with $\eta^{(')}$, only 11 have been measured so far. We will thus assume the SU(3) relations \eqref{zetaetarel}-\eqref{zetaSU3eta}, \eqref{AccSU3}, \eqref{Accpieta} between SCET parameters as in the previous subsection 
and furthermore use the determination of the SCET parameters $\zeta_{(J)}$ and $A_{cc}$ from the $\pi\pi$ and $\pi K$ data given
in Eqs. \eqref{zetafit},  \eqref{Accfit}. The remaining parameters specific to isosinglet modes, $\zeta_{(J)g}$, describing gluonic contributions to $B\to \eta^{(')}$ form factors, and  
$A_{ccg}$, describing the gluonic parts of charming penguin, Fig. \ref{ggsgluon}, are then fixed from a $\chi^2$-fit to observables in the isosinglet modes.

\begin{table*}
\caption{Predicted CP averaged branching ratios ($\times 10^{-6}$, first row) and direct CP asymmetries (second row for each mode) for $\Delta S=0$ and $\Delta S=1$ $B$ decays (separated by 
horizontal line) to isosinglet pseudoscalar mesons. The Theory I and Theory II columns give predictions corresponding to Solution I, II 
sets of SCET parameters. The errors on the predictions are estimates of SU(3) breaking, $1/m_b$ corrections and errors
due to SCET parameters, respectively. No prediction on CP asymmetries is given, if $[-1,1]$ range is allowed at $1\sigma$.}\label{table:predPPiso}
\begin{ruledtabular}
\begin{tabular}{llll} 
Mode  & Exp.  & Theory I & Theory II\\ \hline
$B^-\to \pi^-\eta$		& $4.3\pm0.5 \;(S=1.3)$ 		& $4.9\pm 1.7\pm 1.0\pm 0.5$		& $5.0\pm 1.7\pm 1.2\pm 0.4$\\
						& $-0.11\pm0.08$	& $0.05\pm 0.19\pm 0.21\pm 0.05$	& $0.37\pm 0.19\pm 0.21\pm 0.05$\\
$B^-\to \pi^-\eta'$		& $2.53\pm0.79 \;(S=1.5)$ 	& $2.4\pm 1.2\pm 0.2\pm 0.4$		& $2.8\pm 1.2\pm 0.3\pm 0.3$\\
						& $0.14\pm0.15$		& $0.21\pm 0.12\pm 0.10\pm 0.14$	& $0.02\pm 0.10\pm 0.04\pm 0.15$\\
$\bar B^0\to \pi^0\eta$	& $<2.5$ 				& $0.88\pm 0.54\pm 0.06\pm 0.42$	& $0.68\pm 0.46\pm 0.03\pm 0.41$\\
						& $-$				& $0.03\pm 0.10\pm 0.12\pm 0.05$	& $-0.07\pm 0.16\pm 0.04\pm 0.90$\\
$\bar B^0\to \pi^0\eta'$& $<3.7$ 				& $2.3\pm 0.8\pm 0.3\pm 2.7$ & $1.3\pm 0.5\pm 0.1\pm 0.3$\\
						& $-$				& $-0.24\pm 0.10\pm 0.19\pm 0.24$ & $-$\\	
$\bar B^0\to \eta\eta$	& $<2.0$ 				& $0.69\pm 0.38\pm 0.13\pm 0.58$ & $1.0\pm 0.4\pm 0.3\pm 1.4$\\
						& $-$				& $-0.09\pm 0.24\pm 0.21\pm 0.04$ & $0.48\pm 0.22\pm 0.20\pm 0.13$\\	
$\bar B^0\to \eta\eta'$	& $<4.6$ 				& $1.0\pm 0.5\pm 0.1\pm 1.5$ & $2.2\pm 0.7\pm 0.6\pm 5.4$\\
						& $-$				& $-$ & $0.70\pm 0.13\pm 0.20\pm 0.04$\\
$\bar B^0\to \eta'\eta'$& $<10$ 				& $0.57\pm 0.23\pm 0.03\pm 0.69$ & $1.2\pm 0.4\pm 0.3\pm 3.7$\\
						& $-$				& $-$ & $0.60\pm 0.11\pm 0.22\pm 0.29$\\\hline
$\bar B^0\to \bar K^0\eta'$	& $63.2\pm4.9 \;(S=1.5)$ 	& $63.2\pm 24.7\pm 4.2\pm 8.1$ & $62.2\pm 23.7\pm 5.5\pm 7.2$\\
							& $0.07\pm0.10 \;(S=1.5)$	& $0.011\pm 0.006\pm 0.012\pm 0.002$ & $-0.027\pm 0.007\pm 0.008\pm 0.005$\\
$\bar B^0\to \bar K^0\eta$	& $<1.9$ 		& $2.4\pm 4.4\pm 0.2\pm 0.3$ & $2.3\pm 4.4\pm 0.2\pm 0.5$\\
							& $-$				& $0.21\pm 0.20\pm 0.04\pm 0.03$ & $-0.18\pm 0.22\pm 0.06\pm 0.04$\\
$B^-\to K^-\eta'$	& $69.4\pm2.7$ 				& $69.5\pm 27.0\pm 4.3\pm 7.7$ & $69.3\pm 26.0\pm 7.1\pm 6.3$\\
					& $0.031\pm0.021$	& $-0.010\pm0.006\pm0.007\pm0.005$& $0.007\pm0.005\pm0.002\pm0.009$\\
$B^-\to K^-\eta$	& $2.5\pm0.3$ 				& $2.7\pm 4.8\pm 0.4\pm 0.3$ & $2.3\pm 4.5\pm 0.4\pm 0.3$\\
					& $-0.33\pm0.17\;(S=1.4)$	& $0.33\pm 0.30\pm 0.07\pm 0.03$ & $-0.33\pm 0.39\pm 0.10\pm 0.04$\\						
\end{tabular}
\end{ruledtabular}
\end{table*}

As in the previous subsection only CP averaged decay widths and direct CP asymmetries are used in this determination, while the discussion of
$S_{K_S\eta'}$ is relegated to Section \ref{Ssubsection}. This leaves four observables from $\Delta S=0$ decays: two CP averaged decay widths  and two direct CP asymmetries in $B^-\to\pi^-\eta^{(')}$, and six observables in $\Delta S=1$ decays: three CP averaged decay widths and three direct CP asymmetries in $\bar B^0\to \bar K^0 \eta'$ and $B^-\to K^-\eta^{(')}$ modes. 

In the above modes the functions $\zeta_g$ and $\zeta_{Jg}$ enter in the combinations
$c_1^{(f)}\zeta_g + b_1^{(f)}\zeta_{Jg}$ and $c_4^{(f)}\zeta_g+b_4^{(f)}\zeta_{Jg}$, with the latter being numerically much smaller (cf. \eqref{Cvalues}, \eqref{cmatch}, \eqref{b8}). Since $c_{1}^{(f)}\simeq b_1^{(f)}$ we expect
the combination $\zeta_g + \zeta_{Jg}$ to be relatively well determined from the data, with the orthogonal combination only poorly constrained. We thus define
\begin{align}
\zeta_g^{\pm}&=\zeta_g\pm\zeta_{Jg},
\end{align}
which are then fit from the data. We find two sets of SCET parameters that minimize $\chi^2$:

\underline{Solution I:}
\begin{align}
\zeta_g^{+}&=(-9.9\pm2.4)\times 10^{-2},\label{zetagfit}\\
\zeta_g^{-}&=(-3.5\pm14.6)\times 10^{-2}, \label{zetag-fit}\\
|A_{ccg}|&=(35.8\pm 1.9)\times 10^{-4}~{\rm GeV}, \label{Accgfit}\\
 \arg(A_{ccg})&=-109^\circ\pm 3^\circ, \label{argAccgfit}
\end{align}
with $\chi^2/{\rm d.o.f.}=25.0/(10-4)$, where the largest discrepancy with data is in  ${\cal A}_{\eta K^-}^{CP}$. 
The value of $\chi^2$ is reduced to $\chi^2/{\rm d.o.f.}=7.6/(10-4)$, if theoretical errors due to SU(3) breaking and estimated NLO corrections are added quadratically to experimental errors in the definition of $\chi^2$. The second solution, on the other hand,

\underline{Solution II:}
\begin{align}
\zeta_g^{+}&=(-6.6\pm4.3)\times 10^{-2},\label{zetagfitII}\\
\zeta_g^{-}&=(-11.2\pm28.7)\times 10^{-2}, \label{zetag-fitII}\\
|A_{ccg}|&=(36.2\pm 2.2)\times 10^{-4}~{\rm GeV}, \label{AccgfitII}\\
 \arg(A_{ccg})&=68^\circ\pm 4^\circ,\label{argAccgfitII}
\end{align}
has $\chi^2/{\rm d.o.f.}=40.8/(10-4)$ or $\chi^2/{\rm d.o.f.}=5.4/(10-4)$, if theoretical errors are added in the definition of $\chi^2$. The largest discrepancies with experimental data in this case is in ${\cal A}_{\eta\pi^-}^{CP}$ while the prediction for ${\cal A}_{\eta K^-}^{CP}$ agrees well with data in contrast to
Solution I.

The strong
phases of the gluonic charming penguin in the two solutions lie in opposite quadrants, while the values of $|A_{cc,g}|$ and $\zeta_g^{\pm}$ agree between the two solutions.
The gluonic contribution to the $B\to \eta^{(')}$ form factors, $\zeta_g+\zeta_{Jg}$, is similar in size to $\zeta$ and $\zeta_{J}$ in \eqref{zetafit} as expected from SCET counting, 
Using Eq. \eqref{f_+} we find in the SU(3) limit and at LO in $1/m_b$ and $\alpha_S(m_b)$
\begin{align}\label{fetaq}
f_+^{B\eta_q}(0)&=\left\{
\begin{matrix}
(-2.3\pm4.8)\times 10^{-2},\\
(4.5\pm8.6)\times 10^{-2},
\end{matrix}
\right.
\\
f_+^{B\eta_s}(0)&=\left\{
\begin{matrix}
(-9.9\pm2.4)\times 10^{-2},\\
(-6.6\pm4.3)\times 10^{-2},
\end{matrix}\label{fetas}
\right.
\end{align} 
to be compared with $f_+^{B\pi}(0)=0.176\pm0.007$, that is obtained using the results of $\pi\pi$, $\pi K$ fit \eqref{zetafit}. The upper (lower) rows in \eqref{fetaq}, \eqref{fetas} correspond to values in Solution I (Solution II),
where only experimental errors due to the extracted SCET parameters are shown. 
Because of the large experimental uncertainties, the gluonic contributions to the form factors are still consistent with 
zero at a little above the $1\sigma$ level in Solution II. The gluonic charming penguin $A_{ccg}$ on 
the other hand is shown to be nonzero in both sets of solutions and is of similar size to $A_{cc}$ in \eqref{Accfit} in agreement with SCET counting.

That the gluonic contribution is of leading order in $1/m_b$ has already been recognized in the context of QCD factorization.
In the phenomenological analysis of Ref. \cite{Beneke:2002jn}, the gluonic contributions to the $B\to \eta'$ form factor are proportional to $F_2$, a parameter not known from other sources and given the rather arbitrary values of $F_2=0,0.1$.  With these values, the gluonic contribution accounts for $0\%,40\%$ of the $B\to \eta'$ form factor with constructive interference between the gluonic and the remaining contributions.  This can be compared with our analysis where, in the $B\to \eta'$ form factor, 
destructive interference between $\zeta_{(J)g}$ and $\zeta_{(J)}$ terms is found with the gluonic contribution from $\zeta_{g}+\zeta_{Jg}$ 2.1 (1.4) times larger than the contribution from $\zeta_{}+\zeta_{J}$ in Solution I(II)).

The predicted branching ratios and
CP asymmetries using the above values are compiled in 
Tables \ref{table:predPPiso} and \ref{table:predSPPiso}. 
The errors due to SU(3) breaking and $1/m_b$ or $\alpha_S(m_b)$ corrections are estimated in the same way as in
previous subsection. An error of $20\%$ and a variation on charming penguin strong phase of $20^\circ$ is assigned to relations \eqref{zetaetarel}-\eqref{zetaSU3eta} and \eqref{AccSU3}, \eqref{Accpieta} giving the first error estimate in the Table \ref{table:predPPiso}. The remaining $1/m_b$ and $\alpha_S(m_b)$ errors, listed as second error estimates in Table \ref{table:predPPiso}, are obtained by varying the size and strong phase of leading order amplitudes proportional to $\lambda_u^{(f)}$ or $\lambda_t^{(f)}$  by $20\%$ and $20^\circ$ respectively.

A prominent feature of $B\to K\eta^{(')}$ decays is the large disparity between the branching ratios for $B\to K\eta'$
and $B\to K\eta$ decays. In the SCET framework this is quite naturally explained through a constructive and destructive
interference of different terms
in the amplitudes as has been first suggested in \cite{Lipkin:1990us,Lipkin:1998ew}. 
Specifically, the amplitudes $A_{B\to K \eta^{(')}}$ are related to $A_{B\to K\eta_{q}}$ and $A_{B\to K\eta_{s}}$ through a rotation \eqref{eta'mix}
\begin{align}
A_{\bar B\to \bar K\eta'}&=\cos \phi A_{\bar B\to \bar K\eta_s}+\sin \phi A_{\bar B\to \bar K \eta_q},\\
A_{\bar B\to \bar K\eta}&=-\sin\phi A_{\bar B\to \bar K\eta_s }+\cos\phi A_{\bar B\to \bar K \eta_q},
\end{align}
with $\phi=(39.3\pm1.0)^\circ$, so that $\cos \phi\simeq \sin \phi$. There is therefore a constructive interference in 
$A_{\bar B\to \bar K \eta'}$ and a destructive interference in $A_{\bar B\to \bar K\eta }$ provided that $A_{\bar B\to \bar K\eta_q}\simeq A_{\bar B\to \bar K \eta_s}$, which is exactly what is
found in SCET. In $B\to K \eta^{(')}$ there is a similar hierarchy of terms that was found in $B\to \pi K$ decays so that charming penguin contributions are the largest, while tree contributions are $\lambda_u^{(s)}/\lambda_c^{(s)}\sim 0.04$ suppressed, and QCD penguin and EWP contributions $\sim \max[C_3,\dots,C_{10}]/\alpha_S(2m_c) C_1\sim 0.1$ suppressed. The largest contributions to the amplitudes are thus
\begin{align}
\begin{split}
A_{B^-\to \eta_s K^-}=&\frac{G_F}{\sqrt{2}} m_B^2  \Big\{ \lambda_c^{(s)}\left(A_{cc}^{K \eta_s}+A_{cc,g}^{K\eta_s}\right) +\cdots\Big\}\\
\simeq& A_{\bar B^0\to \eta_s \bar K^0},
\end{split}
\\
\begin{split}
A_{B^-\to \eta_q K^-}=&\frac{G_F}{\sqrt{2}} m_B^2 \Big\{\lambda_c^{(s)}\Big(
 \frac{A_{cc}^{\pi K}}{\sqrt 2}+\sqrt{2}A_{cc,g}^{K\eta_q}\Big) +\cdots\Big\}\\
\simeq& A_{\bar B^0\to \eta_q \bar K^0},
\end{split}
\end{align}
where the $\simeq$ sign denotes equality up to smaller terms represented by ellipses. This gives for the
$A_{B\to K\eta }$ and $A_{B\to K\eta'}$ amplitudes
\begin{align}
\begin{split}\label{etaKsuppression}
A_{B^-\to \eta K^-}=&\frac{G_F}{\sqrt{2}} m_B^2  \lambda_c^{(s)}\cos\phi  
\Big\{(\sqrt{2}-\tan\phi)A_{cc,g}\\
&+\left(\frac{1}{\sqrt 2} -\tan\phi\right)A_{cc} +\cdots\Big\},
\end{split}
\\
\begin{split}\label{eta'Ksuppression}
A_{B^-\to \eta' K^-}=&\frac{G_F}{\sqrt{2}} m_B^2  \lambda_c^{(s)}\cos\phi  
\Big\{(1+\sqrt{2}\tan\phi)A_{cc,g}\\
&+\left(1 +\frac{1}{\sqrt 2}\tan\phi\right)A_{cc} +\cdots\Big\},
\end{split}
\end{align}
where the SU(3) relations \eqref{AccSU3} and \eqref{Accpieta} were used. The same expressions hold for $\bar B^0\to \bar K^0\eta^{(')}$ amplitudes. The coefficient
in front of $A_{cc}$ in the ${B\to K\eta }$ amplitude is $(1-\sqrt{2} \tan\phi)/(\sqrt{2}+\tan \phi)=-0.07$ suppressed
compared to the one in the ${B\to K\eta'}$ amplitude. The relative suppression of the $A_{ccg}$ contributions $(\sqrt{2}- \tan\phi)/(1+\sqrt{2} \tan \phi)=0.28$ is not as strong, so that $B^-\to \eta K^- $ and $\bar B^0\to \eta \bar K^0 $
are dominated by gluonic charming penguin contributions. In the absence of the $A_{ccg}$ term, the corresponding branching
ratios for $B\to \eta K$ would be an order of magnitude smaller still, i.e. of order $O(10^{-7})$ instead of $O(10^{-6})$.

\begin{table*}
\caption{Predictions for the CP violating $S$ parameters. The errors on the predictions are estimates of SU(3) breaking, $1/m_b$ corrections and errors
due to SCET parameters, respectively.}\label{table:predSPPiso}
\begin{ruledtabular}
\begin{tabular}{llll} 
Mode  & Exp.  & Theory I& Theory II \\ \hline
$\bar B^0\to \pi^0\eta$ 	& $-$ 	& $-0.90\pm 0.08\pm 0.03\pm 0.22$ & $-0.67\pm 0.14\pm 0.03\pm 0.81$\\
$\bar B^0\to \pi^0\eta'$ 	& $-$ 	& $-0.96\pm 0.03\pm 0.05\pm 0.11$ & $-0.60\pm 0.08\pm 0.08\pm 1.30$\\
$\bar B^0\to \eta\eta$ 		& $-$ 	& $-0.98\pm 0.06\pm 0.03\pm 0.09$ & $-0.78\pm 0.19\pm 0.12\pm 0.22$\\
$\bar B^0\to \eta\eta'$ 	& $-$ 	& $-0.82\pm 0.02\pm 0.04\pm 0.77$ & $-0.71\pm 0.14\pm 0.19\pm 0.29$\\
$\bar B^0\to \eta'\eta'$ 	& $-$ 	& $-0.59\pm 0.05\pm 0.08\pm 1.10$ & $-0.78\pm 0.09\pm 0.19\pm 0.23$\\\hline
$\bar B^0\to K_S \eta'$ 	& $0.50\pm0.13\;(S=1.4)$ & $0.706\pm 0.005\pm 0.006\pm 0.003$ & $0.715\pm 0.005\pm 0.008\pm 0.002$\\
$\bar B^0\to K_S \eta$ 		& $-$ 	& $0.69\pm 0.15\pm 0.05\pm 0.01$ & $0.79\pm 0.14\pm 0.04\pm 0.01$\\
\end{tabular}
\end{ruledtabular}
\end{table*}

Note that the partial cancellation of charming penguin contributions in $B\to \eta K$ amplitudes occurs regardless of the strong phases carried by $A_{cc}$ and 
$A_{cc,g}$. In particular, the strong phases of $A_{cc}$ and 
$A_{cc,g}$ that were obtained from the fit \eqref{Accfit}, \eqref{argAccgfit}, \eqref{argAccgfitII}, are  near $180^\circ$ and $90^\circ\mod 180^\circ$ respectively,  so that there is little further
cancellation between the  
two terms in \eqref{etaKsuppression}. 
A similar pattern of destructive and constructive interference has also been observed in the framework of 
QCD factorization \cite{Beneke:2002jn,Beneke:2003zv}.

Note that the enhancement of $Br(B\to K\eta')$ over $Br(B\to \pi K)$ is almost entirely due to the additional gluonic charming penguin $A_{ccg}$ that arises from
the annihilation of
the $n$ direction collinear quark with the spectator quark, where simultaneously two new $n$ collinear gluons are emitted, Fig. \ref{ggsgluon}. 
These contributions correspond to singlet penguin amplitude $s$ in the diagrammatic SU(3) approach, see appendix \ref{SU3decomposition}. Explicitly, in the SU(3) limit
\beq
\begin{split}
\frac{A_{B^-\to \eta'K^-}}{A_{\bar B^0\to \pi^+K^-}}&\simeq \Big(\cos \phi+\frac{\sin\phi}{\sqrt 2}\Big) \frac{A_{cc}}{A_{cc}}\\
&+\big(\cos \phi +\sqrt{2}\sin \phi\big) \frac{A_{ccg}}{A_{cc}}+\cdots\\
&\simeq1.22+1.67 \frac{A_{ccg}}{A_{cc}},
\end{split}
\eeq
where ellipses denote numerically smaller terms. In the limit $A_{ccg}\to 0$ thus $Br(B\to K\eta')$ and $Br(B\to \pi K)$ would be of similar size, while in SCET we expect $A_{ccg}\sim A_{cc}$ which provides the observed enhancement. 

In QCD factorization, $A_{ccg}$ is perturbative and is proportional to the $F_2$ contribution in the  
$F_0^{B\to \eta, \eta'}$ form factors of Ref. \cite{Beneke:2002jn}.  As previously discussed, $F_2$ is not known from other sources and the authors of \cite{Beneke:2002jn} assume the arbitrary values $F_2=0,0.1$. 
Other mechanisms that were proposed in 
the literature to explain the large $Br(B\to K \eta')$ are found to be either $\alpha_S(m_b)$ or 
$1/m_b$ suppressed in SCET. The contributions due to $b\to s g g\to s\eta'$ coupling that would arise from integrating out the charm loop, Fig. \ref{charm_gluon}, 
\cite{Ali:1997ex,Beneke:2002jn,Simma:1990nr}
 and could be interpreted as effective charm content of $\eta'$ meson
\cite{Halperin:1997as,Yuan:1997ts,Beneke:2002jn}, lead to a $1/m_b^2$ suppressed operator \eqref{suppr} once matching
to SCET$_{\rm II}$ is performed. A mechanism in which one gluon is emitted from $b$ or $s$ quarks, with 
the other gluon coming from charm loop or from $O_{8g}$ insertion, Fig. \ref {08g},  leads to 
either $\alpha_S(m_b)$ or 
$1/m_b$ suppressed contributions as already discussed below Eq. \eqref{Q9s10s}. The hard spectator contribution $b\to s g^* g^* \to s\eta'$ discussed in \cite{Ahmady:1997fa}, where one of the hard off-shell gluons is emitted from the spectator quark, 
matches onto power suppressed SCET$_{\rm I}$ operators with additional soft and collinear spectator quark fields obtained by integrating out the hard gluon (and other hard degrees of freedom). Similarly, the gluon condensate mechanism of Ref. \cite{Eeg:2005bq} corresponds to a matching onto power suppressed operators with additional soft gluon fields once the intermediate hard-collinear gluon is integrated out. 

Since the smallness of $\Br(B\to \eta K)$ arises
from two large numbers cancelling, the predictions for this mode are rather uncertain, with modest variations on input 
parameters leading to larger relative variations on the observables. This could be used in the future to better constrain the SCET parameters $\zeta_{(J)g}$, $A_{ccg}$. Of special interest are the direct CP asymmetries
in $B^-\to \eta K^-$ and $\bar B^0\to \eta \bar K^0$ that can resolve between the two solutions (cf. Table \ref{table:predPPiso}). Defining the 
``tree" and ``penguin" amplitudes as in \eqref{deftree}, one has
\begin{align}
\begin{split}
T_{\bar B^0\to  \eta_q \bar K^0}=&\frac{G_{F}}{\sqrt 2} m_B^2 \frac{f_{\eta_q}}{\sqrt{2}}\Big[\zeta^{BK}\big(C_2+\frac{1}{N}C_1\big)\\
&+\zeta_J^{BK}\big(C_2+\frac{1}{N}\big(1+\langle x^{-1}\rangle_{\eta_q} \big)C_1 \big)\Big]+\cdots,
\end{split}
\end{align}
\begin{align}
\begin{split}
T_{B^-\to\eta_q K^-}=&T_{\bar B^0\to \eta_q \bar K^0}+\frac{G_{F}}{\sqrt 2} m_B^2 \frac{f_{K}}{\sqrt{2}}\Big[\zeta^{B\eta_q}\big(C_1+\frac{1}{N}C_2\big)\\
&+\zeta_J^{B\eta_q}\big(C_1+\frac{1}{N}\big(1+\langle x_q^{-1}\rangle_K \big)C_2 \big)\Big]+\cdots,
\end{split}
\end{align}
where ellipses denote smaller terms coming from insertions of QCD penguin and EWP operators. 
The ``tree" amplitude $T_{B^-\to\eta_q K^-}$ receives contributions from configurations with two $n$-collinear
gluons, Fig. \ref{matching}, that are part of $\zeta^{B\eta_q}_{(J)}$ SCET parameters \eqref{zetaetarel}. In the diagrammatic approach these terms correspond to often neglected annihilation amplitudes as shown in appendix \ref{SU3decomposition}. We find them to be of LO in $1/m_b$ and should be kept in the analysis (in $\Delta S=1$ decays they are CKM suppressed and are thus numerically significant only for direct CP asymmetries).

At LO in $\alpha_S(m_b)$ the ``tree" amplitudes $T_{\bar B^0\to  \eta_s \bar K^0}$ and $T_{B^-\to\eta_s K^-}$ do not receive  contributions from
tree operators $O_{1,2}^u$, so that
\beq
T_{\bar B^0\to  \eta_s \bar K^0}\ll T_{\bar B^0\to  \eta_q \bar K^0}, \quad T_{B^-\to\eta_s K^-}\ll T_{B^-\to\eta_q K^-},
\eeq
from which one obtains an approximate relation
\beq\label{relsize}
\frac{T_{\bar B\to  \eta \bar K}}{\cos\phi}\simeq \frac{T_{\bar B\to  \eta' \bar K}}{\sin\phi},
\eeq
where $\bar B$ ($\bar K$) can be either $B^-$($K^-$) or $\bar B^0$ ($\bar K^0$). No such simple relation exists between 
$P_{\bar B\to  \eta \bar K}$ and $P_{\bar B\to  \eta' \bar K}$. These are given by \eqref{etaKsuppression} and
\eqref{eta'Ksuppression} (after division by $\lambda_c^{(s)}$) up to numerically smaller terms, and have a hierarchy $P_{\bar B\to  \eta \bar K} \ll P_{\bar B\to  \eta' \bar K}$ as already discussed before. 

Defining
\beq\label{r_f}
r_f e^{i\delta_f}= -2 \Im \left(\frac{\lambda_u^{(s)}}{\lambda_c^{(s)}}\right)\frac{T_{\bar B\to f}}{P_{\bar B\to f}}, 
\eeq
where $-2 \Im \lambda_u^{(s)}/\lambda_c^{(s)}=0.037$, the CP asymmetries are to first order in small parameter $r_f$
\beq\label{ACPexp}
{\cal A}_f^\mathrm{CP}=r_f \sin\delta_f +O(r_f^2).
\eeq
The values of $r_f$ and the strong phase difference $\delta_f$ between ``tree" and ``penguin" amplitudes for $\bar B\to \eta^{(')}\bar K$ decays are given in Table \ref{table:r_f}. While the ``tree" amplitudes in $\bar B\to \eta' \bar K$ and $\bar B\to \eta\bar K$ are of approximately the same size \eqref{relsize}, the ``penguin" amplitudes in $\bar B\to \eta \bar K$ are on the contrary suppressed as discussed above, leading to an order of magnitude larger value for $r_{\eta\bar K}$. Furthermore, in $\bar B\to \eta^{}\bar K$ decays the strong phase difference is predominantly determined by the gluonic charming penguin $A_{ccg}$ due to a much larger suppression of $A_{cc}$, so that
the corresponding CP asymmetries are very sensitive to the value of $\arg A_{ccg}$.

A different type of cancellation occurs in the $\bar B^0\to \pi^0\eta$ amplitude. The two relevant amplitudes are (taking $\langle x^{-1}\rangle_\pi=\langle x^{-1}\rangle_{\eta_q}$)
\begin{align}
\begin{split}
A_{\bar B^0\to \pi^0\eta_s} =&\frac{G_F}{\sqrt 2}m_B^2\big[\frac{c_{2}^{(d)}}{\sqrt 2}f_{\pi}\zeta^{B\eta_s}
+\frac{b_{2}^{(d)}}{\sqrt 2}f_{\pi}\zeta_{J}^{B\eta_s}\\
&-\frac{\lambda_c^{(d)}A_{cc,g}^{\pi\eta}}{\sqrt 2}+\cdots \big],
\end{split}
\\
\begin{split}\label{cancpi0}
A_{\bar B^0\to \pi^0\eta_q} =&\frac{G_F}{\sqrt 2}m_B^2\Big[\frac{c_{2}^{(d)}}{2}(f_{\pi}\zeta^{B\eta_q}-f_{\eta_q}\zeta^{B\pi})\\
&+\frac{b_{2}^{(d)}}{2}(f_{\pi}\zeta_{J}^{B\eta_q}-f_{\eta_q}\zeta_{J}^{B\pi}) \\
&-\lambda_c^{(d)}\Big(A_{cc,g}^{\pi\eta}+A_{cc}^{\pi\pi}\Big)+\cdots\Big],
\end{split}
\end{align}
where the ellipses represent the numerically smaller contributions from $Q_{3d,\dots,6d}$ operators, while
$b_{2}^{(d)}$ is evaluated as $\lambda_u^{(d)}[C_2+(1+\langle x^{-1}\rangle_\pi)C_1/N]+\cdots$.
Since $f_{\eta_q}\simeq f_\pi$
and $\zeta_{(J)}^{B\eta_q}\simeq\zeta_{(J)}^{B\pi}+2 \zeta_{(J)}^{B\eta_s} $, the tree
amplitudes $T_{\bar B^0\to \pi^0\eta^{(')}}$ are predominantly coming from  
$\zeta_{(J)}^{B\eta_s}$, i.e. from gluonic contributions.  Furthermore, the combination $c_2^{(d)}\zeta_g +b_2^{(d)}\zeta_{Jg}$ is a linear combination of $\zeta_g^{+}$ and $\zeta_{g}^{-}$. Since the latter parameter is largely unknown, cf. \eqref{zetag-fit}, \eqref{zetag-fitII}, this leads to relatively large errors on the predicted observables
in $\bar B^0\to \pi^0\eta^{(')}$ decays  as can be seen from Tables \ref{table:predPPiso} and \ref{table:predSPPiso}. Measuring these observables would thus greatly improve our knowledge of $\zeta_g^-$.

The situation is reversed for charming penguin contributions. 
In $\bar B^0\to \pi^0\eta$ the contributions from $A_{cc,g}^{\pi\eta}$ partially cancel, just like in $B\to K\eta$. However, unlike $B\to K\eta$, the nongluonic
contribution $A_{cc}^{\pi\pi}$ is present only in $A_{\bar B^0\to \pi^0\eta_q}$ and not in $A_{\bar B^0\to \pi^0\eta_s}$
and therefore does not cancel in $\bar B^0\to \pi^0\eta$. The same conclusions regarding charming penguins hold for $B^-\to \pi^-\eta$. Note that unlike ${\bar B^0\to \pi^0\eta_q}$, the ``tree" term in $A_{B^-\to \eta_q\pi^-}$, on 
the other hand, is not predominantly gluonic since there is no equivalent cancellation to the one in the ``tree" term of \eqref{cancpi0}.

The predictions are fairly uncertain also for observables in $\bar B^0\to \eta^{(')} \eta^{(')}$ decays, since these
depend on both $\zeta_g^{+}$ and $\zeta_{g}^{-}$, similarly to $\bar B\to \eta^{(')} \pi$ decays. We have
\begin{align}
\begin{split}\label{B0etaeta}
A_{\bar B^0\to \eta^{} \eta^{}}=&A_{\bar B^0\to \eta_q \eta_q}\cos^2\phi +A_{\bar B^0\to \eta_s \eta_s}\sin^2\phi \\
&-A_{\bar B^0\to \eta_q \eta_s} \sin 2\phi ,
\end{split}
\\
\begin{split}
A_{\bar B^0\to \eta^{} \eta'}=&\big( A_{\bar B^0\to \eta_q \eta_q}-A_{\bar B^0\to \eta_s \eta_s}\big) \frac{\sin 2\phi}{2}\\
&+A_{\bar B^0\to \eta_q \eta_s}\cos 2\phi ,
\end{split}
\\
\begin{split}\label{B0eta'eta'}
A_{\bar B^0\to \eta' \eta'}=& A_{\bar B^0\to \eta_q \eta_q}\sin^2\phi+ A_{\bar B^0\to \eta_s \eta_s}\cos^2\phi\\
&+A_{\bar B^0\to \eta_q \eta_s}\sin 2\phi.
\end{split}
\end{align}
The amplitude $A_{\bar B^0\to \eta_s \eta_s}$ receives contributions only from $Q_{6d,7d}$ operators and thus has no
charming penguin contributions. These are present in $A_{\bar B^0\to \eta_q \eta_q}$ and $A_{\bar B^0\to \eta_q \eta_s}$
and therefore also in the amplitudes for the decays into mass eigenstates $\eta^{(')}\eta^({')}$.

\subsection{$S$ parameters in penguin dominated modes}\label{Ssubsection}
The CP violating $S$ parameters in the $\Delta S=1$  decays
$B^0(t)\to K_{S,L}\pi^0$, $B^0(t)\to K_{S,L}\eta^{(')}$ are
of special interest because of the large CKM suppression of ``tree" amplitudes over ``penguin" amplitudes that was already discussed in the previous subsection. 
The decay amplitudes thus to a first approximation do not carry any weak phase and cancel in \eqref{Sf}, leading to the approximate relation $S_f\simeq -\eta_f^\mathrm{CP}\sin 2\beta$ where $\eta_f^\mathrm{CP}$ is the CP of the final state. This leads us to define an effective angle through
\beq
S_f=-\eta_f^\mathrm{CP} \sin 2 \beta^{\rm eff}_f.
\eeq
 In this way the modes with $K_S$ and $K_L$ in the final state have
the same $\beta^{\rm eff}$ (neglecting CP violation in the kaon sector). 

\begin{table}
\caption{Predictions for ``tree" over ``penguin" ratios $r_f$ (first row of each mode, 
in terms of $\times 10^{-2}$
) and strong phase differences $\delta_f$ (second row), defined in Eq. \eqref{r_f}, for
several $\Delta S=1$ $B^0$ and $B^-$ decays (separated by horizontal line).
Theory I (II) column correspond to two sets of SCET parameters Solution I (II) in \eqref{zetagfit}-\eqref{argAccgfitII}.
Since $B^0\to \pi^0 \bar K^0$ does not depend on isosinglet SCET parameters only one prediction is given. The meaning of errors is the same as in Table \ref{table:predSPPiso}.}
\label{table:r_f}
\begin{ruledtabular}
\begin{tabular}{lcc} 
Mode  & Theory I  & Theory II \\ \hline
$\eta' \bar K^0$ 	& $2.9\pm 0.7\pm 0.7\pm 0.4$ 		
&$3.0\pm 0.7\pm 0.7\pm 0.5$\\
					& $(159\pm 10\pm 24\pm 4)^\circ$			& 	$(-117\pm 11\pm 24\pm 6)^\circ$ \\
$\eta \bar K^0$ 	& $21\pm 20\pm 4\pm 3$ 		&	$23\pm 24\pm 5\pm 4$\\
					& $(101\pm 58\pm 20\pm 4)^\circ $			& 	$(-58\pm 62\pm 20\pm 7)^\circ$ \\
$\pi^0 \bar K^0$ 	& $14\pm 4\pm 3\pm 2$ 		&	\\
					& $(24\pm 20\pm 20\pm 6)^\circ $			& 	\\	\hline
$\eta' K^-$ 	& $2.8\pm 0.8\pm 0.5\pm 1.3$ 		
&$0.8\pm 0.6\pm 0.1\pm 1.1$\\
					& $(-22\pm 11\pm 17\pm 4)^\circ$			& 	$(61\pm 11\pm 15\pm 4)^\circ$ \\
$\eta K^-$ 	& $34\pm 31\pm 7\pm 3$ 		
&$42\pm 46\pm 9\pm 5$\\
					& $(105\pm 57\pm 20\pm 3)^\circ$			& 	$(-65\pm 67\pm 20\pm 4)^\circ$ \\		
\end{tabular}
\end{ruledtabular}
\end{table}

In the decays at hand the common final state of $B^0$ and $\bar B^0$ is provided by $K^0-\bar K^0$ mixing, that causes the mass eigenstates $K_{S,L}$ to be an admixture of $K^0$ and $\bar K^0$
states. For instance, we have
\beq
\frac{A_{\bar B^0\to \eta K_{S,L}}}{A_{B^0\to \eta K_{S,L}}}=\mp \frac{p_K}{q_K} \frac{\bar A_{\bar B^0\to \eta\bar K^0}}{A_{B^0\to \eta K^0}},
\eeq
with $q_K/p_K=-{V_{us}^* V_{ud}}/{V_{us} V_{ud}^*}\simeq -1$, where the phases of $|\bar K^0\rangle$, $|K^0\rangle$,  $|\bar B^0\rangle$, $|B^0\rangle$ were chosen so that $CP|\bar K^0\rangle=|K^0\rangle$ and $CP|\bar B^0\rangle=|B^0\rangle$.  This relation can then be used in \eqref{Sf} to obtain the expression 
for $S_f$. As in the previous subsections we decompose the amplitude into ``tree" and ``penguin" parts according to CKM element content
$
 A_{\bar B\to f}=\lambda_c^{(s)} P_{\bar B\to f}+ \lambda_u^{(s)} T_{\bar B\to f}$.
Due to the large CKM hierarchy in the $\Delta S=1$ decays ``penguin" terms dominate over ``tree", with the last term $\sim \lambda^2=0.04$ CKM suppressed compared to the first one, as discussed in Section \ref{isossect}. Expanding in this small ratio we have \cite{Gronau:1989ia}
\beq
\begin{split}
\Delta S_f&\equiv\sin 2 \beta_f^{\rm eff}-\sin 2\beta\\
&=r_f \cos\delta_f \cos 2\beta +O(r_f^2),
\end{split}
\eeq
where the ``tree" over ``penguin" ratio, $ r_f e^{i\delta_f}$, was defined in Eq. \eqref{r_f} 
and already contains the 
ratio of CKM elements.

\begin{table*}
\caption{Predicted CP averaged branching ratios ($\times 10^{-6}$, first row) and direct CP asymmetries (second row for each mode) for $\Delta S=0$ and $\Delta S=1$ $B_s^0$ decays (separated by 
horizontal line). The columns Theory I (II) correspond to two sets of SCET parameters \eqref{zetagfit}-\eqref{argAccgfitII}. Since the decays into nonisosinglet mesons do not depend on parameters in \eqref{zetagfit}-\eqref{argAccgfitII} only one prediction is given. The errors on the predictions are estimates of SU(3) breaking, $1/m_b$ corrections and errors
due to SCET parameters, respectively.}\label{table:predPPB_s}
\begin{ruledtabular}
\begin{tabular}{llll} 
Mode  & Exp  & Theory I & Theory II \\ \hline
$\bar B_s^0\to \pi^-K^+$& $<2.2 f_d/f_s$\footnote[1]{The production fraction ratio of $B_{d,s}^0$ mesons is $f_d/f_s\approx 4$ \cite{Eidelman:2004wy}.} 	
								& $4.9\pm1.2\pm1.3\pm0.3$		& \\
						& $-$	& $0.20\pm0.17\pm0.19\pm0.05 $	& \\
$\bar B_s^0\to \pi^0K^0$& $-$ 	& $0.76\pm0.26\pm0.27\pm0.17$		& \\
						& $-$	& $-0.58\pm0.39\pm0.39\pm0.13 $	&  \\
$\bar B_s^0\to \eta K^0$& $-$ 	& $0.80\pm 0.48\pm 0.29\pm 0.18$ & $0.59\pm 0.34\pm 0.24\pm 0.15$\\
						& $-$	& $-0.56\pm 0.46\pm 0.14\pm 0.06$ & $0.61\pm 0.59\pm 0.12\pm 0.08$\\
$\bar B_s^0\to \eta' K^0$& $-$ 	& $4.5\pm 1.5\pm 0.4\pm 0.5$ & $3.9\pm 1.3\pm 0.5\pm 0.4$\\
						& $-$	& $-0.14\pm 0.07\pm 0.16\pm 0.02$ & $0.37\pm 0.08\pm 0.14\pm 0.04$\\\hline
$\bar B_s^0\to K^- K^+$	& $(9.5\pm 2.0)f_d/f_s$\footnotemark[1]  	
								& $18.2\pm6.7\pm1.1\pm0.5$		& \\
						& $-$	& $-0.06\pm0.05\pm0.06\pm0.02 $	& \\						
$\bar B_s^0\to K^0\bar K^0$& $-$ & $17.7\pm6.6\pm0.5\pm0.6$		& \\
						& $-$	& $<0.1$						& \\
$\bar B_s^0\to \eta \pi^0$& $-$ & $0.014\pm 0.004\pm 0.005\pm 0.004$ & $0.016\pm 0.007\pm 0.005\pm 0.006$\\
						& $-$	& $- $ & $- $\\	
$\bar B_s^0\to \eta' \pi^0$& $-$& $0.006\pm 0.003\pm 0.002^{+0.064}_{-0.006}$ & $0.038\pm 0.013\pm 0.016^{+0.260}_{-0.036}$\\
						& $-$	& $-$ & $-$\\	
$\bar B_s^0\to \eta \eta$& $-$ 	& $7.1\pm 6.4\pm 0.2\pm 0.8$ & $6.4\pm 6.3\pm 0.1\pm 0.7$\\
						& $-$	& $0.079\pm 0.049\pm 0.027\pm 0.015$ & $-0.011\pm 0.050\pm 0.039\pm 0.010$\\	
$\bar B_s^0\to \eta \eta'$& $-$ & $24.0\pm 13.6\pm 1.4\pm 2.7$ & $23.8\pm 13.2\pm 1.6\pm 2.9$\\
						& $-$	& $0.0004\pm 0.0014\pm 0.0039\pm 0.0043$ & $0.023\pm 0.009\pm 0.008\pm 0.076$\\	
$\bar B_s^0\to \eta' \eta'$& $-$& $44.3\pm 19.7\pm 2.3\pm 17.1$ & $49.4\pm 20.6\pm 8.4\pm 16.2$\\
						& $-$	& $0.009\pm 0.004\pm 0.006\pm 0.019$ & $-0.037\pm 0.010\pm 0.012\pm 0.056$\\					
\end{tabular}
\end{ruledtabular}
\end{table*}

The difference between $\sin 2 \beta_f^{\rm eff}$ and $\sin 2\beta$ vanishes if either $r_f$ is zero or if
the strong phase difference $\delta_f$ between ``tree" and ``penguin" amplitudes is equal to $\pm 90^\circ$. The largest deviation, on the other hand, is obtained for 
$\delta_f=0^\circ, 180^\circ$. The deviation itself cannot be very large in the Standard Model because of the already mentioned
CKM suppression of order $\lambda^2=0.04$, unless the ratio $T_f/P_f$ is very large. This can happen in $\eta K_{S,L}$
decay modes, where there is large cancellation in $P_{\bar B^0 \to \eta \bar K^0}$ \eqref{etaKsuppression}. No such cancellations
are possible in $\eta' K_{S,L}$ modes. The large branching ratios 
of $\eta' K$ necessarily imply  small deviations of  $\sin 2 \beta_{\eta' K_{S,L}}^{\rm eff}$ from
$\sin 2\beta$ in the context of Standard Model. 
Using the values of parameters in \eqref{zetafit}, \eqref{Accfit}, and \eqref{zetagfit}-\eqref{argAccgfitII} obtained in the previous two sections
\begin{align}
\Delta S_{\eta'K_{S,L}}&\stackrel{\rm{\scriptsize Th.}}{=}\left\{
\begin{matrix}
(-1.9\pm0.5\pm0.6\pm0.3)\times 10^{-2},\\
(-1.0\pm0.5\pm0.8\pm0.2)\times 10^{-2},
\end{matrix}
\right. \label{SetaPr}
\\
\Delta S_{\eta K_{S,L}}&\stackrel{\rm{\scriptsize Th.}}{=}\left\{
\begin{matrix}
(-3.4\pm15.5\pm5.4\pm1.4)\times 10^{-2},\\
(7.0\pm13.6\pm4.2\pm1.1)\times 10^{-2},
\end{matrix}
\right. \label{Seta}
\end{align}
where the upper (lower) rows correspond to Solution I (II) sets of SCET parameters in \eqref{zetagfit}-\eqref{argAccgfitII}, while 
\begin{align}
\Delta S_{\pi^0 K_{S,L}}&\stackrel{\rm{\scriptsize Th.}}{=}
(7.7\pm2.2\pm1.8\pm1.0)\times 10^{-2},
\label{SpiKS}
\end{align}
where again the first two errors are due to SU(3) breaking and expected $1/m_b, \alpha_s(m_b)$ corrections, while the last is due to experimental uncertainties on SCET parameters extracted from
the $\chi^2$-fit to branching ratios and direct CP asymmetries.

It is illuminating to evaluate the ratios $r_f e^{i\delta_f}$ \eqref{r_f} of ``tree" and ``penguin" terms for these modes. 
The numerical values are gathered in Table \ref{table:r_f}. As expected from the general arguments outlined above,
the ratio $r_f$ is relatively large for $\bar B^0\to \eta K_{S,L}$ due to cancellations in ``penguin" amplitudes. 
The corresponding strong phases $\delta_f$ both for Solution I and Solution II sets of SCET parameters \eqref{zetagfit}-\eqref{argAccgfitII} are not close to $0^\circ$ or $180^\circ$, so the values of $\Delta S_{\eta K_{S,L}}$ 
in \eqref{Seta} are not close to the maximal possible deviations for given values of $r_f$. On the contrary, $\Delta S_{\pi^0 K_{S,L}}$ in \eqref{SpiKS} is already close to maximal positive deviation for fixed value of $r_f$. Phenomenologically probably most interesting is $\Delta S_{\eta' K_{S,L}}$, whose absolute value is much smaller and is below $4\%$ even if $\delta_f$ is taken to be completely unknown. In accordance with general expectations
very small values of $|\Delta S_{\eta' K_{S,L}}|$ have also been found in other approaches to two-body $B$ decays; in 
QCD factorization \cite{Beneke:2005pu,Beneke:2003zv}, in QCD factorization with modeled rescattering 
\cite{Cheng:2005bg} and are
consistent with bounds obtained using SU(3) symmetry \cite{Gronau:2005gz,SU3}. 

\begin{table*}
\caption{Predictions for $(S_f)_{B_s}$ (first row in each mode) and $(H_f)_{B_s}$ (second row) parameters in $B_s$ decays. The columns Theory I (II) correspond to two sets of SCET parameters \eqref{zetagfit}-\eqref{argAccgfitII}. Since the decays into nonisosinglet mesons do not depend on parameters in \eqref{zetagfit}-\eqref{argAccgfitII} only one prediction is given. The errors on the predictions are estimates of SU(3) breaking, $1/m_b$ corrections and errors
due to SCET parameters, respectively. No predictions are made for $(S_f)_{B_s}$ and $(H_f)_{B_s}$ in $\bar B^0_s\to K^0\bar K^0$ and $\bar B^0_s\to \pi^0\eta'$, see text. }\label{table:predSPPBs}
\begin{ruledtabular}
\begin{tabular}{llll} 
Mode  & Exp  & Theory I & Theory II\\ \hline
$\bar B^0_s\to K_S\pi^0$ 	& $-$ 	& $-0.16\pm0.41\pm0.33\pm0.17$	& \\
							& $-$ 	& $0.80\pm0.27\pm0.25\pm0.11$	& \\
$\bar B^0_s\to K_S\eta$ 	& $-$ 	& $0.82\pm 0.32\pm 0.11\pm 0.04$ & $0.63\pm 0.61\pm 0.16\pm 0.08$\\
							& $-$ 	& $0.07\pm 0.56\pm 0.17\pm 0.05$ & $0.49\pm 0.68\pm 0.21\pm 0.03$\\
$\bar B^0_s\to K_S\eta'$ 	& $-$ 	& $0.38\pm 0.08\pm 0.10\pm 0.04$ & $0.24\pm 0.09\pm 0.15\pm 0.05$\\
							& $-$ 	& $-0.92\pm 0.04\pm 0.04\pm 0.02$ & $-0.90\pm 0.05\pm 0.05\pm 0.03$\\\hline
$\bar B^0_s\to K^-K^+$ 		& $-$ 	& $0.19\pm0.04\pm0.04\pm0.01$	& \\
							& $-$ 	& $1-(0.021\pm0.008\pm0.007\pm0.002)$	& \\
$\bar B^0_s\to \pi^0\eta$ 	& $-$ 	& $0.45\pm 0.14\pm 0.42\pm 0.30$ & $0.38\pm 0.20\pm 0.42\pm 0.37$\\
							& $-$ 	& $-0.89\pm 0.07\pm 0.21\pm 0.15$ & $-0.92\pm 0.08\pm 0.17\pm 0.15$\\
$\bar B^0_s\to \eta\eta$ 	& $-$ 	& $-0.026\pm 0.040\pm 0.030\pm 0.014$ & $-0.077\pm 0.061\pm 0.022\pm \
0.026$\\
							& $-$ 	& $1-(0.0035\pm0.0041\pm0.0019\pm0.0015)$& $1-(0.0030\pm0.0048\pm0.0017\pm0.0021)$\\
$\bar B^0_s\to \eta\eta'$ 	& $-$ 	& $0.041\pm 0.004\pm 0.002\pm 0.051$ & $0.015\pm 0.010\pm 0.008\pm \
0.069$\\
							& $-$ 	& $1-(0.0008\pm0.0002\pm0.0001\pm0.0021)$ 	& $1-(0.0004\pm0.0003\pm0.0003\pm0.0007)$\\
$\bar B^0_s\to \eta'\eta'$ 	& $-$ 	& $0.049\pm 0.005\pm 0.005\pm 0.031$ & $0.051\pm 0.009\pm 0.017\pm \
0.039$\\
							& $-$ 	& $1-(0.0012\pm0.0003\pm0.0002\pm0.0017)$	& $1-(0.0020\pm0.0007\pm0.0009\pm0.0041)$\\
\end{tabular}
\end{ruledtabular}
\end{table*} 

The predictions \eqref{Seta} - \eqref{SpiKS} are to be contrasted with the experimental findings, where (neglecting the small difference between 
$S_{J/\Psi K_{S,L}}$ and $\sin 2\beta$ \cite{Boos:2004xp})
\beq
\Delta S_{\eta'K_{S}}\stackrel{\rm{\scriptsize Exp.}}{=}-0.23\pm 0.13.
\eeq
and
\beq
\Delta S_{\pi^0 K_{S,L}}\stackrel{\rm{\scriptsize Exp.}}{=}-0.41\pm 0.26,
\eeq
while no information on $\Delta S_{\eta K_{S}}$ is yet available. 
The difference between $S_{\eta'K_{S}}$, $S_{\pi^0 K_{S}}$  and $\sin 2\beta$ has been even more pronounced in the past, and
has been reduced in the past year to the present level of almost 2 $\sigma$. Since predictions for these two quantities in 
SCET are not prone to large 
uncertainties as shown by the errors in \eqref{SetaPr}, \eqref{SpiKS}, a further reduction of experimental errors with 
unchanged
central values would be a clear signal of beyond the Standard Model physics.

\subsection{$B_s$ decays}\label{Bs}
Using SU(3) symmetry allows us to make predictions for $B_s^0$ decays as well. Predictions made
with the values of SCET parameters \eqref{zetafit}, \eqref{Accfit}, and \eqref{zetagfit}-\eqref{argAccgfitII} for
CP averaged branching ratios and direct CP asymmetries are collected in Table \ref{table:predPPB_s}, while
the predictions for the observables $(S_f)_{B_s}$ \eqref{SfBs} and $(H_f)_{B_s}$ \eqref{Hf} are given in Table \ref{table:predSPPBs}.
The SU(3) breaking on the SCET parameters relations \eqref{zetaetarel}-\eqref{zetaSU3eta} and \eqref{AccSU3} was assumed to be $20\%$ with a $20^\circ$ variation 
on the charming penguin's strong phases. The second errors in Tables \ref{table:predPPB_s}, \ref{table:predSPPBs}, estimate the remaining order $1/m_b$ and $\alpha_S(m_b)$ corrections. These are obtained from a $20\%$ variation on the size and a $20^\circ$ variation on the strong phase of the leading order amplitudes proportional to $\lambda_u^{(f)}$ or $\lambda_t^{(f)}$.

Many observations made about $\bar B^0$ and $B^-$ decays hold also for $\bar B_s^0$ decays. For instance 
$\Delta S=1$ decays $\bar B_s^0\to K \bar K$ and $\bar B_s^0\to \eta^{(')} \eta^{(')}$ are dominated by nonperturbative charming penguins due to a CKM hierarchy just like $B\to \pi K$, $B\to K\eta^{(')}$ decays. Expanding in the CKM suppressed
``tree" over ``penguin" ratio $r_f$ \eqref{r_f} the observables from time dependent decays \eqref{SfBs}, \eqref{Hf}
\beq\label{Sfexp}
(S_f)_{B_s} = \eta_f^\mathrm{CP} \sin 2\epsilon - \eta_f^\mathrm{CP} r_f \cos \delta_f \cos 2\epsilon +O(r_f^2),
\eeq
and
\beq\label{Hfexp}
\begin{split}
(H_f)_{B_s}&= \eta_f^\mathrm{CP} \cos 2\epsilon\Big(1-\frac{r_f^2}{2}\Big) + \eta_f^\mathrm{CP} \sin 2\epsilon \Big[r_f \cos \delta_f \\
&+r_f^2\frac{\Re\left({\lambda_u^{(s)}}/{\lambda_c^{(s)}}\right)}{\Im\left({\lambda_u^{(s)}}/{\lambda_c^{(s)}}\right)}\Big(\cos^2\delta_f-\frac{1}{2}\Big)\Big]+O(r_f^3),
\end{split}
\eeq
while the expression for direct CP asymmetry to first order in $r_f$ is given in \eqref{ACPexp}. Since $\epsilon\sim 1^\circ$ in the Standard Model, $(H_f)_{B_s}$ for the penguin dominated decays $\bar B_s^0\to K \bar K$ and $\bar B_s^0\to \eta^{(')} \eta^{(')}$ is expected to be very close to $1$. In the Standard Model $\sin 2\epsilon\sim 0.035$ and thus $\sin 2\epsilon\sim r_f$, so that the deviations of $(H_f)_{B_s}$ from unity are numerically of order $O(r_f^2)$. For  
$1\gg r_f>2 \sin 2\epsilon$, the $r_f^2$ correction in the first term in \eqref{Hfexp} is actually larger than the second term in \eqref{Hfexp} which starts at linear order in $r_f$. In this case $\eta_f^\mathrm{CP} (H_f)_{B_s}$ is smaller than 1 irrespective of the strong phase difference $\delta_f$.
The deviations of $(H_f)_{B_s}$ from 1 for different modes are listed in Table \ref{table:predSPPBs}.

The charming penguin dominance in the $K \bar K$ modes
\beq
\begin{split}
A_{\bar B_s^0\to K^-K^+}=&\frac{G_F}{\sqrt 2} m_B^2\Big[ \lambda_s^{(s)} A_{cc}^{KK}+\cdots\Big]\\
\simeq & A_{\bar B_s^0\to K^0 \bar K^0},
\end{split}
\eeq
furthermore leads to an approximate relation
\beq
\bar\Gamma_{\bar B_s^0\to K^-K^+}\simeq \bar\Gamma_{\bar B_s^0\to K^0 \bar K^0}.
\eeq
The amplitude for $\bar B_s^0\to K^0 \bar K^0$ does not receive any LO contributions from $Q_{1s,2s}^{(0),(1)}$ operators, cf. Table \ref{table:piK}, so that
at this order in the $\alpha_S(m_B)$ expansion the ``tree" amplitude is entirely due to QCD penguin and EWP operator insertions. 
At NLO in $\alpha_S(m_B)$ there could, however, be contributions proportional to $C_{1,2}\alpha_S(m_B)$ leading to a large
correction to the predicted direct CP asymmetry. We thus give in Table \ref{table:predPPB_s} only an estimated upper bound of this observable. Similarly, no prediction of $(S_f)_{B_s}$, $(H_f)_{B_s}$ for this mode is made in Table \ref{table:predSPPBs}, but conservatively we can expect
\begin{align}
|(S_{K_SK_S})_{B_s}|&<|(S_{K^+K^-})_{B_s}|,\\
|1-(H_{K_SK_S})_{B_s}|&<|1-(H_{K^+K^-})_{B_s}|,
\end{align}

In $\bar B_s^0\to \eta^{(')} \eta^{(')}$ amplitudes the dominating contributions are also given by charming penguins
\begin{align}
A_{\bar B_s^0\to \eta_s\eta_q}=&\frac{G_F}{\sqrt 2} m_B^2\Big[ \sqrt{2}\lambda_s^{(s)} A_{ccg}^{\eta_s\eta_q}+\cdots\Big],\\
A_{\bar B_s^0\to \eta_s\eta_s}=&\frac{G_F}{\sqrt 2} m_B^2 \Big[ \lambda_s^{(s)} 2\Big( A_{ccg}^{\eta_s\eta_s}(s)+A_{cc}^{\eta_s\eta_s}(s)\Big)+\cdots\Big],
\end{align}
with the ellipsis denoting smaller contributions, while in $A_{\bar B_s^0\to \eta_q\eta_q}$ there are no charming penguins. Using \eqref{B0etaeta}-\eqref{B0eta'eta'} we then get for the coefficients in front of $A_{cc}$, $A_{ccg}$ in 
the SU(3) limit (apart from the common multiplicative factor ${G_F} m_B^2/{\sqrt 2}$)
\begin{center}
\begin{tabular}{c|c|c} 
\qquad mode \qquad & \qquad\quad $A_{cc}$\quad\qquad & \quad\qquad\qquad $A_{ccg}$ \quad\qquad\qquad \\ \hline
$\bar B_s^0\to \eta\eta$ &  $\sqrt 2\sin^2\phi$ & $\sqrt 2 \sin^2\phi -\sin 2\phi  $
\\ 
$\bar B_s^0\to \eta\eta'$ & $-\frac{1}{\sqrt 2}\sin2\phi$ & $\cos 2\phi-\frac{1}{\sqrt{2}}\sin 2\phi$
\\
$\bar B_s^0\to\eta'\eta'$ & $\sqrt 2 \cos ^2\phi$ & $\sqrt 2 \cos^2 \phi +\sin 2\phi  $
\end{tabular}
\end{center}
which for $\phi=39^\circ$ gives numerically
\begin{center}
\begin{tabular}{c|c|c} 
\qquad mode \qquad & \qquad $A_{cc}$\qquad & \qquad $A_{ccg}$ \qquad \\ \hline
$\bar B_s^0\to \eta\eta$ &  $-0.42$ & $0.56$
\\ 
$\bar B_s^0\to \eta\eta'$ & $-0.48$ & $-0.69$
\\
$\bar B_s^0\to\eta'\eta'$ & $1.83$ & $0.85$
\end{tabular}
\end{center}
explaining the pattern of branching ratios in Table \ref{table:predPPB_s}.  Note that the branching ratios 
for $\bar B_s^0\to \eta\eta$ and $\bar B_s^0\to \eta'\eta'$ have an additional symmetry factor $1/2$ in \eqref{Gammadef} because of indistinguishable particles in the final state.

Since $B_s\to KK$ and $B_s\to \eta^{(')}\eta^{(')}$ decays are penguin dominated, they represent ideal probes for new 
physics searches. A very useful observable in this respect is $(H_{f})_{B_s}$ \eqref{Hfexp}. In the Standard Model we have a very robust 
prediction that  $(H_{f})_{B_s}\simeq 1$ for penguin dominated modes $B_s\to KK$ and $B_s\to \eta^{(')}\eta^{(')}$, where
the deviations from this relation are of order $O(r_f^2)$ as discussed above, where $r_f$ itself is of order $\lambda^2\sim 0.04$ \eqref{r_f}. The predicted values in SCET for different modes are given in Table \ref{table:r_fBs}.  The deviations from $(H_{f})_{B_s}= 1$ are therefore expected to be very small, well below percent level for $B_s\to \eta^{(')}\eta^{(')}$ modes. If virtual corrections from the beyond Standard Model particles modify either the $B_s-\bar B_s$ 
mixing phase or introduce new phases in $b\to s$ penguins, a hint of which may have already been experimentally seen in $B\to \pi K$, $B\to \eta' K$ modes, $(H_{f})_{B_s} $ can easily deviate from 1 by a correction of $O(1)$. Another attractive feature of 
$(H_{f})_{B_s}$ is that it can be measured in untagged $B_s$ decays \cite{Dunietz:1995cp}.   

\begin{table}
\caption{Predictions for ``tree" over ``penguin" ratios $r_f$ (first row of each mode, 
in terms of $\times 10^{-2}$) and strong phase differences $\delta_f$ (second row), defined in Eq. \eqref{r_f}, for
penguin dominated $\Delta S=1$ $B_s^0$ decays.
Theory I (II) column correspond to two sets of SCET parameters Solution I (II) in \eqref{zetagfit}-\eqref{argAccgfitII}.
Since $\bar B_s^0\to K^+K^-$ does not depend on isosinglet SCET parameters only one prediction is given. The meaning of errors is the same as in Table \ref{table:predSPPiso}.}
\label{table:r_fBs}
\begin{ruledtabular}
\begin{tabular}{lcc} 
Mode  & Theory I  & Theory II \\ \hline	
$K^-K^+$ 	& $16.2\pm 3.5\pm 3.3\pm 1.0$ 		
&\\
					& $(-159\pm 18\pm 21\pm 5)^\circ$			& 	\\
$\eta \eta$ 	& $10\pm 5\pm 2\pm 2$ 		&	$12\pm 7\pm 2\pm 2$\\
					& $(53\pm 25\pm 20\pm 4)^\circ $			& 	$(-6\pm 26\pm 20\pm 4)^\circ$ \\
$\eta \eta'$ 	& $0.5\pm 0.4\pm 0.2\pm 5.1$ 		& $3.1\pm 1.0\pm 0.5\pm 0.5$	\\
					& $(175\pm 17\pm 48\pm 4)^\circ $			& $(47\pm 18\pm 18\pm 4)^\circ$ 	\\	
$\eta' \eta'$ 	& $1.6\pm 0.5\pm 0.4\pm 3.6$ 		
&$4.0\pm 1.1\pm 1.0\pm 3.6$\\
					& $(145\pm 13\pm 25\pm 9)^\circ$			& 	$(-112\pm 12\pm 25\pm 9)^\circ$ \\		
\end{tabular}
\end{ruledtabular}
\end{table}

A special case among the $\Delta S=1$ $B_s$ decays are the decays $B_s^0\to \pi^0\eta^{(')}$.  Since these are $\Delta I=1$ transitions, there are no QCD penguin contributions to these decays, 
including no contributions from the nonperturbative charming penguins. The theoretical control over these modes
is thus much greater, on par
with $B^-\to \pi^- \pi^0$, where charming penguins are also absent. 
Furthermore, $B_s^0\to \pi^0\eta^{(')}$ decays are color suppressed, which makes them 
a perfect probe of the SCET prediction that color suppression is lifted due to $Q_{2s}^{(1)}$ contributions. Unfortunately
their branching ratios are of the order $\sim 10^{-7}-10^{-8}$ because of the CKM suppression of the ``tree" amplitude. The absence of QCD penguins in $B_s^0\to \pi^0\eta^{(')}$ also leads to a generic prediction that $(S_{f})_{B_s}$ and $1-(H_{f})_{B_s}$ can be of $O(1)$. Namely, the ``penguin" amplitude in $B_s^0\to \pi^0\eta^{(')}$ is coming exclusively from EWP operators, which have small Wilson coefficients, making them of similar size as the CKM suppressed ``tree" amplitudes, giving $r_f\sim O(1)$ in \eqref{Sfexp} and \eqref{Hfexp}. Furthermore, because there are no contributions from charming penguins, while the other contributions factorize at LO in $1/m_b$, the strong phase differences vanish at leading order in $\alpha_s(m_b)$ and $1/m_b$, $\delta_f=(0^\circ\mod 180^\circ)$, so that ${\cal A}_f^{\mathrm{CP}}$ \eqref{ACPexp} in $B_s^0\to \pi^0\eta^{(')}$ vanishes at this order. This will be lifted at NLO.

Scanning the allowed range of the at present very poorly constrained parameter $\zeta_{g}^{-}$ \eqref{zetag-fit}, \eqref{zetag-fitII}, we find that a cancellation between different terms in $A_{B_s^0\to \pi^0\eta'}$ is possible for special values of $\zeta_{g}^{-}$. We thus give only a $1\sigma$ range for the decay width, while no prediction on $(S_{f})_{B_s}$ and $1-(H_{f})_{B_s}$ for $B_s^0\to \pi^0\eta'$ can be made. 

The $\Delta S=0$ decays have ``tree" and ``penguin" contributions of similar size. Furthermore, both $\bar B_s^0\to K^0\eta^{(')}$ and
$\bar B_s^0\to \pi^-K^+, \pi^0 K^0$ do receive contributions from nonperturbative charming penguins, leading to 
sizeable direct CP asymmetries. A peculiar case is the $\bar B_s^0\to K^0\eta$ decay, where a similar cancellation occurs as in $\bar B^0\to \bar K^0 \eta$ between
charming penguin contributions due to $\eta-\eta'$ mixing. Namely, up to numerically smaller terms from $Q_{3d,\dots,6d}$ insertions we have in the SU(3) limit
\begin{align}
A_{\bar B_s^0\to \bar K^0\eta_s}&=\frac{G_F}{\sqrt{2}} m_B^2\Big[\lambda_c^{(d)} \big(A_{cc}+A_{ccg}\big)+\dots\Big],\\
\begin{split}
A_{\bar B_s^0\to \bar K^0\eta_q}&=\frac{G_F}{\sqrt{2}} m_B^2\Big[\lambda_c^{(d)} \big(\frac{1}{\sqrt 2}A_{cc}+\sqrt{2} A_{ccg}\big)\\
&+\frac{f_{\eta_q}}{\sqrt 2}\big(\zeta c_2^{(d)}+\zeta_J b_2^{(d)}\big)+\cdots\Big],
\end{split}
\end{align}
giving a $(1-\sqrt{2} \tan\phi)/(\sqrt{2}+\tan\phi)=-0.07$ suppressed coefficient in front of 
$A_{cc}$ in $\bar B_s^0\to K^0\eta^{}$ compared to the coefficient in front of $A_{cc}$ in $\bar B_s^0\to K^0\eta'$.
Similarly, the coefficient in front of $A_{ccg}$ in $\bar B_s^0\to K^0\eta^{}$ is $(1- \tan\phi/\sqrt{2})/(1/\sqrt{2}+\tan\phi)=0.28$ suppressed compared to the one in $\bar B_s^0\to K^0\eta'$. No 
such cancellations are present in ``tree" amplitudes. Because
$\bar B_s^0\to K^0\eta^{(')}$ decays are $\Delta S=0$, with no CKM hierarchy between ``tree" and ``penguin" amplitudes, 
the effect of cancellations on predicted branching ratios is less drastic than in $\bar B^0\to \bar K^0 \eta^{(')}$, as can be
seen from results in Table \ref{table:predPPB_s}.

\section{Conclusions}
We have provided expressions for decay amplitudes of $\bar B^0, B^-$ and $\bar B_s^0$ mesons
to two light pseudoscalar or vector mesons, including isosinglet mesons $\eta, \eta', \omega, \phi$, using
Soft Collinear Effective Theory at LO in $1/m_b$. For the decays into $\eta,\eta'$ mesons the contributions where 
a spectator quark is annihilated by the weak operator, while two collinear gluons are created, is found to be of 
the leading order in $1/m_b$. This leads to a new jet function in SCET$_{\rm I}\to$ SCET$_{\rm II}$ matching for the operators that contribute only to isosinglet final states. 

In the phenomenological analysis we work at LO in $\alpha_S(m_b)$ and $1/m_b$. In particular, no expansion in $\alpha_S(\sqrt{\Lambda m_b})$ at intermediate scale is made. Instead, following Refs. \cite{Bauer:2004tj,Bauer:2005kd} a set of new nonperturbative SCET functions is 
introduced, which are then extracted from data using a $\chi^2$-fit. The numerical analysis is performed for the decays into
two pseudoscalars. From separate $\chi^2$-fits to $B\to\pi\pi, \pi K$ data and to $B\to \pi \eta^{(')}, K\eta^{(')}$ data
SCET parameters for  nonisosinglet final states and for isosinglet final states are extracted, respectively. 
Due to scarce data for isosinglet final states, SU(3)
symmetry is imposed on the SCET parameters. Even so, two sets of solutions are found for the SCET parameters $\zeta_{(J)g}$,
$A_{ccg}$ that enter only in the amplitudes for decays into isosinglet final states. 

Predictions for branching ratios, direct and indirect CP asymmetries in $\bar B^0, B^-$ and $\bar B_s^0$ decays together with estimates of theoretical errors due to SU(3) breaking, $1/m_b$ and $\alpha_S(m_b)$ corrections and
errors due to extracted values of SCET parameters are listed in Tables \ref{table:predPP}-\ref{table:predSPPiso} and Tables \ref{table:predPPB_s}-\ref{table:predSPPBs}.
We find:
\begin{itemize}
\item 
A discrepancy between theoretical and experimental values for the $R$ and $R_n$ ratios of decay widths \eqref{R}, \eqref{Rn}, \eqref{RTh}, \eqref{RnTh} at a level of $2\sigma$ if both $\pi\pi$ and $\pi K$ data are used simultaneously to determine the SCET parameters. No statistically significant deviations are
found if only a fit to $\pi K$ data is made, but the resulting values of SCET parameters strongly violate flavor SU(3). 
\item
The enhancement of $\mathrm{Br}(B\to K\eta')$ over $\mathrm{Br}(B\to K\eta)$ is naturally explained in SCET due to destructive and constructive interference governed by $\eta-\eta'$ mixing as first proposed by Lipkin \cite{Lipkin:1990us,Lipkin:1998ew}.  The enhancement of
$\mathrm{Br}(B\to K\eta')$ over $\mathrm{Br}(B\to \pi K)$ is due to the ``gluonic" charming penguin $A_{ccg}$, where the $n$ collinear quark annihilates with the spectator quark producing two $n$ collinear gluons. 
The other mechanisms for enhancing $\mathrm{Br}(B\to K\eta')$ discussed in the
literature are found to be suppressed by at least one order of $1/m_b$.  
\item
The SCET parameters $\zeta_{(J)g}$ and $A_{ccg}$ corresponding to annihilation diagrams with two simultaneously emitted collinear gluons are found to be both parametrically and numerically of the same order as the other leading order contributions in accordance
with SCET counting. The combination $\zeta_{Jg}-\zeta_g$ is at present very poorly constrained, but can be constrained
by future measurements of decay widths and CP asymmetries in $B\to \pi^0\eta^{(')}$ and CP asymmetries in $B\to K^0\eta^{(')}$ decays. 
The orthogonal combination $\zeta_{Jg}+\zeta_g$ which also enters into $B\to \eta^{(')}$ form factors is already much better contrained and the corresponding values of the $B\to \eta_{q,s}$ form factors are given in Eqs. \eqref{fetaq} and \eqref{fetas}. In the future, with more abundant data on isosinglet decays, the assumption of SU(3) flavor symmetry that was used in the present analysis can be relaxed.
This will significantly reduce the theoretical uncertainties associated with many of our predictions.
\item

The deviation $\Delta S$ of the $S$ parameter from $\sin 2 \beta$
in $\bar B^0(t)\to K_S\eta'$ decay, which was not used as a constraint in the $\chi^2$-fit, is predicted to lie in the range $[-0.026,0]$ at $1\sigma$ and to be below
$4\%$ even if no information on
the strong phases is used.  Further constraints on $\Delta S$ in $\bar B^0\to K_S\eta, K_S\pi^0$ decays are given in subsection \ref{Ssubsection}.
\item
Predictions are made for the branching ratios in $B_s\to PP$ decays, as well as for related direct CP asymmetries, $(S_f)_{B_s}$ parameters and the coefficient $(H_f)_{B_s}$ multiplying $\sinh(\Delta \Gamma t/\Gamma)$ in the time dependent decay width.  A robust prediction,
$(H_f)_{B_s}=1$, for penguin dominated $\Delta S=1$ decays holds in the Standard Model up to corrections that are at the permil level for $B_s\to \eta^{(')}\eta^{(')}$ decays and at the percent level for $B_s\to KK$ decays. This prediction follows from general arguments independent of the SCET framework. In beyond the Standard Model scenarios, the relation $(H_f)_{B_s}=1$ can generically receive corrections of $O(1)$, making them very interesting probes of new physics.
\end{itemize}

We thank Ira Rothstein for initiating this work and for his helpful comments, 
Michael Gronau, Chul Kim, Manfred Paulini, Dan Pirjol and Amarjit Soni for comments on 
the manuscript and Kre\v simir Kumeri\v cki, Adam Leibovich and Kornelija Passek-Kumeri\v cki for enlightening discussions. This work was supported in part by the United States Department of Energy 
under Grants No.\ DOE-ER-40682-143 and DEAC02-6CH03000. 

\appendix
\section{Notation}\label{Notation}
Choosing $n^\mu=(1,0,0,-1)$ and $\bar n^\mu=(1,0,0,1)$, so
that $\sla n =\gamma^0+\gamma^3$ and $ \sla\bar n=(\gamma^0-\gamma^3)$, while in shorthand notation $n\cdot k=k_+$ and $\bar n \cdot k=k_-$ for any four-vector $k^\mu$, and choosing for the Levi-Civita tensor normalization to be 
\beq
\epsilon^{0123}=-1,
\eeq
while
$\gamma_5=i\gamma^0\gamma^1\gamma^2\gamma^3,$
so that for instance $\Tr[\gamma_5\gamma^\mu\gamma^\nu\gamma^\alpha\gamma^\beta]=4 i \epsilon^{\mu\nu\alpha\beta}$, and 
furthermore defining
\beq
\epsilon_\perp^{\mu\nu}=\frac{1}{2}\epsilon^{\mu\nu\alpha\beta}\bar n_\alpha n_\beta
\eeq
so that $\epsilon_\perp^{12}=-\epsilon_\perp^{21}=1$ and $\gamma_5=\frac{i}{8}[\sla \bar n,\sla n]\epsilon_{\perp\mu\nu}
\gamma_\perp^\mu\gamma_\perp^\nu$, this leads to the following relations
\begin{align}
\sla n \gamma_\perp^\mu\gamma_\perp^\nu P_{L,R}&=(g_\perp^{\mu\nu}\pm i \epsilon_\perp^{\mu\nu})\sla n P_{L,R},\\
\sla \bar n \gamma_\perp^\mu\gamma_\perp^\nu P_{L,R}&=(g_\perp^{\mu\nu}\mp i \epsilon_\perp^{\mu\nu})\sla \bar n P_{L,R},
\end{align} 
where $g_{\perp}^{\mu\nu}=g^{\mu\nu}-\frac{1}{2}(n^\mu\bar n^\nu+\bar n^\mu n^\nu)$
and $P_{L,R}=(1\mp\gamma_5)/2$.

\section{The structure of jet functions}\label{app:jet}
In this appendix we show that in the matching \eqref{jet-funcs} only four jet functions appear to all orders in 
$\alpha_S(\sqrt{\Lambda m_b})$. Let us first discuss
the case where in 
\beq\label{Tprod}
T\big[(\bar \xi_nW)_{z \omega}i g {\cal B}_{n,\,-\bar z \omega}^{\perp\,\alpha}P_{R,L}\big]^{ia}(0)\big[i g\sla{\cal B}_n^\perp W^\dagger \xi_n\big]^{jb}(y),
\eeq
the gluon fields are contracted, leading to $J_\perp$ and $J$ jet functions \eqref{jet-funcs}. At higher orders in 
$\alpha_S(\sqrt{\Lambda m_b})$ the corrections will come from insertions of leading order SCET Lagrangian 
\beq
{\cal L}_{\xi\xi}^{(0)}=\bar \xi_{n,p'}\left(in\cdot D_c+i \sla D_c^\perp \frac{1}{i \bar n\cdot D_c}i \sla D_c^\perp\right)
\frac{\sla \bar n}{2} \xi_{n,p},
\eeq
and the purely gluonic ${\cal L}_{cg}^{(0)}$. Thus 
the most general form of the operator with  two external collinear quark fields that \eqref {Tprod} matches onto is (up to 
operators with nontrivial color structure)
\beq
\delta^{ab}\big[\bar u_n^c \prod_i  (\sla\bar n \gamma_\perp^{\mu_i} \gamma_\perp^{\mu_{i+1}} \sla n )P_{R,L}\big]^{i} 
\big[\gamma_\perp^\beta \prod_j(\sla n \sla \bar  n \gamma_\perp^{\mu_j} \gamma_\perp^{\mu_{j+1}} )u_n^c\big]^{j},
\eeq
with $\gamma_\perp^\beta$ coming from $\sla{\cal B}_n^\perp$, $\sla n$ coming from the collinear quark propagator, and pairs of $\gamma_\perp$ matrices coming from 
${\cal L}_{\xi \xi}^{(0)}$. After Fierz transformation then (up to an overall factor) 
\beq\label{Fierzed}
\begin{split}
\delta^{ab}\sum_I (\bar u_n \Gamma_I u_n)  
\big[&\gamma_\perp^\beta \sla n \sla \bar  n \prod_j (\gamma_\perp^{\mu_j} \gamma_\perp^{\mu_{j+1}})\times \\
&\times\Gamma_I' \prod_i   (\gamma_\perp^{\mu_i} \gamma_\perp^{\mu_{i+1}} )\sla \bar n \sla n P_{R,L}\big]^{ji},
\end{split}
\eeq
where 
\beq
\begin{split}
\Gamma_I\otimes \Gamma_I'=& \sla \bar n\otimes \sla n-\sla \bar n \gamma_5\otimes \sla n \gamma_5 -\sla \bar n \gamma_\perp^\nu \otimes \sla n \gamma_{\perp \nu}=\\
=& \sla \bar n (1\mp\gamma_5) \otimes \sla n -\sla \bar n \gamma_\perp^\nu \otimes \sla n \gamma_{\perp\nu}.
\end{split}
\eeq
In the last equality the action of $\gamma_5$ on $P_{R,L}$ has been used. The two Dirac structures give rise to the operators with jet functions $J$ and $J_\perp$ in \eqref{jet-funcs}. For instance the 
 $\sla \bar n (1\mp\gamma_5) \otimes \sla n $ Dirac structure leads to the operator
\beq
\big[\bar q_{n,\, x\omega}\sla \bar nP_{L,R}q'_{n,\, -\bar x\omega}\big]\big({\sla n}\gamma_\perp^\alpha P_{R,L},\big)^{ji},
\eeq
which follows from the relation $\sla n \sla \bar n \sla n \propto\; \sla n$ and the fact that all the Lorentz indices in \eqref{Fierzed} are contracted pairwise (contraction with external $p_\perp$ momentum is subleading) except for the index $\alpha$ which is carried by the remaining $\gamma_\perp$. Namely, the $\gamma_\perp$ matrices with 
contracted indices can be permuted to be next to each other so that the end result is a pure number times $\gamma_\perp^\alpha$. 

The  $\sla \bar n \gamma_\perp^\nu \otimes \sla n \gamma_{\perp\nu}$ Dirac structure, on the other hand, leads to the operator 
\beq
\big[\bar q_{n,\, x\omega}\sla \bar n\gamma_\nu^\perp q'_{n,\, -\bar x\omega}\big] \big(\frac{\sla n}{2}P_{R,L}\gamma_\perp^\alpha\gamma_\perp^\nu\big)^{ji}.
\eeq
This immediately follows from the relation
[odd]$\gamma_\perp^\nu$[even] $\propto \gamma_\perp^\alpha \gamma_\perp^\nu$ to be proven below. Here [odd] and [even] denote products of odd and even number of $\gamma_\perp$ matrices, with $\gamma_\perp^\alpha$ either in [odd] or [even], while all other indices are contracted. 

Before we proceed let us show by induction that [odd]$\gamma_\perp^\nu $[odd]=0, where the indices of  $\gamma_\perp$ matrices in [odd] are all contracted pairwise. This is true at lowest order, $\gamma_\perp^\beta \gamma_\perp^\nu \gamma_{\perp\beta}=0$. Now assume that the relation holds for $N-2$ $\gamma_\perp$ matrices and look at the case of $N$ $\gamma_\perp$ matrices. Since in [odd] there is an odd number of $\gamma_\perp$ there must be at least one pair of $\gamma_\perp$ matrices that has contracted indices and sits on the opposite sides of $\gamma_\perp^\nu $. Moving these matrices next to $\gamma_\perp^\nu $ using $\gamma_\perp^{\sigma}\gamma_\perp^{\delta}=2 g_\perp^{\sigma \delta}- \gamma_\perp^{\delta}
\gamma_\perp^{\sigma}$ leads to terms of form [odd]$\gamma_\perp^\nu $[odd] with $N-2$ matrices, which are zero by assumption, and a term [even]$\gamma_\perp^\beta \gamma_\perp^\nu \gamma_{\perp\beta}$[even], which is also zero. Similarly, one can show that [even]$\gamma_\perp^\nu $[even]$\propto \gamma_\perp^\nu $. This holds trivially at lowest order with [even] empty. Let us assume that it holds for $N-2$ $\gamma_\perp$ matrices and move to $N$ $\gamma_\perp$ matrices. If in [even]$\gamma_\perp^\nu $[even] there are no cross contractions then [even] is just a number. If there are cross contractions, then as before the corresponding two
matrices can be moved next to $\gamma_\perp^\nu $. This leads to terms [even]$\gamma_\perp^\nu $[even] with $N-2$ matrices, which are proportional to $\gamma_\perp^\nu $ by assumption, and a term  [odd]$\gamma_\perp^\beta \gamma_\perp^\nu \gamma_{\perp\beta}$[odd], which is zero.

We can now show by induction that [odd]$\gamma_\perp^\nu$[even] $\propto \gamma_\perp^\alpha \gamma_\perp^\nu$. The relation is trivially satisfied when [even] is an empty set and [odd]$=\gamma_\perp^\alpha$.
Let us assume that the relation holds also for $N-2$ $\gamma_\perp$ in [odd] and [even] and move to $N$ $\gamma_\perp$. We distinguish two cases, (i) $\gamma_\perp^\alpha$ is in [odd] and (ii) $\gamma_\perp^\alpha$ is in [even]. For option (i) the matrix $\gamma_\perp^\alpha$ can be moved to the far left using $\gamma_\perp^{\mu_i}\gamma_\perp^\alpha=2 g_\perp^{ \mu_i\alpha}- \gamma_\perp^\alpha\gamma_\perp^{\mu_i}$. The terms with $g_\perp^{\mu_i\alpha}$ have $N-2$ $\gamma_\perp$ and are proportional to 
$\gamma_\perp^\alpha \gamma_\perp^\nu $ by assumption, leaving a term $\gamma_\perp^\alpha$[even]$\gamma_\perp^\nu $[even]$\propto \gamma_\perp^\alpha \gamma_\perp^\nu$ (since  [even]$\gamma_\perp^\nu $[even]$\propto \gamma_\perp^\nu$). In the case (ii)  $\gamma_\perp^\alpha$ is moved to the right. This leads to terms with $N-2$ gamma matrices, which are proportional to $\gamma_\perp^\alpha\gamma_\perp^\nu$ by assumption, and a term
[odd]$\gamma_\perp^\nu$[odd]$\gamma_\perp^\alpha$, which is zero as shown above. 

Finally let us discuss the matching in \eqref{Tprod} with fermion fields contracted. Generally, this leads to an operator 
of the form
\beq\label{Tfermion}
\delta^{ab}\big[\gamma_\perp^\beta \sla n\prod_i (\sla \bar n \gamma_\perp^{\mu_i} \gamma_\perp^{\mu_{i+1}} \sla n )P_{R,L}]^{ji}
{\cal B}_{\perp n -\bar x\omega}^{\mu A} {\cal B}_{\perp n -x\omega}^{\nu A},
\eeq
where one of the Lorentz indices is equal to $\alpha$, while the others are contracted pairwise. This general operator
can always be written as a linear combination of operators
\begin{align}
\label{Op1}
&\delta^{ab}\big[\gamma_\perp^\alpha \sla n P_{R,L}]^{ji}
{\cal B}_{\perp n -\bar x\omega}^{ A} \cdot {\cal B}_{\perp n -x\omega}^{ A},\\
&\delta^{ab}\big[ \sla {\cal B}_{\perp n -\bar x\omega}^{ A} \sla n P_{R,L}]^{ji}
 {\cal B}_{\perp n -x\omega}^{A\alpha}, \label{Op2}\\
&\delta^{ab}\big[ \sla {\cal B}_{\perp n - x\omega}^{ A} \sla n P_{R,L}]^{ji}
 {\cal B}_{\perp n -\bar x\omega}^{A\alpha}. \label{Op3}
\end{align}
This is trivially satisfied for only $\gamma_\perp^\beta$ in \eqref{Tfermion} without additional $\gamma_\perp$ insertions. It is also true in general, since
the Dirac structure  
[any]$\gamma_\perp^\nu$[any]$\gamma_\perp^\mu$[any]$\gamma_\perp^\alpha$[any], where [any] 
is a product of an arbitrary number of $\gamma_\perp$ matrices, with the unshown Lorentz indices contracted, is a sum of 
$\gamma_\perp^\nu\gamma_\perp^\mu\gamma_\perp^\alpha$ and the similar terms with permuted Lorentz indices. Decomposition
in terms of operators \eqref{Op1}-\eqref{Op3} then follows from 
\beq
\gamma_\perp^\nu\gamma_\perp^\mu\gamma_\perp^\alpha=g_\perp^{\mu\nu}\gamma_\perp^\alpha-g_\perp^{\nu\alpha}\gamma_\perp^\mu+g_\perp^{\mu\alpha}\gamma_\perp^\nu.
\eeq
Since \eqref{Op2} and \eqref{Op3} differ merely by $x \leftrightarrow\bar x$ interchange, only one of the two is needed after 
integration over $x$ in \eqref{jet-funcs}.

\section{SU(3) decomposition}\label{SU3decomposition}
In the limit $m_s, m_{u,d}\ll \Lambda$ a useful approach is to use the transformation 
properties of the weak Hamiltonian  \eqref{HW} under flavor SU(3) and decompose the decay amplitudes in terms of reduced matrix elements \cite{zeppenfeld}. The diagrammatic approach of Refs. \cite{Gronau:1994rj,Dighe:1995bm}
is equivalent to the SU(3) decomposition of the amplitudes (for recent applications see e.g. \cite{Baek:2004rp,Chiang:2003pm,Soni:2005ah}). 
It is usually followed, however,
by further dynamical assumptions, with annihilation and exchange topologies neglected \cite{Chiang:2003pm}. For 
nonisosinglet final states this assumption can be justified using SCET, since the reduced matrix elements corresponding to the two topologies are found to be $1/m_b$ suppressed \cite{Bauer:2004tj,Bauer:2005kd,Bauer:2004ck}. As we will show
in this appendix, the annihilation and exchange topologies, on the other hand, cannot be neglected for isosinglet final states since
they come as leading order contributions in $1/m_b$ expansion in SCET.
In this appendix we also provide the translation between our LO SCET results \eqref{LO} 
and the 
diagrammatic language. All the results will be given assuming exact SU(3), using thus the relations \eqref{zetaSU3}, \eqref{zetaSU3eta}, \eqref{AccSU3}, \eqref{Accpieta} along with a similar relation for the decay
constants $f_M=f_\pi=f_K=f_{\eta_q}=f_{\eta_s}$. 

\begin{table*}
\caption{The SU(3) decomposition of $\Delta S=0$ (above horizontal line) and $\Delta S=1$ decays (below horizontal line)
into final states with $\eta_0$. Each amplitude should be divided by the common denominator in the "Factor" column, so that for instance $A_{B^-\to\pi^-\eta_0}=(2{\cal C}_3+{\cal C}_6+3{\cal C}_{15}+2{\cal A}_6+6{\cal A}_{15}+3{\cal E}_3+3{\cal D}_6+9{\cal D}_{15})/{\sqrt{3}}=(c+t+2p+t_s+p_s)/{\sqrt{3}}$. The 
"Diagrammatic" column shows the decomposition in the diagrammatic approach notation, with ${\cal A}_{3,6,15}$ neglected. }\label{SU3table}
\begin{ruledtabular}
\begin{tabular}{lrrrrrrrrrrcl} 
Mode  & ${\cal C}_3$ & ${\cal C}_6$&${\cal C}_{15}$& ${\cal A}_3$ & ${\cal A}_6$&${\cal A}_{15}$&${\cal E}_3 $&${\cal D}_3$&${\cal D}_6$&${\cal D}_{15}$& Factor & Diagrammatic \\ \hline
$B^-\to\pi^-\eta_0$ & $2 $ & $1$ & $3$ & $0$ &$2$ & $6$ &$3$ & $0$ & $3$ & $9$ & $\sqrt{3}$ & $t+c+2p+t_s+p_s$
\\
$\bar B^0\to\pi^0\eta_0$ & $-2 $ & $-1$ & $5$ & $0$ &$-2$ & $10$ &$-3$ & $0$ & $-3$ & $15$ & $\sqrt{6}$ & $-2p+c_s-p_s $
\\
$\bar B^0\to\eta_8\eta_0$ & $-2 $ & $3$ & $-3$ & $0$ &$6$ & $-6$ &$-3$ & $0$ & $9$ & $-9$ & $3\sqrt{2}$ & $-(2c+2p+c_s+p_s) $
\\
$\bar B^0\to\eta_0\eta_0$ & $2 $ & $0$ & $0$ & $6$ &$0$ & $0$ &$6$ & $18$ & $0$ & $0$ & $3$ & $ 2(c+p+c_s+p_s+s_0) $
\\
$\bar B_s^0\to K^0\eta_0$ & $2 $ & $-1$ & $-1$ & $0$ &$-2$ & $-2$ &$3$ & $0$ & $-3$ & $-3$ & $\sqrt{3}$ & $c+2p+p_s $
\\
\hline
$B^-\to K^-\eta_0$ & $2 $ & $1$ & $3$ & $0$ &$2$ & $6$ &$3$ & $0$ & $3$ & $9$ & $\sqrt{3}$ & $ t'+c'+2p'+t'_s+p'_s$
\\
$\bar B^0\to \bar K^0 \eta_0$ & $2 $ & $-1$ & $-1$ & $0$ &$-2$ & $-2$ &$3$ & $0$ & $-3$ & $-3$ & $\sqrt{3}$ & $ c'+2p'+p'_s $
\\
$\bar B_s^0\to \pi^0 \eta_0$ & $0 $ & $-2$ & $4$ & $0$ &$-4$ & $8$ &$0$ & $0$ & $-6$ & $12$ & $\sqrt{6}$ & $ c'+c'_s$
\\
$\bar B_s^0\to \eta_8 \eta_0$ & $4 $ & $0$ & $-6$ & $0$ &$0$ & $-12$ &$6$ & $0$ & $0$ & $-18$ & $3\sqrt{2}$ & $ c'+4p'-c'_s+2p'_s $
\\
$\bar B_s^0\to \eta_0 \eta_0$ & $2 $ & $0$ & $0$ & $6$ &$0$ & $0$ &$6$ & $18$ & $0$ & $0$ & $3$ & $2(c'+p'+c'_s+p'_s+s'_0) $
\\
\end{tabular}
\end{ruledtabular}
\end{table*}

The effective weak Hamiltonian  \eqref{HW} transforms under flavor SU(3) as $\bar 3\otimes 3\otimes \bar 3=\bar 3\oplus
\bar 3\oplus 6\oplus \overline{15}$, so that it can be decomposed in terms of a vector $H^i(3)$, a traceless tensor antisymmetric in upper indices, $H^{[ij]}_k(6)$, and a traceless tensor symmetric in the upper indices, $H^{(ij)}_k(15)$.
We further define a vector of $B$ fields $B_i=(B^+,B^0, B_s^0)$ and a matrix of light pseudoscalar fields
\beq\label{M}
M^i_j=\begin{pmatrix}
\frac{\pi^0}{\sqrt{2}}-\frac{\eta_8}{\sqrt{6}}&\pi^- & K^-\\
\pi^+&-\frac{\pi^0}{\sqrt{2}}-\frac{\eta_8}{\sqrt{6}}&\bar{K^0}\\
K^+&K^0 &\sqrt{\frac{2}{3}}\eta_8
\end{pmatrix}+\frac{\eta_0}{\sqrt 3} \openone,
\eeq
where the SU(3) singlet $\eta_0\sim (u\bar u+d\bar d+s\bar s)/\sqrt 3$ and octet
$\eta_8\sim (2 s\bar s-u\bar u-d \bar d)/\sqrt 6$ admixtures of $\eta_{q,s}$ are used, which is the natural choice in 
unbroken SU(3). The most general Hamiltonian that has the same transformation properties under SU(3) as the weak 
Hamiltonian in \eqref{HW}
is then a sum of terms that contribute to both SU(3) singlet and octet final states
\beq\label{nonisodecomp}
\begin{split}
&{\cal A}_3 B_i H^i(3)M_k^jM_j^k+{\cal C}_3 B_i M^i_j M^j_k H^k(3)\\
+&{\cal A}_6B_i H^{ij}_k(6)M_j^lM_l^k+{\cal C}_6 B_i M^i_j H^{jk}_l(6) M_k^l \\
+&{\cal A}_{15} B_i H^{ij}_k(15) M^l_j M_l^k +{\cal C}_{15} B_i M^i_j H^{jk}_l (15) M_k^l,
\end{split}
\eeq
and terms that are nonzero only if SU(3) singlet is in the final state
\beq\label{isodecomp}
\begin{split}
&{\cal E}_3 B_i M^i_j H^j(3) M^k_k+{\cal D}_3 B_i H^i(3) M^j_j M_k^k\\
+&{\cal D}_6 B_i H^{ij}_k(6) M^k_j M^l_l +D_{15} B_i H^{ij}_k (15) M^k_j M^l_l,
\end{split}
\eeq
where for $\Delta S=0$ decays
\begin{align}
H^2(3)&=1,\label{H3}\\
H^{12}_1(6)&=-H^{21}_1(6)=H^{23}_3(6)=-H^{32}_3(6)=1,\\
\begin{split}
2H^{12}_1(15)&= 2 H^{21}_1(15)=-3H^{22}_2(15)=\\
&=-6H^{23}_3(15)=-6H^{32}_3(15)=6,\label{H15}
\end{split}
\end{align}
with the remaining entries zero, while for $\Delta S=1$ decays the nonzero entries in $H^i(3)$,
$H^{ij}_k(6)$, $H^{ij}_k(15)$ are obtained from \eqref{H3}-\eqref{H15} with the replacement $2\leftrightarrow 3$.
Note that since the final state $|PP\rangle$ is symmetric, there are only 9 reduced matrix elements in $B\to PP$ \cite{zeppenfeld}. Namely, the coefficients ${\cal C}_6, {\cal A}_6, {\cal D}_6$ in the above decompositions \eqref{nonisodecomp}, \eqref{isodecomp} always appear in the combinations ${\cal C}_6 -{\cal A}_6$ and ${\cal D}_6+{\cal A}_6$.

In the 
diagrammatic approach the linear combinations of reduced matrix elements ${\cal C}_{3,6,15}$ are redefined as $t,c,p$ amplitudes
\begin{align}
t&=2{\cal C}_6+4{\cal C}_{15},\\
c&=-2{\cal C}_6+4{\cal C}_{15},\\
p&={\cal C}_3-{\cal C}_6-{\cal C}_{15},
\end{align}
while ${\cal A}_{3,6,15}$, leading to amplitudes $e,a,pa$ in the diagrammatic notation, are usually neglected. This dynamical assumption can be justified using SCET, where  ${\cal A}_{3,6,15}$ are found to be $1/m_b$ suppressed \cite{Bauer:2004tj,Bauer:2005kd,Bauer:2004ck}, while the remaining amplitudes are at LO in $1/m_b$ and $\alpha_S(m_b)$ 
\begin{align}
t&=\frac{G_F}{\sqrt 2} m_B^2 f_M\big[ b_1^{(d)}\zeta_J+ c_1^{(d)}\zeta\big],\label{tdiagr}\\
c&=\frac{G_F}{\sqrt 2} m_B^2 f_M\big[(b_2^{(d)}-b_3^{(d)})\zeta_J + (c_2^{(d)}-c_3^{(d)})\zeta\big],\\
p&=\frac{G_F}{\sqrt 2} m_B^2 \big[f_M( b_4^{(d)}\zeta_J+ c_4^{(d)}\zeta)+\lambda_c^{(d)} A_{cc}\big].\label{pdiagr}
\end{align}
The Wilson coefficients $b_i^{(d)}$ are here understood to be already convoluted with light pseudoscalar LCDA. In the SU(3) limit this amounts to a replacement 
$m_b/\omega_2=-m_b/\omega_3=\langle x^{-1}\rangle_\pi= \langle x^{-1}\rangle_K=\langle x^{-1}\rangle_\eta\simeq 3$ in
Eq. \eqref{b8}. The amplitudes for $\Delta S=1$ transitions are obtained from \eqref{tdiagr}-\eqref{pdiagr} with $b_i^{(d)}\to b_i^{(s)}$, $c_i^{(d)}\to c_i^{(s)}$.

The complete SU(3) decomposition of amplitudes for decays not containing $\eta_0$ is given in Ref. \cite{Dighe:1995bm} both in terms of reduced matrix elements ${\cal A}_{3,6,15},{\cal C}_{3,6,15}$ as well as in terms of diagrammatic amplitudes and will thus not be repeated here. (In  Ref. \cite{Dighe:1995bm} a different phase conventions was used, so the replacements $\pi^0\to -\pi^0, \pi^-\to -\pi^-, K^-\to -K^-$ need to be made, while the amplitudes into two indistinguishable states should be multiplied by $\sqrt{2}$ to have our normalization of the amplitudes. In addition
we find the "Factor" in Table 2. of \cite{Dighe:1995bm} for $B_s \pi^0\eta_8$ to be $-\sqrt{3}$.) The complete SU(3) decomposition of amplitudes for decays into SU(3) singlets, on the other hand, is provided in Table \ref{SU3table}.

The reduced matrix elements ${\cal E}_3$ and ${\cal D}_{3,6,15}$ describing the decays into SU(3) singlet final 
states are found in SCET to be all nonzero already at leading order in $1/m_b$ and $\alpha_S(m_b)$. 
Defining the singlet "diagrammatic" amplitudes
\begin{align}
t_s&=6({\cal D}_6+2 {\cal D}_{15}),\\
c_s&=-6({\cal D}_6-2 {\cal D}_{15}),\\
p_s&=3({\cal C}_6-{\cal C}_{15}-{\cal D}_6-{\cal D}_{15}+{\cal E}_3),\\
s_0&=9({\cal D}_3+{\cal D}_6-{\cal D}_{15}),
\end{align}
we find at leading order in $1/m_b$ and $\alpha_S(m_b)$
\begin{align}
t_s=&\frac{G_F}{\sqrt 2} m_B^2 3f_M\big[b_1^{(d)}\zeta_{Jg}+c_1^{(d)}\zeta_g\big],\nonumber
\\
c_s=&\frac{G_F}{\sqrt 2} m_B^2 3f_M\big[(b_2^{(d)}-b_3^{(d)})\zeta_{Jg}+(c_2^{(d)}-c_3^{(d)})\zeta_g\big],\nonumber
\end{align}
\begin{align}
\begin{split}
p_s=&\frac{G_F}{\sqrt 2} m_B^2 3\big[f_M\big((b_5^{(d)}-b_6^{(d)})\zeta_{J}+(c_5^{(d)}-c_6^{(d)})\zeta_{}\big)\nonumber
\\
&+f_M\big(b_4^{(d)}\zeta_{Jg}+c_4^{(d)}\zeta_{g}\big) +\lambda_c^{(d)} A_{ccg} \big],\nonumber
\end{split}
\\
s_0=&\frac{G_F}{\sqrt 2} m_B^2 9f_M\big[(b_5^{(d)}-b_6^{(d)})\zeta_{Jg}+(c_5^{(d)}-c_6^{(d)})\zeta_g\big],\label{SCETts}
\end{align}
where $s_0$ contributes only to $\eta_0\eta_0$ decays. The $\Delta S=1$ amplitudes $t_s',c_s',p_s',s_0'$ are
obtained by replacing $d\to s$ in \eqref{SCETts}. 
Note that these amplitudes are arising from gluon
content of isosinglet final states and correspond to diagrams on Fig. \ref{matching} b) and \ref{matching} d) and on Fig. \ref{ggsgluon}, 
with $p_s$ receiving also nongluonic contributions.

In the applications of SU(3) decomposition using the diagrammatic approach it is frequently assumed that
only one reduced matrix element, ${\cal E}_3$, is nonzero, while ${\cal D}_{3,6,15}$ are assumed to be suppressed 
\cite{Dighe:1995bm,Chiang:2003pm}. This corresponds to taking $p_s$ nonzero, while neglecting $t_s,c_s$ and $s_0$ (commonly $s=p_s/3$ is introduced instead of $p_s$). In the SCET result \eqref{SCETts} this would correspond to a limit $\zeta_{(J)g}\ll \zeta_{(J)}$, while from
SCET scaling of the operators we expect $\zeta_{(J)g}\sim \zeta_{(J)}$. At present both possibilities are still allowed by experiment, Eqs. \eqref{zetagfit}, \eqref{zetag-fit}, \eqref{zetagfitII}, \eqref{zetag-fitII}, but nonzero values of $\zeta_{g}+\zeta_{Jg}$ are preferred. 
More precisely, in Solution I $\zeta_{Jg}+\zeta_g$ cannot be zero, while it is still consistent with zero at $\sim 1.5\sigma$ in Solution II (the poorly constrained orthogonal combination $\zeta_{Jg}-\zeta_g$ is consistent with zero in both cases).



\end{document}